\documentclass[traditabstract]{aa}  
\usepackage{nameref}
\usepackage[switch]{lineno}
\usepackage{graphicx}
\usepackage{natbib}
\usepackage{amsmath} 

\newcommand{\kms}{$\rm{\,km \,s}^{-1}$}
\usepackage[english]{babel}
\usepackage[utf8]{inputenc}
\usepackage{amssymb}
\usepackage{xcolor}
\makeatletter

\newcommand{\Rmnum}[1]{\expandafter\@slowromancap\romannumeral #1@}
\makeatother
\usepackage[nottoc]{tocbibind}
\usepackage{txfonts}
\usepackage{verbatim} 
\usepackage[colorlinks=true,linkcolor=bluea,allcolors=blue]{hyperref}
\usepackage[font=normalsize,labelfont={bf}]{caption}

\begin{document} 

\subtitle{
  Warm molecular and ionized gas kinematics in the Type-2 quasar J0945+1737}
  \author{G. Speranza\inst{1,2}
  		\and C. Ramos Almeida\inst{1,2} 
		\and J. A. Acosta-Pulido\inst{1,2}
		\and R. A. Riffel\inst{3}
		\and C. Tadhunter\inst{4}
		\and J. C. S. Pierce\inst{5}
		\and A. Rodríguez-Ardila\inst{6,7}
		\and M. Coloma Puga\inst{2}
	    \and M. Brusa\inst{8,9}
	    \and B. Musiimenta\inst{8,9}
		\and D. M. Alexander\inst{10}
		\and A. Lapi\inst{11}
		\and F. Shankar\inst{12}
		\and C. Villforth\inst{13}
		}
\institute{Instituto de Astrofísica de Canarias, Calle Vía Láctea, s/n, E-38205, La Laguna, Tenerife, Spain;\\
\email giovanna.speranza@iac.es 
\and Departamento de Astrofísica, Universidad de La Laguna, E-38206 La Laguna, Tenerife, Spain
\and Departamento de Física, CCNE, Universidade Federal de Santa Maria, Santa Maria 97105-900, RS, Brazil
\and Department of Physics and Astronomy, University of Sheffield, Sheffield S3 7RH, UK
\and Centre for Astrophysics Research, University of Hertfordshire, College Lane, Hatfield AL10 9AB, UK
\and Laboratorio Nacional de Astrofisica,
R. dos Estados Unidos, CEP 37504-364, Itajub - MG, Brazil
\and Instituto Nacional de Pesquisas Espaciais,
Av. dos Astronautas, CEP 12227-010, So Jos dos Campos - SP, Brazil
\and Dipartimento di Fisica e Astronomia "Augusto Righi", Università di Bologna, via Gobetti 93/2, 40129 Bologna, Italy
\and INAF - Osservatorio di Astrofisica e Scienza dello Spazio di Bologna, via Gobetti 93/3, 40129 Bologna, Italy
\and Centre for Extragalactic Astronomy, Department of Physics, Durham University, South Road, Durham, DH1 3LE, UK
\and SISSA, Via Bonomea 265, 34136 Trieste, Italy
\and Department of Physics and Astronomy, University of Southampton, Highfield, SO17 1BJ, UK
\and University of Bath, Department of Physics, Claverton Down, Bath, BA2 7AY, UK
}

   \date{}

  \abstract {We analyze Near-Infrared Integral Field Spectrograph (NIFS) observations of the type-2 quasar (QSO2) SDSS J094521.33+173753.2 to investigate its warm molecular and ionized gas kinematics. This QSO2 has a bolometric luminosity of 10$^{45.7}$ erg s$^{-1}$ and a redshift of z = 0.128. The K-band spectra provided by NIFS cover a range of 1.99-2.40 $\mu$m where low-ionization (Pa$\alpha$ and Br$\delta$), high ionization ([S XI]$\lambda$1.920 $\mu$m and [Si~VI]$\lambda$1.963 $\mu$m) and warm molecular lines (from H$_2$ 1-0S(5) to 1-0S(1)) are detected, allowing us to study the multi-phase gas kinematics. Our analysis reveals gas in ordinary rotation in all the emission lines detected and also outflowing gas in the case of the low- and high-ionization emission lines. In the case of the nuclear spectrum, which corresponds to a circular aperture of 0.3\arcsec~(686 pc) in diameter, the warm molecular lines can be characterized using a single Gaussian component of full width at half maximum (FWHM)= 350--400 \kms, while Pa$\alpha$, Br$\delta$, and [Si~VI] are best fitted with two blue-shifted Gaussian components of FWHM$\sim$800 and 1700 \kms, in addition to a narrow component of $\sim$300 \kms. We interpret the blue-shifted broad components as outflowing gas, which reaches the highest velocities, of up to $-$840 \kms, in the south-east direction (PA$\sim$125$^{\circ}$), extending up to a distance of $\sim$3.4 kpc from the nucleus. The ionized outflow has a maximum mass outflow rate of $\dot{\text{{M}}}_{\text{{out, max}}}$=42-51 M$_\odot$ yr$^{-1}$, and its kinetic power represents 0.1\% of the quasar bolometric luminosity. 
 VLA data of J0945 show extended radio emission (PA$\sim$100$^{\circ}$) that is aligned with the clumpy emission traced by the narrow component of the ionized lines up to scales of several kpc, and with the innermost part of the outflow (central $\sim$0.4\arcsec= 915 pc). Beyond that radius, at the edge of the radio jet, the high-velocity gas shows a different PA, of $\sim$125\degr. This might be indicating that the line-emitting gas is being compressed and accelerated by the shocks generated by the radio jet. 
 }


\maketitle

\begin{figure*}
\centering
\includegraphics[width=0.439\textwidth]{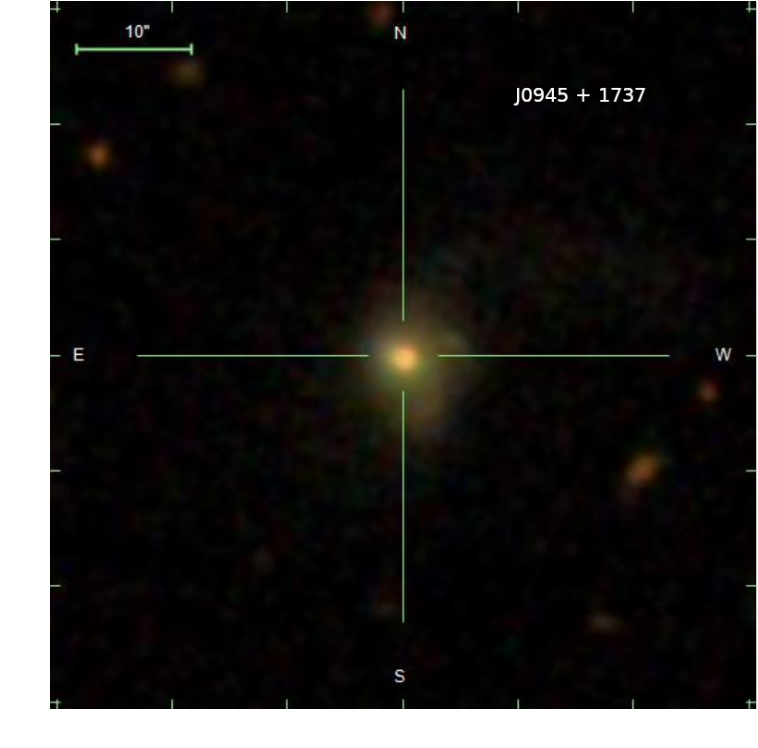}
\includegraphics[width=0.445\textwidth]{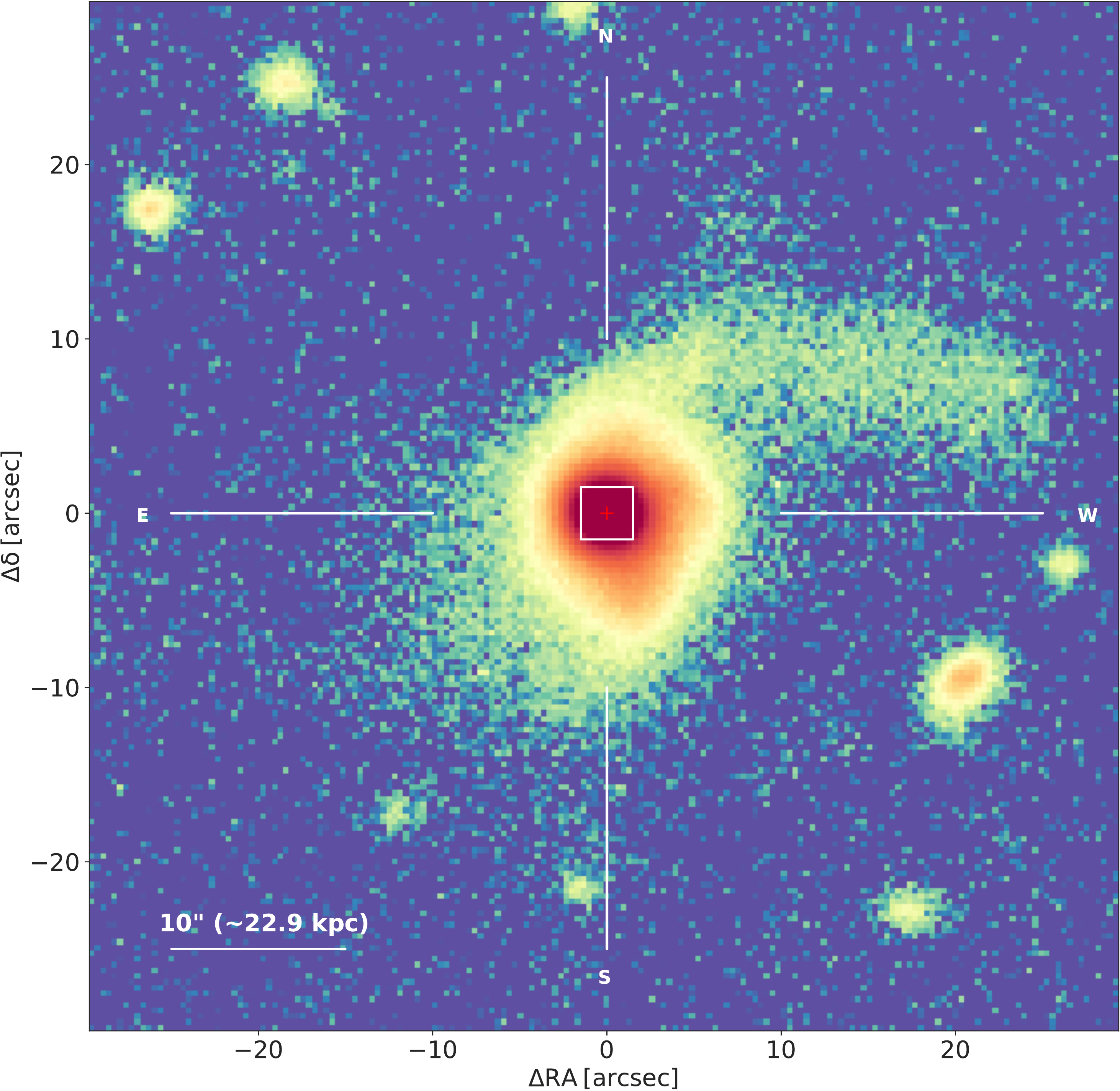}
\caption {Optical images of J0945. Left panel shows the colour-combined SDSS image with the corresponding scale. Right panel shows a deeper R-band image obtained with the WFC on the INT. The field-of-view (FOV) of both panels is 60\arcsec $\times$\,60\arcsec~($\sim 137 \times 137$ kpc$^2$).  The white square on the right panel represents the FOV of Gemini/NIFS (3\arcsec $\times$ 3\arcsec $\sim 6.9 \times 6.9$ kpc$^2$).}
\label{fig:combined_im}
\end{figure*}

\section{Introduction}
\label{introduction}

The growth of supermassive black holes (SMBHs) located at the centres of massive galaxies, i.e., active galactic nuclei (AGN), has been proved to have an impact in galaxy evolution. In particular, AGN feedback now represents a fundamental ingredient for regulating star formation in galaxies by heating, removing, and/or disrupting the gas available to form stars (e.g., \citealt{Fabian12,Peng15,2021Natur.594..187M}). Moreover, the implementation of AGN feedback is now included in cosmological simulations of galaxy formation, as it is required, among other things, to reproduce the high-mass end of the luminosity function of galaxies (\citealt{Granato04,Springel05,Croton06,Lapi06,2014ApJ...782...69L}).

  There are two main channels through which AGN feedback acts: the radio- or kinetic-mode, and the quasar- or radiative-mode. The first type of feedback is generally, but not always,  associated with powerful radio sources characterized by radiatively inefficient accretion, where the jets inject energy in the galaxy halos, preventing gas from cooling (\citealt{2007ARA&A..45..117M}). At higher accretion efficiencies, the quasar-mode is expected to be dominant, with AGN-driven winds of ionized, neutral, and molecular gas being produced by radiation pressure on the circumnuclear gas (\citealt{Fabian12}).

In type-2 quasars (QSO2s; i.e., having L$_{[\text{O~III}]}>10^{8.5}\text{L}_{\odot}$; \citealt{Reyes08}), quasar-mode feedback is expected to dominate. However, it would be an over-simplification to consider such a clear distinction between the two types of feedback. In fact, low-power jets (P$_{\rm jet}\sim$10$^{44}$ erg~s$^{\rm -1}$) might drive multi-phase outflows in radio-quiet AGN, including QSO2s (see \citealt{Jarvis19,Morganti21,2022MNRAS.tmp..170G} and references therein), and radiativelly-driven winds are observed in radio-loud sources (e.g., \citealt{Speranza21}).
To disentangle whether jets, radiatively-driven winds, or both are responsible for driving gas outflows in quasars, it is necessary to study each case separately, focusing on the stellar, gas and radio morphologies (using e.g., data in the optical, near-infrared and radio) and on various outflow properties, including geometry, extent, density, and kinetic power. 

In this context, QSO2s represent excellent laboratories to search for outflows and evaluate their impact on the host galaxies. Thanks to nuclear obscuration of the AGN continuum and the broad components of the permitted lines produced in the broad-line region (BLR), we can easily identify broad lines associated with outflowing gas and also characterize the underlying stellar populations (\citealt{Bessiere17, Bessiere22}). 

Several studies have been conducted focusing on ionized outflows, confirming that they are almost ubiquitous in QSO2s at z$\la$0.8 (\citealt{Liu13,2013MNRAS.433..622M,Harrison14,2014MNRAS.440.3202V,2014MNRAS.442..784Z,Fischer18}). 
However, the ionized component represents just one gas phase of the outflows (\citealt{2018NatAs...2..176C}). To accurately quantify the impact of outflows on the host galaxies and understand how feedback works in AGN it is important to consider all the outflow phases (see e.g. \citealt{2021MNRAS.505.5753F}). In this context, the near-infrared (NIR), and in particular the K-band ($\sim$2.0--2.4 $\mu$m), brings to light observations of nearby targets from which the warm molecular and ionized gas phases can be traced simultaneously. This can be done by means of several H$_2$ lines, hydrogen recombination lines, and the coronal line [Si~VI]$\lambda$1.963 $\mu$m, which is unequivocally associated with AGN activity given its high ionization potential (167 eV).
Despite this, to the best of our knowledge,  these NIR spectral features have
been studied in only four nearby QSO2s (z $\leq$ 0.1) to date: F08572+3915:NW (\citealt{Rupke13}), Mrk\,477 (\citealt{Villar15}), J1430+1339, also know as the Teacup, (\citealt{Ramos17}), and J1509+0434 (\citealt{Ramos19}). These works showed that the outflow properties are different in the two gas phases. For example, ionized outflows reach higher velocities than the warm molecular outflowing gas. In addition, there are cases, like the Teacup (\citealt{Ramos17}), in which the warm molecular outflow is not detected, unlike its cold molecular counterpart (\citealt{Ramos21}). This likely happens because the bulk of the warm molecular gas is usually following circular rotation in the galaxy plane, whereas ionized gas is more easily disrupted, leaving the galaxy plane and showing non-circular motions (see e.g., \citealt{Riffel21} and references therein).  


Here we explore for the first time the NIR spectrum of the QSO2 SDSS J094521.33+173753.2 (J0945 hereafter) using seeing-limited integral field observations. More details on this source are provided in Section \ref{source}.


\section{SDSS J094521.33+173753.2 (J0945)}
\label{source}

J0945 is one of the 48 QSO2s with z$<$0.14 and L$_{\rm [OIII]}>10^{8.5}L_{\sun}$ (L$_{\rm bol}>10^{45.6}$ erg~s$^{-1}$, using the bolometric correction of 3500 from \citealt{2004ApJ...613..109H}) in the Quasar Feedback (QSOFEED\footnote{http://research.iac.es/galeria/cra/qsofeed/}) sample (\citealt{Ramos19,Ramos21}). This sample is drawn from \citet{Reyes08}, one of the largest compilations of narrow emission line AGN. J0945 has a bolometric luminosity L$_\text{{bol}}$ = 10$^{45.7}$ erg~s$^{-1}$, as determined from spectral energy distribution fitting (\citealt{Jarvis19}) and a redshift z=0.128. 

The galaxy appears to be disk-dominated and it shows a tidal tail extending up to $\sim$25\arcsec ($\sim$57 kpc) north-west of the nucleus. This structure is not observed in the colour-combined SDSS image (see left panel of Fig.~\ref{fig:combined_im}), but it is clearly visible in the deeper image obtained with the Wide Field Camera (WFC) on the Isaac Newton Telescope (INT), at the Roque de Los Muchachos Observatory, in La Palma (see right panel of Fig.~\ref{fig:combined_im} and \citealt{2022MNRAS.510.1163P}). There is also a shorter tidal tail extending in the opposite direction, to the south-east.
The galaxy is moderately inclined, i$\sim$44$^{\circ}$, as measured from the isophotal major and minor axes from r-band SDSS DR6 photometry (2007). 

J0945 shows a single nucleus based upon Hubble Space Telescope (HST) observations (\citealt{Cui01}) performed using the Wide Field Planetary Camera 2 (WFPC2) in the I-band filter F814W. Nonetheless, \citet{Storchi18} reported a disrupted galaxy morphology from optical images obtained with the HST Advanced Camera for Surveys (ACS).  They used the narrow-band filters FR716N and FR931N, centered on the redshifted [O~III]$\lambda$5007 and H$\alpha$+[N~II] emission, as well as a broader filter (either F775W or FR647M), centered in the continuum between these two sets of lines. These images show a cloud of line-emitting gas at  $\sim$5\arcsec~(11.5 kpc) north-west of the nucleus, which \citet{Storchi18} interpreted as either reminiscent of a merger or produced by a previous outflow episode. Based on the [O~III] line-emitting gas, they report a radius of 8.4 kpc for the extended narrow-line region (ENLR), with a PA$\sim90^{\circ}$.

\citet{Jarvis20} reported a stellar mass M$_{\star} \approx 10^{10.1} \text{M}_{\odot}$ and a star formation rate (SFR) $\approx$73 M$_{\odot} \,\text{yr}^{-1}$ for J0945 (uncertainties of $\sim$0.3 and 0.42 dex, respectively). They were measured from SED fitting,  using different templates, including the AGN contribution, and the ``Code Investigating GALaxy Emission'' (CIGALE\footnote{\url{https://cigale.lam.fr/}; \citealt{Burgarella05, Noll09, Boquien19})}. This stellar mass is much lower than that estimated from the K-band magnitude of the QSO2, of M$_{\star} \approx 10^{11.0} \text{M}_{\odot}$ (see \citealt{Ramos21} for details on the stellar mass determination\footnote{ Both \citet{Jarvis20} and \citet{Ramos21}  adopted the \citet{Chabrier03} initial mass function (IMF).}). Despite the large uncertainties ($\sim$0.3 dex in the case of the SED fitting value and $\sim$0.2 dex in the case of the K-band value), this might be indicative of either an overestimation of the AGN component in the SED fit or a strong AGN contribution to the K-band flux. The stellar mass estimated from the K-band flux is in better agreement with the average values found for the QSOFEED sample (10$^{11.1\pm{0.2}} \text{M}_{\odot}$) and from the sample of 86 QSO2s at z$<$0.5 studied by \citet{2019ApJ...873...90S}, the latter obtained from SED fitting. On the other hand, the SFR derived from SED fitting is in agreement with the value estimated from the far-infrared luminosity of the QSO2, of 84 M$_{\odot} \,\text{yr}^{-1}$.

A black hole mass of 10$^{8.27\pm0.79}$M$_{\sun}$ and an Eddington ratio of 10$^{-0.27\pm0.80}$ were reported by \citet{2018ApJ...859..116K} for J0945 using SDSS spectra. These values are similar to the median values of M$_{\rm BH}$=10$^{8.2}M_{\sun}$ and $L_{\rm bol}/L_{\rm Edd}$=10$^{-0.7}$ reported by \citet{2018ApJ...859..116K} from 669 of the QSO2s in \citet{Reyes08}. These QSO2s are near-Eddington to Eddington-limit obscured AGN in the local universe.


According to its 1.4 GHz and [O~III] luminosities ($\text{L}_{1.4\text{GHz}}$ = 10$^{24.3}~\text{W Hz}^{-1}$ and L$_{\rm [O~III]}=10^{42.7}$ erg~s$^{-1}$, from \citealt{2021MNRAS.tmp..644J}), J0945 is classified as radio-quiet  (\citealt{Xu99,Lal10}). However, in some radio-quiet sources, compact radio jets are observed at
high angular resolution (e.g. \citealt{Gallimore06, Baldi18}). Indeed, for J0945, Very Large Arrey (VLA) imaging at $\sim0.25$\arcsec~resolution revealed a jet-like structure not related to star formation (HR-A and HR-B in \citealt{Jarvis19}). This extended radio structure has a steep spectral index ($\alpha=-0.8$) that indicates that most likely corresponds to synchrotron emission. In fact, J0945 is a ``radio-excess source'' according to its position in the infrared luminosity--radio luminosity diagram (see Fig. 3 in \citealt{Jarvis19}). These authors reported, for J0945 and other QSO2s, that the radio and [O~III]$\lambda$5007 emission are often co-spatial. 
From this they concluded that compact radio jets with modest radio luminosities can influence the ionized gas kinematics and hence, constitute an important feedback mechanism in radio-quiet AGN, including J0945. 

 
Regarding the emission line kinematics, high velocity wings were detected in the [O~III]$\lambda$5007 emission lines in different datasets. Using a non-parametric analysis, \citet{Harrison14} reported broad and asymmetric profiles (e.g., velocities  higher than $-$1500 \kms~for the second velocity percentile, v$_{02}$, and widths of W$_{80}\sim$ 1000 \kms) using Gemini-south GMOS integral field data. They observed these blue-shifted wings across the whole field-of-view (FOV; $\approx 8 \times 11$ kpc),  but brighter in the central region and preferentially located along the north–south axis. Using the same data, \citet{Oliveira21} performed a parametric fitting of the emission lines that revealed that three Gaussian components are needed to model the [O~III] profiles. They detected a narrow component associated with the ordinary rotating gas, and two blue-shifted (velocities larger than $-$200 \kms) and broad (full width at half maximum, FWHM$\sim$700 \kms and $\sim$2100\kms) components. In addition, they find that the velocities of the broad components increase along the north–south axis (see Fig.~A.1 in \citealt{Oliveira21}), in agreement with \citet{Harrison14}.  
 


With the aim of studying the ionized and warm molecular gas kinematics of J0945, an interesting example of radio-quiet QSO2 with extended radio emission,
here we explore new seeing limited (0.29\arcsec= 660 pc) K-band integral field spectroscopic data taken with the Gemini-North Telescope. The paper is organized as follows: in Section~\ref{observations} we describe the NIR observations and data reduction. In Section~\ref{analysis} we describe the analysis of the nuclear and extended emission and present the results, including the measured outflow properties and molecular gas content. In Section~\ref{discussion} we discuss the implications of the results, which are summarized in Section~\ref{conclusion}. 
Throughout this work we assume the following cosmology: H$_0 = 70.0 ~ \mbox{km s}^{-1}~ \mbox{Mpc}^{-1}$, $\Omega_M = 0.3$ and $\Lambda = 0.7$. This results in a corrected scale of 2.287 \mbox{kpc arcsec}$^{-1}$ for J0945.


\section{NIFS observations and data reduction}
\label{observations}



The target was observed on 2018 Dec 2nd with the Near-Infrared Integral Field Spectrograph (NIFS), installed on the Gemini-north telescope (program ID: GN-2018B-Q-314-125; PI: Ramos Almeida). It has a FOV of 3$\arcsec \times 3 \arcsec$ (6.9 kpc $\times$ 6.9 kpc at the redshift of the source), divided into 29 slices with an angular sampling of 0.103$\times$0.042 arcsec$^2$. The observations were performed using the K grism, which covers a spectral range of 1.99-2.40 ${\mu}$m. The nominal spectral resolution at the central wavelength of 2.20 $\mu$m is R$\sim$5290 (56.7 \kms). However, from the measurement of the FWHM of ArXe lamp lines we obtain a  spectral resolution of $\sim 3.4 \, \AA$ ($\sim$46 \kms~at 2.20 $\mu$m), that we use as our instrumental broadening. 
Due to the important and rapid variation of the sky emission in the NIR, the observations were split into four cycles of exposures following a jittering O-S-O 
pattern for on-source (O) and sky (S) frames. Since each exposure in the cycle was 500 s, the total on-source time was 4000 s with clear observing conditions.  

To measure the seeing FWHM we used the photometric standard star observed just before and immediately after the target. We measured a seeing FWHM=0.27\arcsec~for the stars observed before J0945 and FWHM=0.32\arcsec~for the stars observed afterwards. These values show how the seeing maintained an excellent and stable value during the entire observation of J0945. We consider the average value between them the angular resolution of the data, and the range of variation as the uncertainty, i.e. FWHM= 0.29$\pm$0.03\arcsec~($\sim$660 pc). 

For the reduction of the data, we used tasks contained in
the NIFS package, which is part of the \texttt{GEMINI.IRAF} package, as well
as generic \texttt{IRAF} tasks. The standard
procedure includes trimming of the images, flat-fielding,
cosmic ray rejection, sky subtraction, wavelength calibration, and s-distortion
correction. In order to remove the telluric absorption from the galaxy
spectrum, we observed the telluric standard star HIP 15925 just
before the K-band observations. The galaxy spectrum was divided
by the normalized spectrum of the telluric standard star using
the \texttt{NFTELLURIC} task of the \texttt{NIFS.GEMINI.IRAF} package.
To flux-calibrate the galaxy spectrum, a black-body function was interpolated to the spectrum of the telluric standard. 

The K-band data cubes were constructed with an angular sampling of 0.05$\times$0.05 arcsec$^2$ for each individual exposure.
These individual data cubes were then combined using a sigma clipping algorithm in order to eliminate bad
pixels and remaining cosmic rays by mosaicing the dithered spatial
positions. 
For further details on the NIFS data reduction process we refer the reader to \citet{2020MNRAS.496.4857R} and references therein.

\begin{figure}
\centering
\includegraphics[width=0.49\textwidth]{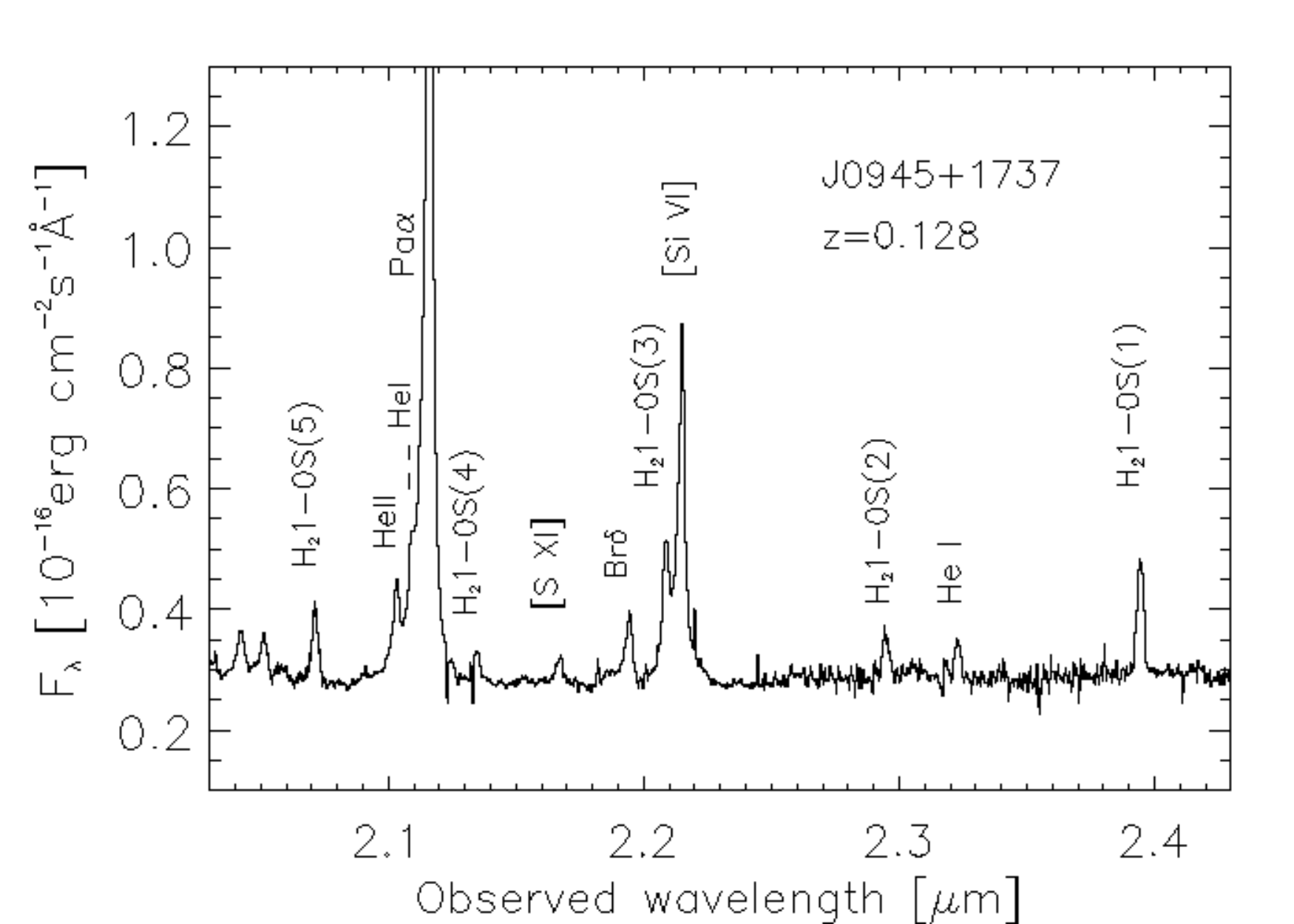}
\caption {NIR nuclear spectrum of J0945+1737 extracted in a circular aperture of $\sim$0.3$\arcsec$ ($\sim$0.69 kpc) in diameter. Low-ionization emission lines (Pa$\alpha$ and Br$\delta$), high-ionization emission lines ([S XI] and [Si VI]) and molecular lines (H$_2$) are detected.}
\label{fig:spectrum}
\end{figure}

\section{Results}  \label{analysis}




\subsection{The nuclear K-band spectrum} \label{nucleus}

\begin{figure*}
\centering
\includegraphics[width=0.503\textwidth]{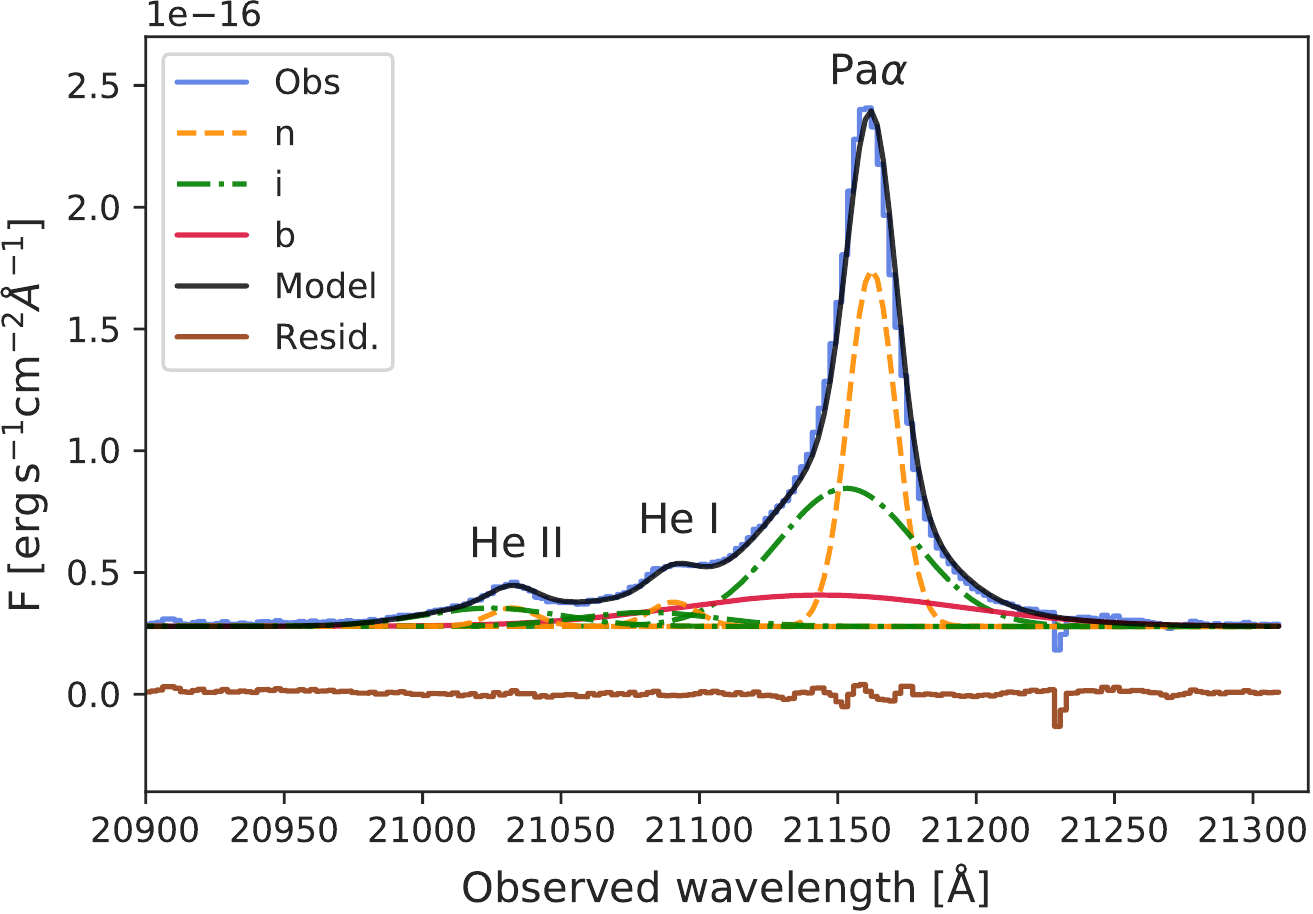}
\includegraphics[width=0.49\textwidth]{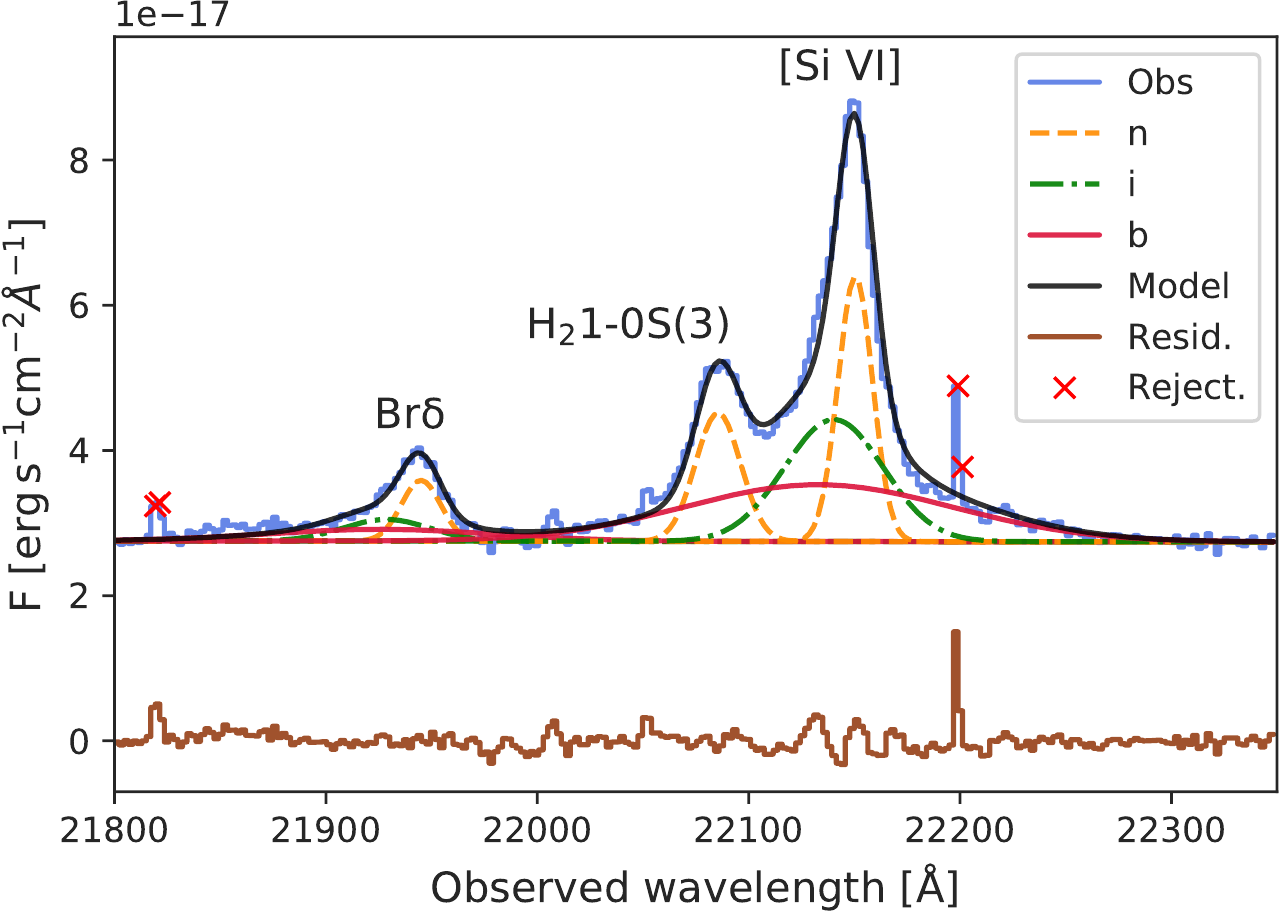}
\caption {Examples of fits of low-, high-ionization and H$_2$ lines detected in the nuclear spectrum, extracted with an aperture of 0.3\arcsec. The observed spectra are shown in blue, the model including all the components in black, and the residuals from the fit in brown. Left panel: Pa$\alpha$ profile modelled using three Gaussians corresponding to a narrow (n), intermediate (i) and broad (b) component (orange dashed, green dot-dashed and red solid line respectively). He~I and He~II are detected in the blue wing of Pa$\alpha$, which are fitted with a narrow and intermediate components. Right panel:  Br$\delta$ and [Si~VI] are fitted with the same kinematic components as Pa$\alpha$. The profile of H$_2$1-0S(3) is well reproduced with just one Gaussian. Red crosses correspond to points removed from the fit of the profiles.
}
\label{fig:nuc}
\end{figure*}

In order to study the nuclear emission of J0945, we extracted a spectrum in a circular aperture of the same diameter as the seeing FWHM = 0.29$\pm$0.03$\arcsec$ ($\sim$0.66 kpc), centered at the
peak of the continuum emission. The spectral range covered by the K-grism of NIFS enables us to simultaneously study different phases of the gas at the redshift of the target (z=0.128). In  Fig.~\ref{fig:spectrum} we can see low-ionization emission lines including Pa$\alpha$, Br$\delta$, He~II$\lambda$1.8637
${\mu}$m, He~I$\lambda$1.8691 $\mu$m, and He~I$\lambda$2.0587 $\mu$m; the high-ionization/coronal emission lines [S~XI]$\lambda$1.920 ${\mu}$m, [Si~VI]$\lambda$1.963 ${\mu}$m, and 
the molecular lines from H$_2$1-0S(5) to 1-0S(1).  
These H$_2$ lines trace warm molecular gas (T > 1000 K), which is only a small fraction of the total molecular gas content (\citealt{Dale05, Mazzalay13,Emonts17}).  
The most prominent emission line of the K-band nuclear spectrum is Pa$\alpha$, followed by the coronal line [Si~VI]. The molecular lines are also prominent, as can be seen from Fig.~\ref{fig:spectrum}. 
 H$_2$1-0S(3) is strongly blended with [Si~VI] even at this spectral resolution. 

In this work we aim at studying the kinematics of the line-emitting gas, looking for broad components that might be related with outflows in different phases. To do so, we need to characterize the emission line profiles, and first subtract the adjacent continuum. For this purpose, we performed a linear fitting of two continuum bands: one on the blue side and another on the red side of each emission line. The width of each of these bands and the distance from the line peak depends on the proximity of other emission lines. After removing the contribution from the continuum we model each emission line with single or multiple Gaussians using an in-house developed Python program based on the \textit{modeling} module of \textit{Astropy} (\citealt{Astropy13,Astropy18}). 
 Before performing the fits we removed points clearly detached from the line profile,  i.e., those with $\ga$5 times the standard deviation of the residuals (see right panel of Fig. \ref{fig:nuc} for an example). To do so, we first did a preliminary fit of the observed spectrum in order to evaluate which points fell above the above mentioned threshold. Then we repeated the fit avoiding those points, which correspond to hot pixels in the detector. 
In Table \ref{tab:nuc} we list the parameters that characterize each of the fitted Gaussian components, i.e. the FWHM, the velocity shift relative to the narrow component of Pa$\alpha$ ($\Delta$v$_s$) and the integrated flux. We use Pa$\alpha$ as our reference because it is the most prominent emission line. Using the centroid of its narrow component (21162.27 $\pm\,0.04\,{\mu}$m), we measure a redshift of 0.1283, which is slightly higher than the redshift measured from the SDSS spectrum (z=0.1281). 
The uncertainties of the parameters listed in Table  \ref{tab:nuc} were measured by means of a Monte Carlo simulation. We generate mock spectra by varying the flux of each spectral element of the emission line profile by adding random values extracted from a normal distribution with an amplitude given by the noise of the pixels. The uncertainties of the parameters are then computed as 1$\sigma$ of each parameter distribution computed from 100 mock spectra.

\begin{table}
\centering
\small
\begin{tabular}{lccc}
\hline
\hline
Line   &  FWHM  & $\Delta$v$_s$ & Flux $\times10^{-15}$  \\
      &  [km s$^{-1}$]   & [km s$^{-1}$]   &   [erg cm$^{-2}$s$^{-1}$]  \\
\hline
$[\text{O~III}]$   (n) & $~~288\pm1$     & $~~~~27\pm1$   & $46.25\pm0.28$ 
 \\
$[\text{O~III}]$    (i) & $~~746\pm4$     & $-127\pm3$   & $53.58\pm0.61$
\\
$[\text{O~III}]$    (b) & $~~1972\pm18$     & $-233\pm8$   & $43.00\pm0.92$ 
\\
H$\beta$   (n) & $~~~301\pm5$     & $~~~~27\pm4$   & $~~4.40\pm0.11$ 
 \\
H$\beta$ (i) & $~~774\pm1$     & $~~-97\pm1$   & $~~5.27\pm0.28$
\\
H$\beta$ (b) & $1704\pm1$     & $-235\pm1$   & $~~4.38\pm0.35$ 
\\

Pa$\alpha$ (n) & $~~288\pm2$     & $~~~~~~~0\pm1$   & $~~3.17\pm0.05$ \\
Pa$\alpha$ (i) & $~~~~836\pm33$    & $-130\pm5$   & $~~3.56\pm0.27$  \\
Pa$\alpha$ (b) & $~~~~~~1703\pm1161$ & $~~~-260$ & $~~1.64\pm1.24$ \\
He~I        (n) & $~~288$ & $~~~~23\pm7$ & $~~0.21\pm0.01$ \\
He~I        (i) & $~~~~836$ & $-114\pm3$ & $~~0.37\pm0.05$ \\
He~II       (n) & $~~288$ & $~~~~66\pm8$ & $~~0.16\pm0.02$ \\
He~II       (i) & $~~~~836$ & $~~-50\pm3$ & $~~0.47\pm0.06$ \\
Br$\delta$        (n) & $~~~~297\pm48$    & $~~~~~~~59\pm21$    & $~~0.19\pm0.04$  \\
Br$\delta$        (i) & $~~~~~~657\pm158$   & $~~~-164\pm43$  & $~~0.15\pm0.07$ \\
Br$\delta$        (b) & $~~~~1730\pm150$  & $~~-200$ & $~~0.22\pm0.07$ \\
$[\text{Si~VI}] $ (n) & $~~~~261\pm15$    & $~~~~~30\pm3$      & $~~0.75\pm0.06$  \\
$[\text{Si~VI}]$  (i) & $~~~~~746$         & $~~-107\pm44$  & $~~0.98\pm0.17$ \\
$[\text{Si~VI}]$  (b) & $~~~~2030\pm115$  & $~~-207\pm65$  & $~~1.25\pm0.24$ \\
H$_2$1-0S(1)      (n) & $~~~~355\pm24$    & $~~~~~62\pm8$    & $~~0.59\pm0.05$ \\
H$_2$1-0S(2)      (n) & $~~~~404\pm50$    & $~~~~-55\pm30$    & $~~0.21\pm0.05$ \\
H$_2$1-0S(3)      (n) & $~~~~342\pm26$    & $~~~~67\pm6$    & $~~0.47\pm0.05$ \\
H$_2$1-0S(4)      (n) & $~~~~400\pm28$    & $~~~~~~~~0\pm13$    & $~~0.15\pm0.02$ \\
H$_2$1-0S(5)      (n) & $~~~~388\pm58$    & $~~~~~~12\pm22$    & $~~0.33\pm0.07$ \\
\hline
\end{tabular}
\caption{Emission line properties measured from the SDSS optical spectrum and the NIFS K-band spectrum. Column description: (1) emission line name followed by the kinematic component: narrow (n), intermediate (i) or broad (b); (2) corresponding FWHM corrected from instrumental broadening; (3) velocity shift ($\Delta$v$_s$) relative to the central $\lambda$ of the narrow component of Pa$\alpha$; (4) integrated flux. Measurements without errors correspond to parameters that have been fixed.}
\label{tab:nuc} 
\end{table}

In the case of Pa$\alpha$, Br$\delta$ and [Si~VI], the best fits were obtained by modelling the line profiles with three Gaussians, as shown in Fig.~\ref{fig:nuc}. The number of components was chosen by adding additional Gaussians until there is no improvement in the resulting residuals. These three kinematic components were also shown to be necessary to reproduce the [O III] lines detected in GMOS optical spectra (see \citealt{Oliveira21} and Section~\ref{literature} for discussion).

In the case of the NIFS spectrum, we fitted a narrow component of FWHM$\sim$300 \kms, an intermediate component of $\sim$660-840 \kms, and a broad component of $\sim$1700-2000 \kms. The intermediate component is blue-shifted by 130--160 \kms~from the narrow component of Pa$\alpha$, and the broad component by 200-260 \kms.
In the case of the molecular lines, we find that only a narrow component of FWHM$\sim$340-400 \kms~  is needed to characterize their profiles (see right panel of Fig.~\ref{fig:nuc} for an example). 

Since some of the detected emission lines show complex profiles due to blending with other lines, 
 we used the optical SDSS spectrum as a reference to fit these NIR lines. In particular, we used the [O~III]$\,\lambda5007~\AA$ and H$\beta$ emission lines, which are well separated from other optical emission lines. We thus used the information derived from the fit of H$\beta$ for the characterization of the Pa$\alpha$ and Br$\delta$ profiles, and of [O~III] for [Si~VI], since they should share the same kinematics. Using the methodology described above, we also found that three components are needed to characterize the [O~III] and H$\beta$ profiles. These fits are shown in Appendix~\ref{Appendix A}, and the corresponding parameters are included in Table \ref{tab:nuc}. Having this information about the line profiles was necessary to fit for example, the Pa$\alpha$ line, because of the presence of He~I and He~II on its blue wing (see left panel of Fig.~\ref{fig:nuc}). We fixed $\Delta$v$_s$ of the broad component of Pa$\alpha$ to be the same of H$\beta$ (and the same for Br$\delta$), relative to its narrow component, and we modelled He~I and He~II with narrow and intermediate components having the same FWHM as Pa$\alpha$. We could not fit a broad component to the profiles of He~I and He~II because they are too faint. In addition, because of the blend between the [Si~VI] and  H$_2$1-0S(3) lines, we had to impose the FWHM of the intermediate component of the [Si~VI] to be the same as of [O~III]. 
Aside from these restrictions, we found consistency (within the uncertanties) between the parameters that we measure from the NIR emission lines and from [O~III] and H$\beta$ (see Table \ref{tab:nuc}).
It is noteworthy that the SDSS optical spectrum corresponds to a physical size of 3$\arcsec$ in diameter, which is significantly larger than the 0.3$\arcsec$ aperture that we used to extract the K-band nuclear spectrum. However, by extracting a NIR spectrum in a 3$\arcsec$ aperture and performing the same fits to the emission lines, we found the same parameters within the uncertainties. This indicates that the 3$\arcsec$ aperture spectrum is dominated by the nuclear emission of the quasar.

We identify the narrow component with gas in the narrow-line region (NLR). The intermediate and broad components correspond to turbulent gas that appears to be outflowing, as it is blue-shifted relative to the narrow component. Thus, we detect outflowing components in the low- and high-ionization lines, but not in the warm molecular lines, as in the case of the Teacup QSO2 (\citealt{Ramos17}). 
We discard the possibility of a broad-line region (BLR) origin of the broad components because they are detected in forbidden emission lines such as [O~III] and [Si~VI], and moreover, they are significantly blue-shifted from the narrow component (from $-$130 up to $-$260 \kms). Previous evidence for an ionized outflow in J0945 was already reported from optical Gemini-south GMOS (\citealt{Harrison14, Oliveira21}) and HST data (\citealt{Storchi18}).




\subsection{Electron density of the ionized gas}
\label{density} 
 
Once we have characterized the nuclear emission line kinematics, we can use the emission lines present in the optical SDSS spectrum of J0945 to measure the electron density (n$_{\rm e}$) of the gas associated with each of the three components considered in our analysis.   

\begin{figure}
\centering
\includegraphics[width=0.49\textwidth]{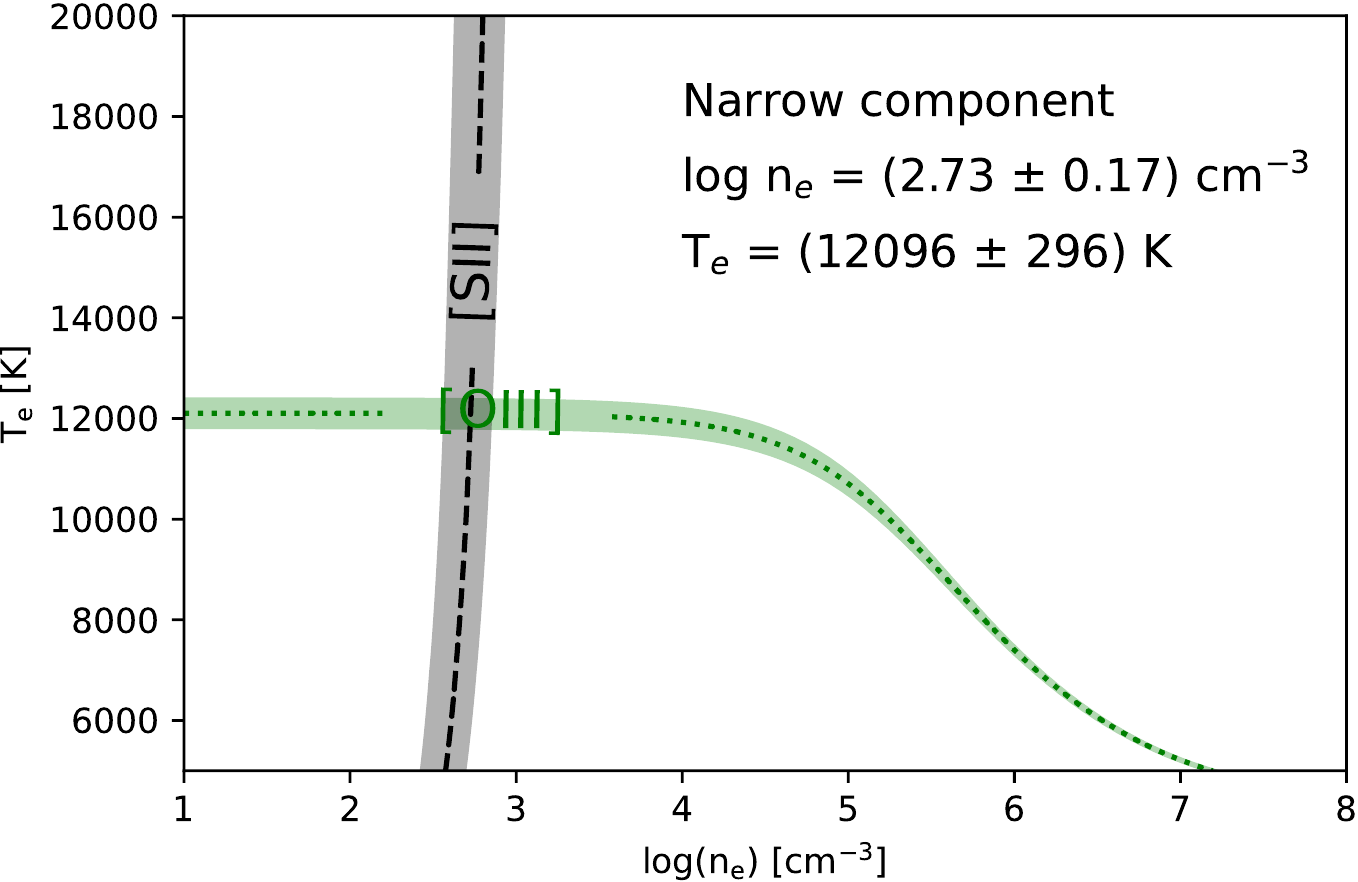}
\includegraphics[width=0.49\textwidth]{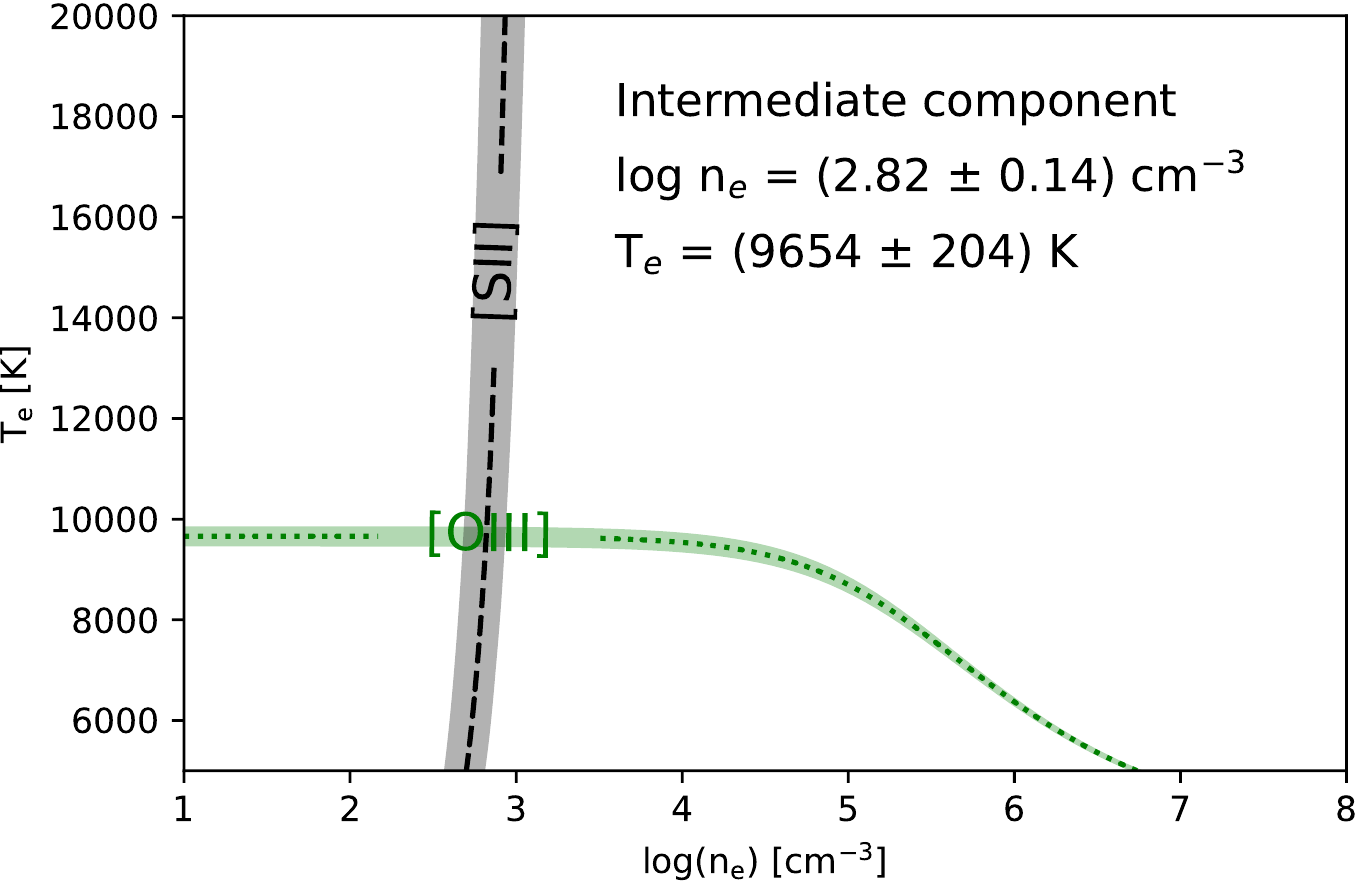}
\includegraphics[width=0.49\textwidth]{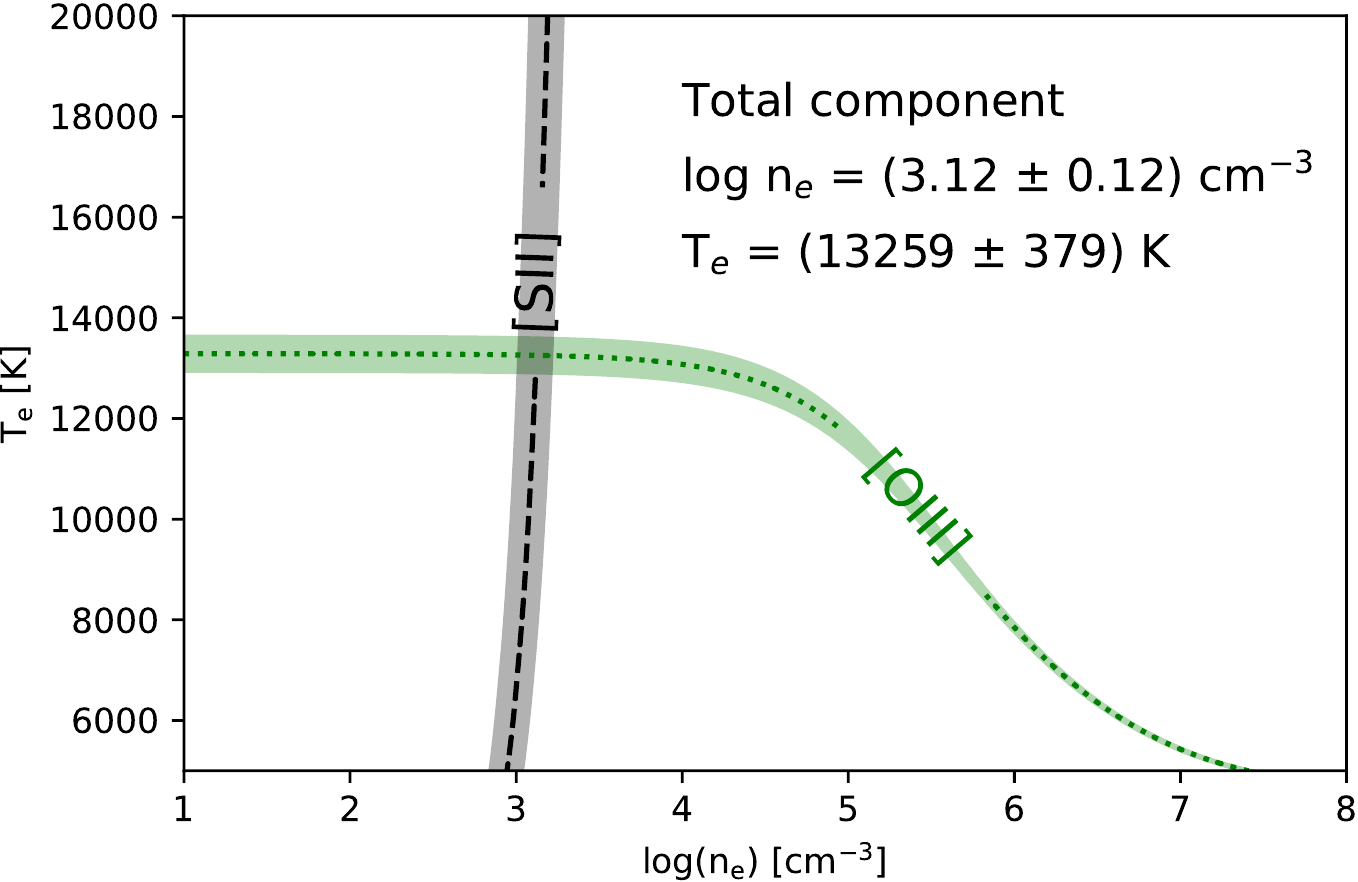}
\caption {Diagnostic diagrams of the electron temperature (T$_{\rm e}$) and density (n$_{\rm e}$), computed from the [S~II]$\lambda \lambda 6716,6731$ doublet and the [O~III]$\lambda 4363$ to [O~III]$\lambda 5007$ ratio. We show the diagrams corresponding to the narrow, intermediate and total fluxes obtained using the \textit{Pyneb} tool.}
\label{fig:denpy}
\end{figure}

 We measured n$_{\rm e}$ using two different methods. First, using the density-sensitive optical doublet ratio of [S~II]$\lambda\lambda$6716,6731~\citep{Osterbrock06}. Taking advantage of the optical SDSS spectrum of J0945 we can measure the total fluxes and the fluxes of each kinematic component by fitting the doublet with a narrow, intermediate and broad components (see left panel of Fig.~\ref{fig:fitpy} in Appendix \ref{Appendix A}). Then, to obtain n$_{\rm e}$ we used the \textit{Pyneb} tool by \citet{Luridiana15}, which in addition to the [S~II] ratio, requires the [O~III]$\lambda$4363 to [O~III]$\lambda$5007 ratio as diagnostic for the electron temperature (T$_{\rm e}$). See right panel of Fig.~\ref{fig:fitpy} in Appendix \ref{Appendix A} and top panel of Fig.~\ref{fig:nucop} in Appendix \ref{Appendix A} for the corresponding fits.

Fig.~\ref{fig:denpy} shows the plot of T$_{\rm e}$ vs n$_{\rm e}$ for the narrow, intermediate and total fluxes measured from the [S~II] and [O~III] optical doublets.
The crossing point between the dashed and dotted lines provide us with the value of T$_{\rm e}$ and n$_{\rm e}$. For the narrow and intermediate components we measure log n$_{\rm e}$ = 2.73 $\pm$ 0.17 and 2.82 $\pm$ 0.14 cm$^{-3}$, respectively (see Table \ref{tab:dens}). Fig.~\ref{fig:denpy} does not include the broad component because the [S II] doublet is sensitive to relatively low densities (2 $\lesssim$ log n$_{\rm e} \lesssim$ 3.5 cm$^{-3}$; \citealt{Rose18}) and, according to our measurements, the broad component would have a density log n$_e \gtrsim$ 4 cm$^{-3}$. This is in line with expectations for outflow densities to be higher than the densities traced by the [S~II] lines (see e.g. \citealt{Rose18, Davies20} and references therein). However, for the intermediate component, the outflow density is relatively low, as it has been also observed in nearby AGN (see e.g. \citealt{2019A&A...622A.146M}).

 \begin{table*}
  \caption{Properties of the different kinematic components  of Pa$\alpha$.}
  \centering
\begin{tabular}{c c c c c c c c c}
\hline
Comp.   &   log n$_{\rm e}$*      &  M$_{\text{gas}} \times 10^{7}$ &  $\dot{\text{{M}}}_{\text{{out}}}$ &  $\dot{\text{{M}}}_{\text{{out, max}}}$  &   log $\dot{\text{{E}}}_{\text{{kin}}}$ &  log $\dot{\text{{E}}}_{\text{{kin, max}}}$ & $\xi$ & $\xi_{\text{max}}$ \\
            & $[\text{cm}^{-3}]$       &   $[\text{M}_{\odot}]$  & $[\text{M}_{\odot}~\text{yr}^{-1}]$ & $[\text{M}_{\odot}~\text{yr}^{-1}]$  &   $[\text{erg}~\text{s}^{-1}]$ &   $[\text{erg}~\text{s}^{-1}]$ &  [\%] & [\%] \\
            
     (1)   &  (2)  &   (3) &  (4)  &  (5) &    (6) &   (7)&   (8)  & (9) \\
\hline
n  & $(2.73\pm0.17)_{\text{[S~II]}}$ &   $4.80\pm1.57$  & -- & -- & -- & -- & -- & -- \\
i  & $(2.82\pm0.14)_{\text{[S~II]}}$  &   $6.65\pm1.83$  & $7.9\pm2.2$ &  $51\pm14$ & 40.85 & 42.62 &  0.0014 & 0.082\\
tot & $(3.12\pm0.12)_{\text{[S~II]}}$  &   $11.47\pm2.41^{\star}$  & --& -- & -- & -- & --& --\\
\hline

Comp.   &   log n$_{\rm e}$**      &  M$_{\text{gas}} \times 10^{7}$ &  $\dot{\text{{M}}}_{\text{{out}}}$ &  $\dot{\text{{M}}}_{\text{{out, max}}}$  &   log $\dot{\text{{E}}}_{\text{{kin}}}$ &  log $\dot{\text{{E}}}_{\text{{kin, max}}}$ & $\xi$ & $\xi_{\text{max}}$ \\
            & $[\text{cm}^{-3}]$       &   $[\text{M}_{\odot}]$  & $[\text{M}_{\odot}~\text{yr}^{-1}]$ & $[\text{M}_{\odot}~\text{yr}^{-1}]$  &   $[\text{erg}~\text{s}^{-1}]$ &   $[\text{erg}~\text{s}^{-1}]$ &  [\%] & [\%] \\
\hline
n  & $(2.9\pm0.3)_{\text{TA}}$ &   $~~3.22\pm0.41$  & -- & --  & --  & -- & --& --\\
i  & $(2.9\pm0.2)_{\text{TA}}$  &   $~~5.55\pm0.29$  & $6.6\pm0.4$ & $42\pm3$ & 40.77 & 42.54&  0.0018 & 0.109 \\
b  & $(4.1\pm0.2)_{\text{TA}}$  &   $~~0.13\pm0.02$  & $0.31\pm0.04$ & $~~2.1\pm0.9$ & 40.06 & 41.84 & 0.00036 & 0.022\\
tot & $~~(3.2\pm0.15)_{\text{TA}}$  & $~~~~8.90\pm0.50 ^{\star}$   & --& --  & -- & -- & -- & --\\
\hline
\end{tabular}
\tablefoot{(1) kinematic component: narrow (n), intermediate (i), broad (b), and total (tot); (2) electron density; 
(3) gas mass;
 (4) mass outflow rate computed as 3$\,\times\,\text{v}_{\text{out}}\times \text{M}_{\text{out}}$/R$_{\text{out}}$,  where v$_{\text{out}} = -130$ \kms~and $-260$ \kms~for the intermediate and broad component, and R$_{\text{out}}$=3.37 kpc; 
 (5) maximum mass outflow rate,  with v$_{\text{max}} = -840$ \kms~and $-1700$ \kms~for the intermediate and broad component;
 (6) outflow kinetic energy as $\dot{\text{{E}}}_{\text{kin}} = \dot{\text{M}}_{\text{out}}/2 \times(\text{v}_{\text{out}}^{2} + 3\sigma^2)$; 
 (7) maximum outflow kinetic energy; 
 (8) coupling efficiency, i.e. $\xi$= $\dot{\text{{E}}}_{\text{{kin}}}/\text{L}_{\text{bol}}$;
 (9) maximum coupling efficiency.
 *Electron densities computed with \textit{Pyneb} using the [S~II]$\lambda\lambda 6716,6731$ doublet  ([S~II]). **Electron densities computed using the trans-auroral lines  (TA). $^{\star}$The total mass corresponds to the sum of the different components. 
}
\label{tab:dens}
\end{table*}

\begin{figure}
\centering
\includegraphics[width=0.49\textwidth]{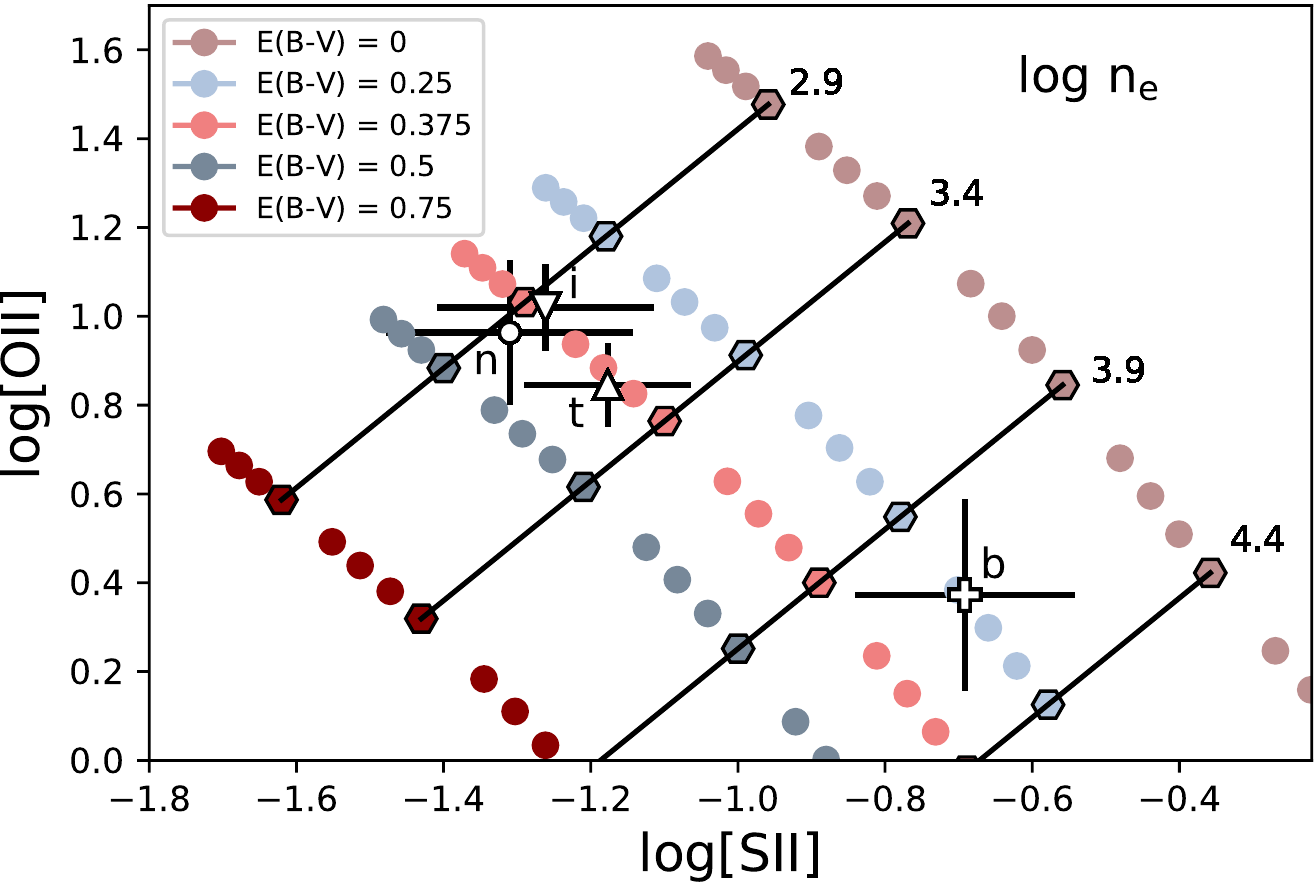}
\caption {Diagnostic diagram from \citet{Rose18} to simultaneously measure the reddening E(B-V) and n$_{\rm e}$ of the ionized gas. The open symbols with error bars correspond to our flux measurements of the narrow (n), intermediate (i), total (t) and broad (b) components (circle, down- and up-pointing triangles, and plus symbol, respectively). Colour symbols correspond to the grid of photoionized models computed with  \textit{Cloudy}.}
\label{fig:denstrans}
\end{figure}

To obtain the n$_{\rm e}$ associated with the broad component and contrast the [S~II]-based values obtained for the intermediate and narrow components, we used another method based on the trans-auroral lines. This technique uses the flux ratios of the [S II]$\lambda\lambda$6716,6731 and [O II]$\lambda\lambda$3726,3729 doublets as well as of the trans-auroral [O II]$\lambda\lambda$7319,7331 and [S II]$\lambda\lambda$4068,4076 lines. The trans-auroral ratios have the advantage of being sensitive to higher density gas than the classical [S~II] and [O~II] doublet ratios (\citealt{Holt11,Rose18,Ramos19}). Thus, we also fitted these optical lines with a narrow, intermediate and broad components (see Fig.~\ref{fig:fittrans} in Appendix \ref{Appendix A}). By comparing the [O~II] and [S~II] ratios (TR([O~II]) = F(3726+3729)/F(7319+7331) and TR([S~II]) = F(4068+4076)/F(6717+6731)) with a grid of photoionization models computed with  \textit{Cloudy} (\citealt{Ferland13}), we can determine the n$_{\rm e}$ and E(B-V) of the gas simultaneously. Fig.~\ref{fig:denstrans} shows the ratios of the narrow, intermediate, broad, and total fluxes measured for J0945 plotted on top of the diagnostic diagram from \citet{Rose18}. We found that the reddening is the same for the different components, E(B-V)$\sim$0.37. The only exception is the broad component, for which we measure a slightly lower value of E(B-V) = 0.25 $\pm$ 0.1.



  For the narrow and intermediate components we find the same electron densities, of log n$_{\rm e}$=2.9 cm$^{-3}$. These values are fully consistent with the measurements from the [S~II] doublet. For the broad  component we find a much higher value, of log n$_{\rm e}\sim$(4.1$\pm$0.2) cm$^{-3}$. This value is also consistent with the lower limit derived from the [S~II] lines, and also with the values measured for the outflowing components of nearby ULIRGs (\citealt{Rose18, Spence18}) and type-2 AGN (\citealt{Baron19}).

 \subsection{The extended line emission} \label{voronoi}

\begin{figure*}[!h]
\centering
\includegraphics[width=0.65\textwidth]{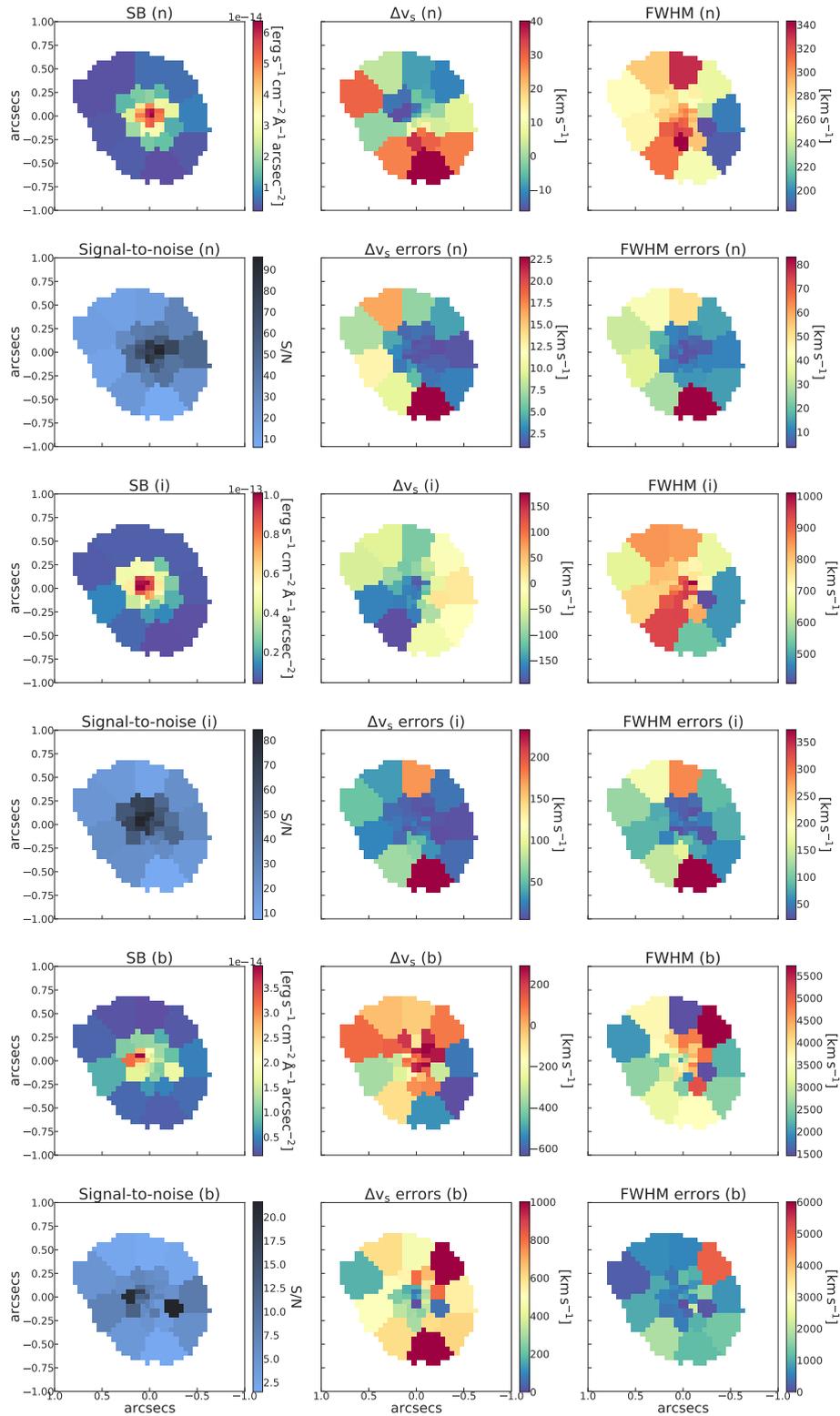}
\captionsetup{width=.65\linewidth}
\caption {Pa$\alpha$ maps built after applying Voronoi tessellation. From top to bottom we show the emission line maps for the narrow (first two rows), intermediate (middle rows) and broad component (last two rows). For each component we show the surface brightness (SB),  velocity shift ($\Delta$v$_{\rm s}$), and FWHM maps (from left to right), and underneath, the corresponding S/N and error maps. North is up and east to the left.}
\label{fig:vorpa}
\end{figure*}

\begin{figure*}[!h]
\centering
\includegraphics[width=0.8\textwidth]{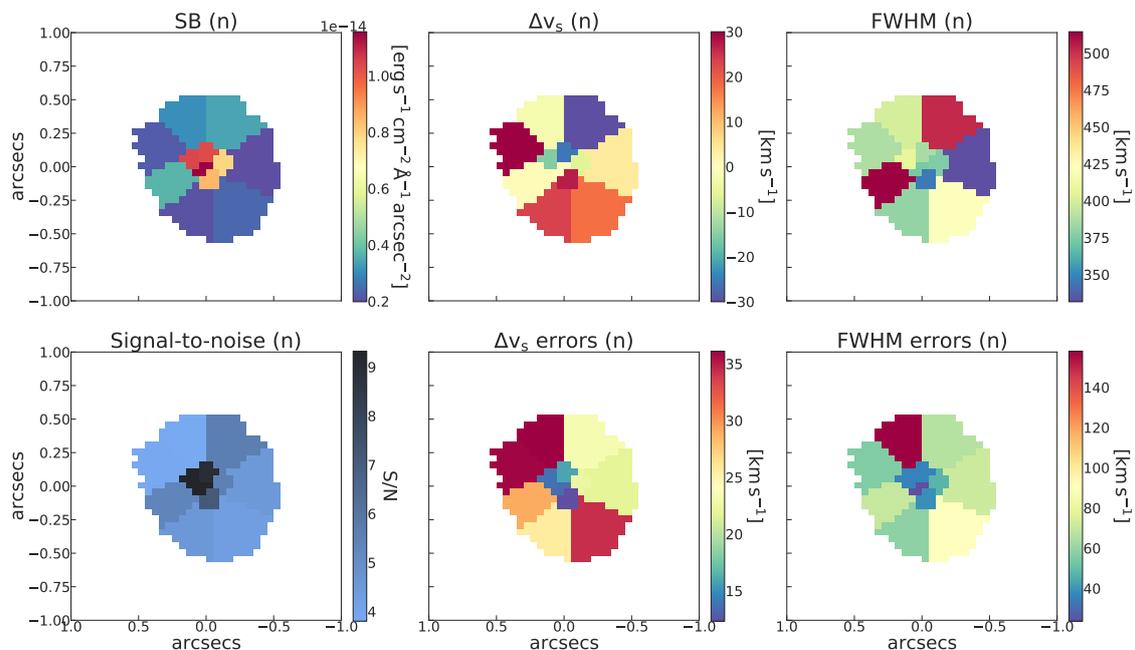}
\captionsetup{width=.8\linewidth}
\caption {Same as in Figure~\ref{fig:vorpa}, but relative to the H$_2$1-0S(1) emission line, and including only one kinematic component.}
\label{fig:vorh1}
\end{figure*}

In the previous section we described the properties of the ionized and warm molecular lines detected in the nuclear spectrum of J0945. Here we take advantage of the spatial information provided by integral field observations to study the distribution and kinematics of Pa$\alpha$, [Si VI] and the  H$_2$1-0S(3) and H$_2$1-0S(1)  lines (i.e. the most prominent emission lines). As discussed in Section \ref{nucleus}, the K-band spectrum extracted in a large aperture of 3\arcsec~diameter, which is the extent of the NIFS FOV, is dominated by the nuclear emission, and indeed, most of the emission line flux comes from the very central spaxels of the datacube, making it difficult to characterize the emission line profiles beyond the central region. 
In order to increase the signal-to-noise (S/N) in the outer regions of the NIFS datacube, we performed a Voronoi tessellation following the procedure described by \citet{Cappellari03}. Essentially, the method consists on binning the two-dimensional data to a constant S/N ratio per bin for each emission line. These bins are generally referred to as voxels. 
Thus, we imposed a S/N=15 for the Pa$\alpha$, Br$\delta$ and [Si~VI] emission lines and S/N=10 for the molecular lines, where the signal and the noise are extracted directly from the data cube. 

Once we applied the tessellation, we fitted the emission lines listed above for each voxel, following the same procedure applied to the nuclear spectrum, described in Section~\ref{nucleus}. Using the measured properties of the emission lines in every voxel, we produce surface brightness (SB, i.e., flux divided by the area of each voxel), velocity, and FWHM maps for each kinematic component. Two examples of these maps are shown in Figs.~\ref{fig:vorpa} and~\ref{fig:vorh1}. 
As we explained in Section~\ref{nucleus}, the nuclear Pa$\alpha$, Br$\delta$, and [Si~VI] line profiles are best characterized by three components (narrow, intermediate, and broad). Here we also fitted these three components for all the voxels, and just one component in the case of the molecular lines. 
For each kinematic component, in Figs.~\ref{fig:vorpa} and~\ref{fig:vorh1} we show six panels.
The first row of each figure corresponds to the SB, velocity, and FWHM maps of the narrow component. The second row includes the corresponding S/N map, and the velocity and FWHM maps. The same six panels are shown for the intermediate and broad components (3rd and 4th, and 5th and 6th rows, respectively, in Figs.~\ref{fig:vorpa}). 
The error maps were built by running Monte Carlo simulations for each voxel, following the same procedure described in Section~\ref{nucleus}. 
The S/N maps were computed relative to the flux of each kinematic component. The noise was measured from the regions where we fitted the continuum, adjacent to each emission line, and then propagated through the line width.


In the case of Pa$\alpha$, from the top panels of Figure \ref{fig:vorpa} we can see that the narrow component peaks in the center of the flux map, and it shows negative velocities in the north and positive in the south, which is consistent with gas rotation, with velocities between $-$20 \kms~and 40 \kms. However, it is noteworthy the presence of a detached voxel showing redshifted velocities at x$\approx$0.5\arcsec~(1.14 kpc) and y$\approx$0.25\arcsec~(0.57 kpc). In this voxel, which is north-east from the nucleus, the gas reaches a velocity of 28 $\pm$6 \kms, which is discrepant with the surrounding voxels. 
After inspecting the emission line fit for this voxel, we are confident that its positive velocity is real and indeed, it could be a signature of gas that has been disrupted by the galaxy interaction/merger that J0945 is undergoing (see Section~\ref{morphology} for further details). Observing the direction of the galaxy’s arms/tidal tails shown in Fig.~\ref{fig:combined_im}, we argue that, if they are trailing, the rotation of J0945 would be counterclockwise. This, in combination  with the velocity field shown in the top panel of Fig.~\ref{fig:vorpa}, would imply that the west is the far side of the galaxy. However, the perturbed rotation due to the interaction might be indicative of a three-dimensional rather than a disk-like system, so we cannot be completely certain about the far/near side.

From the velocity map of the intermediate component (i.e. third row of Fig.~\ref{fig:vorpa}), we can see how the gas velocity increases towards the south-east direction up to $-195\pm32$\kms, having a cone-like morphology. The same velocities and morphology were reported by \citet{Oliveira21} for this kinematic component as measured from optical integral field data (see the fourth row and second column of their Fig.~A4.). We also see negative velocities to the north-east, but not as fast. Indeed, by looking at the flux map of the intermediate component, we see higher fluxes in the center and to the south-east. The FWHM also reaches the highest values there, of $\sim$900--1000 \kms. This strengthens our hypothesis of this being outflowing gas. In the case of the [Si~VI] line (i.e. high-ionization emission-line gas), we find similar results for the intermediate component (see Fig.~\ref{fig:vorsi} in Appendix~\ref{Appendix B}).   If the east is the near side of the galaxy, since we only observe the blue-shifted component of the ionized outflow, it would have to subtend a small angle relative to the galaxy disc, as otherwise it would show redshifted velocities in projection. Finally, the broad component reaches even higher velocities, of up to $\sim -$600 \kms~to the south-west, with FWHM$\sim$2500-3000 \kms. However, the outflow geometry and kinematics are not as well defined as in the case of the intermediate component, which has higher S/N (see Section \ref{comparison}).


Finally, in Fig.~\ref{fig:vorh1} we show the emission-line maps of H$_2$1-0S(1). We just fitted a narrow component for all the voxels, which shows negative velocities to the north and positive to the south, consistent with the ionized gas kinematics of the narrow component. For the molecular line we also found the discrepant red voxel that appears north-east of the velocity maps of the narrow components of Pa$\alpha$. We find consistent results for the H$_2$1-0S(3) emission line, whose corresponding maps are shown in Fig.~\ref{fig:vorh3} in Appendix~\ref{Appendix B}.

\subsubsection{Tidally disrupted emitting-line gas} \label{morphology}



\begin{figure}
\centering
\includegraphics[width=0.49\textwidth]{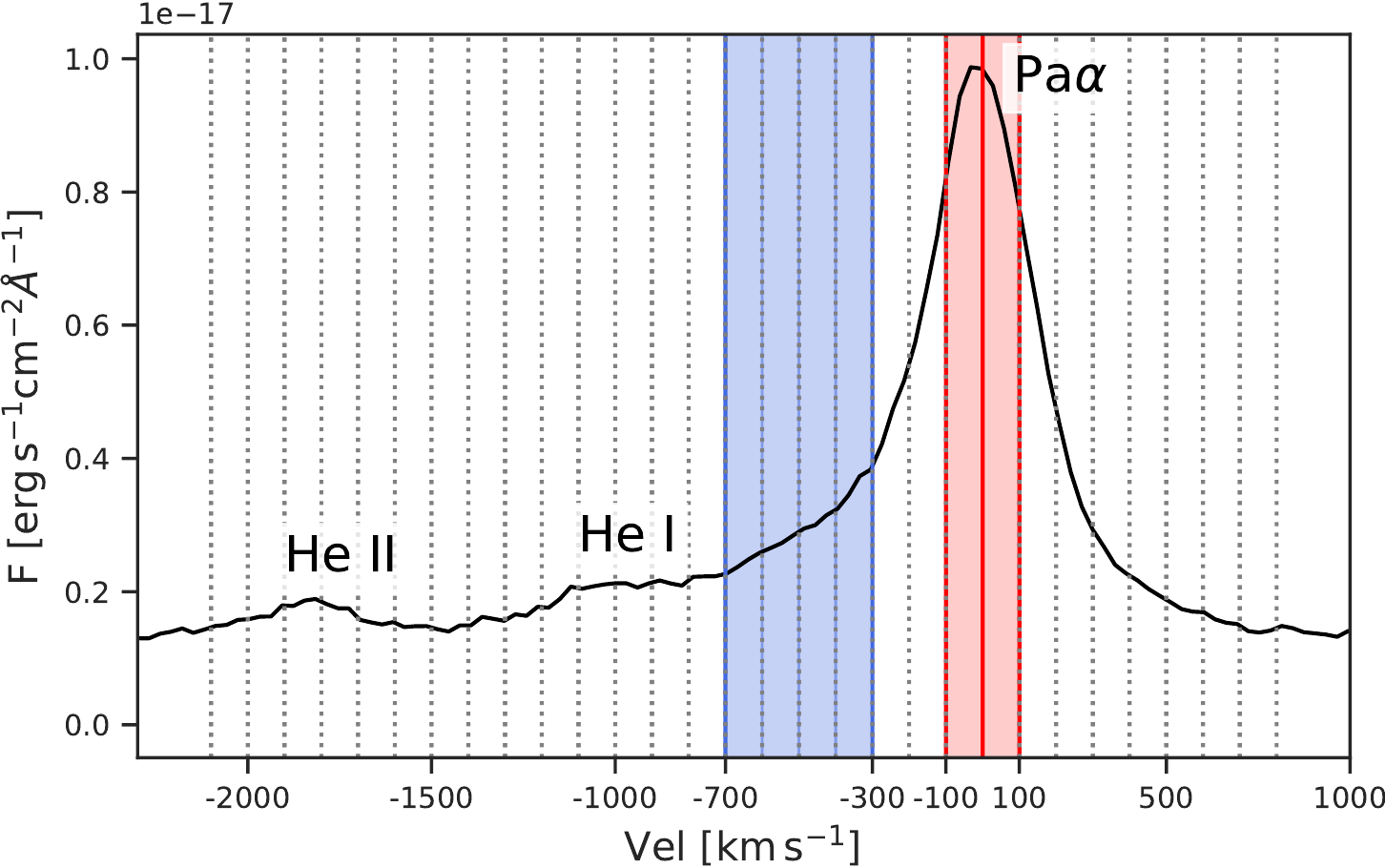}
\caption {Nuclear spectrum of Pa$\alpha$ extracted in an aperture of 0.3$\arcsec$ in diameter.  The dotted vertical lines delimit velocity intervals of 100 \kms. In red we highlight the emission between $-$100 and 100 \kms, and in blue the blue-shifted emission associated with outflowing gas and not including the helium lines, i.e., between $-$700 and $-$300 \kms.}
\label{fig:slice_100}
\end{figure}

In addition to the Voronoi tessellation and emission-line fitting performed in Section \ref{voronoi}, we carried out a complementary analysis for a further inspection of the results. 
In Fig. \ref{fig:cutmap} in Appendix \ref{Appendix C} we show the continuum-subtracted Pa$\alpha$ maps extracted in consecutive velocity intervals of 100 \kms~(from $-$2100 to 900 \kms), centred at the maximum of the line profile in the nuclear spectrum (see Fig. \ref{fig:slice_100}). These velocity cuts permit us to characterize the orientation and extent of the Pa$\alpha$ emission in the core and the wings of the line, as well as the He I and He II emission.


\begin{figure}
\centering
\includegraphics[width=0.49\textwidth]{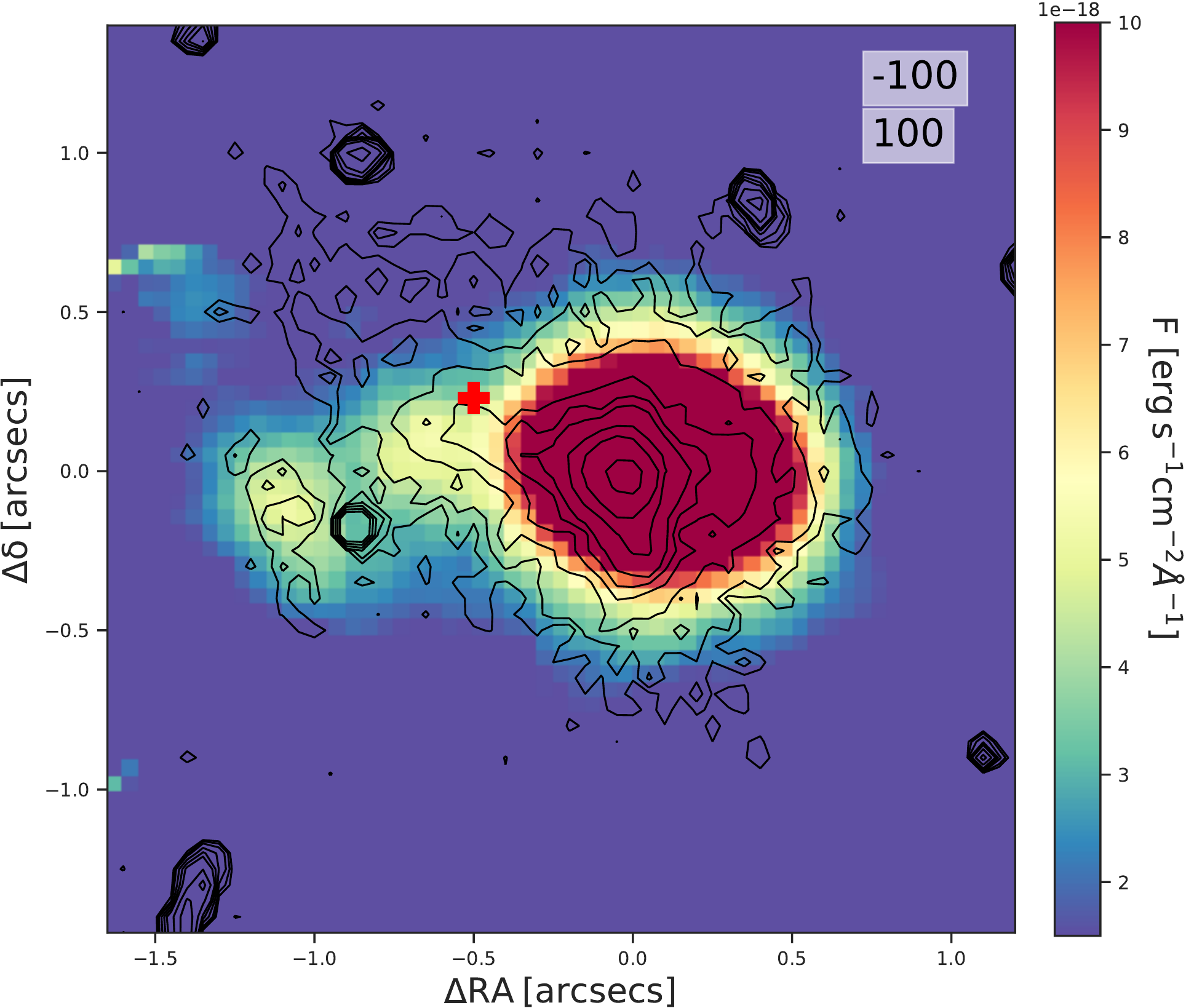}
\caption { Continuum-subtracted Pa$\alpha$ flux map of J0945 extracted in the velocity interval between $-$100 and 100 \kms~from the NIR data cube, with HST optical continuum (FR647M filter) contours starting at 3$\sigma$ superimposed in black. The position of the red voxel is indicated with a red cross. }
\label{fig:interaction}
\end{figure}

The analysis  of the maps presented in Section \ref{voronoi} revealed relevant features of the line-emitting gas kinematics. We found tentative evidence of a kinematic signature of a galaxy interaction or merger after inspection of the narrow emission line maps (in both ionized and warm molecular gas). This signature is the red voxel detected to the north-east, where the underlying rotating distribution appears blue-shifted (see first row and second column panels in Figs.~\ref{fig:vorpa} and~\ref{fig:vorh1}). To further investigate this, in Fig.~\ref{fig:interaction} we show the Pa$\alpha$ emission with velocities between $-$100 and 100 \kms,
i.e., around the emission peak (see Fig.~\ref{fig:slice_100}). This range of velocity, which mainly corresponds to the narrow component, ensures that we are considering just gas in ordinary rotation and not associated with the outflow.

 Fig.~\ref{fig:interaction} shows a tail of Pa$\alpha$ emission to the east, extending up to $\sim$1.3\arcsec~(3 kpc), which then bents and extends even further to the north-east. The red voxel identified in Section \ref{voronoi} is indicated with a red cross. This extended emission can be clearly seen in the scan maps shown in Fig. \ref{fig:cutmap} in Appendix \ref{Appendix C}, and it shows a good correspondence with the optical continuum emission determined from HST data  (i.e., stellar continuum). We downloaded the HST ACS data of J0945 used in \citet{Storchi18} from the Hubble Legacy Archive\footnote{https://mast.stsci.edu/portal/Mashup/Clients/Mast/Portal.html}. In Fig.~\ref{fig:interaction} we show the contours of the continuum emission, as determined from the image in the FR647M filter. There is good agreement between the optical continuum contours and the Pa$\alpha$ emission, especially to the east.  The morphological signatures of mergers are usually studied by looking at the stellar continuum (e.g., {\citealt{2022MNRAS.510.1163P}}). Thus, the discrepant red voxel  most likely corresponds to tidally disrupted gas associated with the merger or interaction that produced the disturbed optical morphology, both on the small-scales probed by the NIFS data, and on the large-scale shown in Fig.~\ref{fig:combined_im}. Signatures of this galaxy interaction were previously reported by \citet{2021A&A...650A..84V} based on an HST/WFPC2 F814W continuum image.

 \subsubsection{Outflow properties} \label{properties}

 
 The emission-line fitting of the nuclear spectrum shown in Section \ref{nucleus}, along with the Pa$\alpha$ and [Si VI] maps shown in Section \ref{voronoi}, revealed the presence of turbulent, high-velocity low- and high-ionization gas in J0945. This turbulent gas is split in two kinematic components (i.e., the intermediate and broad components in the previously presented fits), and we identify them with outflowing gas. In order to characterize these outflow components by means of the mass outflow rate ($\dot{\text{{M}}}_{\text{{out}}}$) and kinetic power ($\dot{\text{{E}}}_{\text{{kin}}}$), which are important quantities
for investigating the mechanisms that drive them and their impact on the surrounding environment, we need to measure the 
outflow mass (M$_\text{out}$). To do that, it is essential to have a reliable measurement of the outflow electron density (n$_{\rm e}$; see Section \ref{density}) and extension (R$_{\text{{out}}}$).

To determine R$_{\text{{out}}}$, we used the continuum-subtracted Pa$\alpha$ flux map extracted in the velocity interval between $-$700 and $-$300 \kms~(highlighted in blue in Fig. \ref{fig:slice_100}). This velocity range ensures that we are mapping the outflow emission by reducing contamination from the narrow Pa$\alpha$ component, He I and He II. This velocity range is dominated by emission from the intermediate and broad components (see left panel of Fig. \ref{fig:nuc}).

\begin{figure}
\centering
\includegraphics[width=0.49\textwidth]{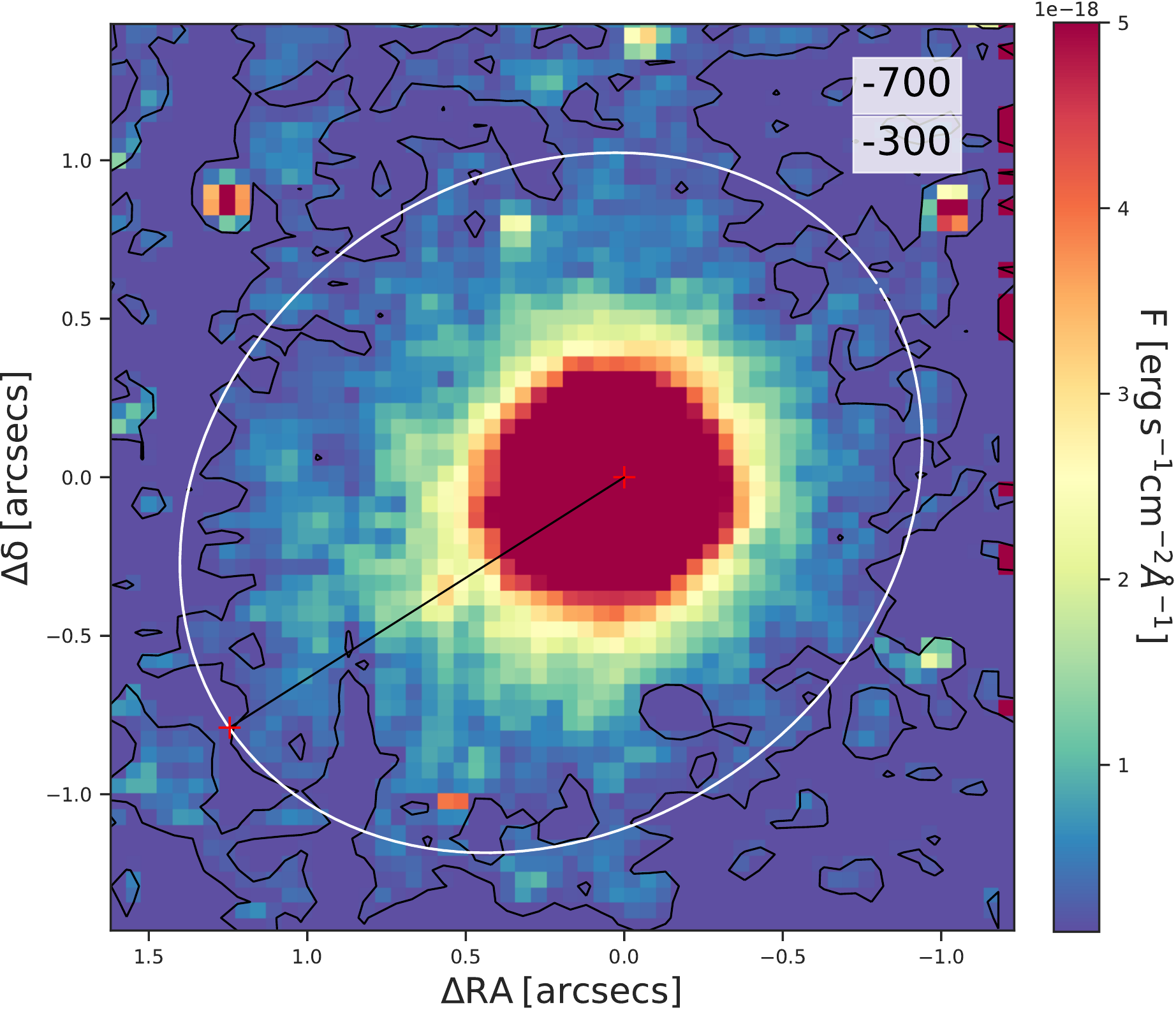}
\caption {Continuum-subtracted Pa$\alpha$ flux map of J0945 extracted in the
velocity interval between $-$700 and $-$300 \kms~from the NIR data cube. The white ellipse defines the region that encompasses the bulk of the emission above 3$\sigma$ (black contours). The two red crosses are the Pa$\alpha$ emission peak and the most distant point from it on the ellipse. We consider the distance between crosses (black straight line) as the maximum extension of the outflow (R$_{\rm out}$).}
\label{fig:outflow}
\end{figure}

The flux map, with 3$\sigma$ contours overlaid, is shown in Fig. \ref{fig:outflow}. The emission is clearly elongated to the south-east, where the most blue-shifted velocities and highest velocity dispersions are measured (see Section \ref{voronoi}). From this outflow map, we can estimate the radial size of the outflow from the FWHM of the outflow map (FWHM$_\text{obs}$=0.33\arcsec), as in \citet{Ramos17}. Following \citet{Rose18}, we consider an outflow resolved, attending to its radial size  (i.e., FWHM$_{\text{obs}}$) when
\begin{equation}
\text{R}_{\text{{out}}} > \text{FWHM}_{\text{seeing}} + 3\sigma_{\text{seeing}}  = 0.29\arcsec + 3\times0.03\arcsec = 0.38\arcsec,
\end{equation}

 where FWHM$_{\text{seeing}}$ is the seeing size and $\sigma_{\text{seeing}}$ its uncertainty (see Section \ref{observations}).

According to this definition, the outflow would be unresolved in J0945. However, from Figure \ref{fig:outflow} and from the maps shown in Section \ref{voronoi}, it is clear that the outflowing gas is extended and well resolved. We believe that the high contrast between the nuclear and the extended emission,  shown in Figure \ref{fig:outflow}, makes the radial size not suitable for obtaining a reliable estimation of the outflow size,  at least in the case of this QSO2. Thus, measuring outflow sizes using the FWHM of outflow flux maps might lead to underestimations when the extended emission is faint.




We then measured the maximum outflow radius (R$_{\rm out}$) using the Pa$\alpha$ emission at $\geqslant$ 3$\sigma$ (see black contours in Fig. \ref{fig:outflow}). Using the \textit{Photutils} package of Python (\citealt{Bradley20}), we fit an ellipse to this $	\geqslant$3$\sigma$ emission and we define R$_{\rm out}$ as the distance between the peak of the Pa$\alpha$ emission and the most distant point from it in the ellipse (indicated in Fig. \ref{fig:outflow}). From this we obtain R$_{\text{{out}}}$ = 1.47$\arcsec$ ($\sim$3.37 kpc) with a PA=125$^{\circ}$. 
Since we cannot disentangle the emission from the intermediate and broad components in Fig. \ref{fig:outflow}, in the following we will consider R$_{\text{{out}}}$ as the outflow extent for both outflow components. In the case of the [Si~VI] line, which is blended with H$_2$1-0S(3), we cannot measure the outflow extension as we did for Pa$\alpha$.

Using the values of n$_{\rm e}$ reported in Table~\ref{tab:dens} and the integrated flux of the intermediate and broad components from the Voronoi maps, we measure the mass of the ionized outflow traced by Pa$\alpha$ using Eq.~1 in \citet{Rose18}:

\begin{equation}
\centering
\text{M}_{\rm out} = \frac{\rm L_{\rm H_\beta} \rm m_{\rm p}}{\alpha_{\rm H_\beta} ^{\rm eff}\, \rm h \, \rm \nu_{\rm H_\beta}\, \rm n_e} 
\end{equation}

 where L$_{\rm H\beta}$ is the H$\beta$ luminosity, $\alpha^{\text{eff}}_{\text{H}_{\beta}}$ the effective Case B recombination coefficient (see \citealt{Osterbrock06}), h the Planck's constant, m$_{\text{p}}$ the proton mass, and $\nu_{\text{H}_{\beta}}$ the frequency of H$\beta$.
Since it is written in terms of H$\beta$ luminosity, we converted our Pa$\alpha$ luminosity using the Case B recombination coefficient ( L$_{\rm H\beta}$ = L$_{\rm Pa\alpha}$/0.332), after correcting L$_{\rm Pa\alpha}$ from the extinction. To do that we used the E(B-V) values determined in Section~\ref{density}  and the model by \citet{Fitzpatrick99} using A$_\text{v}/$E(B-V)=3.1, as in \citet{Cardelli89}.
 The outflow masses are reported in Table~\ref{tab:dens}. The intermediate component has a mass of (5.5-6.6)$\times$10$^7$ M$_{\odot}$ considering the different electron densities measured with the two methods described in Section~\ref{nucleus}. Partly due to the higher gas density, the broad component has a lower mass of (0.13$\pm$0.02)$\times$10$^7$ M$_{\odot}$. We also measured the gas mass of the narrow component by using the corresponding electron densities, obtaining a value of (3.2-4.8)$\times$10$^7$ M$_{\odot}$. Thus, according to our analysis, approximately half of the total ionized gas mass in the central $\sim$6 kpc of J0945 would be outflowing.

Finally,  using the measured R$_{\text{{out}}}$ of $\sim$3.4 kpc, we assume a spherical or multi-cone geometry for the intermediate and broad components of the outflow to calculate $\dot{\text{{M}}}_{\text{{out}}}$ (see Eq.~B.2 in \citealt{Fiore17}).  Corresponding kinetic powers ($\dot{\text{{E}}}_{\text{{kin}}}$) are obtained using Eq.~4 in \citet{Rose18}.
We also estimate upper limits for these outflow properties using the maximum velocities calculated as v$_{\text{max}}$ = v$_{\text{out}}$ + 2$\sigma$ (\citealt{Rupke13, Fiore17,Puglisi21}), where $\sigma$ = FWHM/2.355. We used the values of {\bf $\Delta$v$_{\rm s}$ (i.e., v$_{\rm out}$)} and FWHM measured from the nuclear spectrum, which are reported in Table~\ref{tab:nuc} for the intermediate and broad component of Pa$\alpha$. We obtain v$_{\text{max}} = (- 840 \pm 28)$ \kms~for the intermediate component and v$_{\text{max}} = (- 1706 \pm 681)$ \kms~for the broad component. The corresponding mass outflow rates and kinetic powers are shown in Table~\ref{tab:dens}, together with the corresponding coupling efficiencies ($\xi$= $\dot{\text{{E}}}_{\text{{kin}}}/\text{L}_{\text{bol}}$).

\subsection{The warm molecular gas} \label{molecular}

From the H$_2$ lines that we detect in the K-band spectrum of J0945, we can estimate the warm molecular gas mass using the following relation from  \citet{Mazzalay13}:

\begin{equation}
\text{M}_{\text{H}_2} \simeq 5.0875 \times 10^{13} \left( \frac{\text{D}}{\text{Mpc}} \right)^{2} \left( \frac{\text{F}}{\text{erg} \,\text{s}^{-1}\text{cm}^{-2}} \right) 10 ^{0.4 \text{A}_k}
\end{equation}
where D is the luminosity distance, of 600.1 Mpc, $\text{A}_\text{k}\sim0.086$ mag is the extinction derived from the reddening determined from the \citet{Fitzpatrick99} model, using a factor of 3.1 as in \citet{Cardelli89} and the E(B-V) values determined in Section~\ref{density}, and F=(3.0$\pm$1.5)$\times 10^{-15} \,\text{erg} \,\text{s}^{-1}\text{cm}^{-2}$ is the total flux obtained from the map of H$_2$1-0S(1), shown in Fig.~\ref{fig:vorh1}. 
Using these values we obtained M$_{\text{H}_2} = (5.9 \pm 3.1) \times 10^4 \text{M}_{\odot}$. This value is much larger than the nuclear gas masses of between 4 and 74 M$_{\odot}$ measured by \citet{Mazzalay13} for a sample of six nearby low-luminosity AGN using data from VLT/SINFONI. A larger average gas mass, of 850 M$_{\odot}$, was reported by \citet{Riffel21} for a larger sample of nearby Seyfert 2 galaxies observed with Gemini/NIFS as part of the AGNIFS survey. Finally, our mass is similar to those measured for other QSO2s in the QSOFEED sample. \citet{Ramos17} reported a warm gas mass of 3 $\times 10^{3}$ M$_{\odot}$ for the Teacup (J1430+1339) using data from VLT/SINFONI, and based on data from GTC/EMIR, \citet{Ramos19} measured a gas mass of 1.9 $\times 10^4$ M$_{\odot}$ for J1509+0434.

As we already  mentioned in Section~\ref{nucleus}, the warm H$_2$ is just a small fraction of the total molecular gas residing in  galaxies. The  bulk of the molecular gas is cold and can be traced, for example, using the carbon monoxide transitions observed in the (sub)millimeter range (see e.g., the recent compilations by \citealt{Lamperti20, Esposito22}). J0945  was observed with the Atacama Pathfinder EXperiment (APEX) in CO(2-1). From this observation, \citet{Jarvis20} reported a molecular gas mass of (1.00 $\pm$ 0.22) $\times 10^{10}\text{M}_{\odot}$ by  assuming  R$_{21}$= L$^{\prime}_{\text{CO(2-1)}}$/L$^{\prime}_{\text{CO(1-0)}}$ = 0.8 and an $\alpha_{\rm CO}$=4.1 M$_{\sun}(\rm K~km~s^{-1}~pc^2)^{-1}$. Using this mass we derive  a warm-to-cold gas mass ratio of  M$_{\text{H}_2}^{\rm warm}$/M$_{\text{H}_2}^{\rm cold}$ = (5.9$\pm3.3) \times 10^{-6}$.
This ratio is in the midlle of the range of values reported by 
\citet{Dale05} for starburst galaxies and buried AGN  ($\simeq10^{-7}-10^{-5}$), and lower than the values measured for LIRGs and ULIRGs in the local universe (3 -- 6) $\times 10^{-5}$ (\citealt{Emonts14, Pereira16, Pereira18}). We note, however, that the mass of warm molecular gas corresponds to the central $\sim$6 kpc of the galaxy, whereas the APEX value corresponds to the whole galaxy.

\section{Discussion}  \label{discussion}

In this work we explore the kinematics of the warm molecular and ionized gas in the type-2 quasar J0945. Thanks to the capabilities of the NIFS instrument, apart from the analysis of the nuclear spectrum, which we extracted in an aperture of 0.69 kpc diameter, we are able to explore the multi-phase gas kinematics of the extended emission in the central 6.8 $\times$ 6.8 kpc$^2$. 
In this section we discuss our findings, putting them in context with previous results and comparing with ancillary observations. 




\subsection{Comparison with the [O~III] kinematics} \label{literature}
\label{comparison}

\citet{Harrison14} measured the ionized outflow properties in J0945, but using optical spectra observed with the GMOS (in) IFU and performing a non-parametric analysis  of the [OIII] lines. With this method they measured velocity widths based on the 10th and 90th percentiles of velocities (W$_{80} = \text{v}_{90}-\text{v}_{10}$). W$_{80}$ includes the 80\% of the emission line flux and it is approximately 1.09 $\times$ FWHM. For J0945 they measure W$_{80} = 1100$ \kms, that is in between the FWHM values measured for our intermediate and broad components (836 and 1703 \kms, respectively). Moreover, they considered a velocity offset based on the 5th and 95th percentiles of velocity ($\Delta\text{v} = (\text{v}_{05}+\text{v}_{95}$)/2) that it is similar to our velocity shift from the narrow component (v$_{\rm s}$). In Fig.~A1 in \citet{Harrison14} the velocity offset shows negative  velocities across the whole FOV of $\sim$8$\times$11 kpc$^{2}$. In the central 6.8$\times$6.8 kpc$^2$ (i.e., the NIFS FOV), they observed that the velocity increases preferentially along the south-east direction up to $-$300 \kms, in agreement with our results from the Voronoi maps (see Fig.~\ref{fig:vorpa}). Beyond the NIFS FOV, \citet{Harrison14} showed that the velocities reach the highest values at 2\arcsec~north and south from the nucleus. Therefore, the NIR spectra reveal signatures of an ionized outflow that is consistent with the one detected by \citet{Harrison14} in the optical. The parametric analysis we carried out allows us to distinguish between two different components of the outflow, the intermediate and the broad. These components, apart from being characterized by different blueshifts and widths, have also different densities, of $\sim$700 cm$^{-3}$ and $\sim$12600 cm$^{-3}$, respectively. 

\citet{Oliveira21} analyzed the same data as in \citet{Harrison14}, but performing a parametric fitting of the  [O~III] lines. They also found that three components are needed to characterize the emission line profiles and they argue that J0945 shows signatures of a second outflow component, in agreement with our results  for the ionized gas. They reported the presence of an additional narrow component that extends away from the nucleus, at $\ga$2 kpc to the east. We do not find this fourth component because we imposed that the Pa$\alpha$ and [Si~VI] lines detected in all the voxels always have the same three components that we find in the nuclear spectrum.

 If we compare the characteristics of the two blue-shifted broad components detected by \citet{Oliveira21} with the ones reported here, we find that the velocities that we measure for the intermediate component (see third row of Fig.~\ref{fig:vorpa}) are comparable with the velocity map shown in the fourth row of Fig.~A4 by \citet{Oliveira21}. In both cases the velocities of the outflowing gas increase in the south-east region, reaching values between  $-200$ and $- 250$ \kms~at $\sim$1.7 kpc from the center. 
In the case of the broad component the similarity is less clear due to its much lower flux, one order of magnitude lower than the intermediate component. The [O~III] broad component reported by \citet{Oliveira21} reaches lower velocities (up to -300 \kms) than those measured for Pa$\alpha$ (up to -600 \kms~from the Voronoi map), although entirely consistent within the errors. 


Regarding the outflow extent, using the methodology described in Section \ref{properties}, we measure a radius $\sim$3.4 kpc with PA$\sim$125$\degr$ (see Fig.~\ref{fig:outflow}). We consider this as the radius of the two outflow components because we cannot
disentangle the emission from the intermediate and broad components in Fig. \ref{fig:outflow}. \citet{Oliveira21} measured a smaller outflow radius, of <1 kpc, because they only considered the extent of the broadest component. However, they also found that the intermediate component reaches the highest velocities in the south-east region extending up to $\sim$3 kpc from the center, in agreement with our results. 

Therefore, we conclude that the ionized outflow kinematics measured from optical and NIR data are similar and the outflows are co-spatial, at least in the case of the intermediate component.


Since the two outflow components measured here have different densities (n$_e \sim 700$ and 12600 cm$^{-3}$) and velocities (v$_{\rm max} \sim$ -840 and -1700 \kms), they likely correspond to two distinct outflow phases. This is further supported by the fact that they show different geometries. In the case of the intermediate component, we only see the approaching side of the outflow, with the receding side most likely hidden by the host galaxy. For the broad component, we also see redshifted velocities in the line maps (see fifth row of Fig.~\ref{fig:vorpa}), although slower than the approaching gas and having large uncertainties. It these redshifted velocities are real, this outflow would be more coplanar with the host galaxy, making it possible to observe its receding side.

 Finally, we can compare the FWHMs measured for the two outflow components with those from different AGN samples, compiled by  \citet{Bischetti17} in their Fig.~5. J0945 shows comparable values of the FWHM for the intermediate and broad component (840 and 1700 \kms, respectively) as other QSO2 samples of similar L$_{\text{[O~III]}}$ at z<1, as e.g., \citealt{Villar11} and \citealt{Liuza13}. Fig.~5 in \citealt{Bischetti17} shows that the higher the AGN luminosity, the larger the FWHM of the outflows, although with large dispersion.



\subsection{QSO2s observed in the K-band} \label{compNIR}


 In our work, we detect just the ionized counterpart of the outflow in both low- and high-ionization emission lines, but not its warm molecular counterpart.  This was also the case for the Teacup QSO2, where \citet{Ramos17} detected broad and blue-shifted components with FWHM$\sim$1800 and 1600 \kms~in the Pa$\alpha$ and [Si~VI] emission lines, respectively, but not in the H$_2$ lines. The results for the Teacup and J0945 are in contrast with those found for the QSO2s J1509 and F08572+3915:NW. For J1509, \citet{Ramos19} reported clear signatures of outflowing gas in the warm molecular phase. In this case the molecular outflow is slower than the ionized outflow (i.e., $-$130 \kms~for H$_2$ vs $-$330 \kms~for Pa$\alpha$), and they have radial extensions of 1.46$\pm$0.20 kpc and 1.34$\pm$0.18 kpc, respectively. For F08572+3915:NW, a warm molecular outflow of 400 pc radius, directed along the galaxy minor axis, was reported by \citet{Rupke13}. They found maximum H$_2$ outflow velocities between $-$1300 and $-$1700 \kms, while for the ionized gas the maximum velocities are $\sim$-3350 \kms.  
 
    The AGN luminosity and/or the total warm molecular gas mass of the QSO2s are factors that could be driving the presence or not of warm molecular outflows. Focusing first on the AGN luminosity, and considering that these four QSO2s have practically the same values (see Table \ref{tab:mol}), L$_{\rm bol}$ does not seem to be a key factor driving the presence warm molecular outflows, despite the low number of QSO2s.

\begin{table}
\centering
\begin{tabular}{lccc}
\hline
\hline
QSO2  &  log L$_{\rm bol}$ &   H$_2$ outflow  &M$_{\text{H}_2}^{\text{tot}}$  \\
      & [erg~s$^{-1}$] &   detected                 &   [M$_{\odot}$]  \\
\hline
J0945+1737      &  45.7 (a)& No   & 5.9$\times10^4$ (d) \\
J1430+1339      &  45.5 (a)& No   & 1.0$\times10^4$ (e) \\
J1509+0434      &  45.7 (b)& Yes  & 1.9$\times10^4$ (b) \\
F08572+3915:NW  &  45.6 (c)& Yes  & $>$5.2$\times10^4$ (f) \\
\hline
\end{tabular}
\caption{QSO2s with NIR H$_2$ detections reported in the literature. Columns 2, 3, and 4 list the AGN luminosity, detection or not of a warm molecular outflow, and total H$_2$ mass measured in apertures of $\sim$2 kpc, except for F08572+3915:NW, for which the outflow mass is given as a lower limit. References: (a) \citet{Jarvis19}; (b) \citet{Ramos19}; (c) \citet{2020A&A...635A..47H}; (d) This work; (e) \citet{Ramos17}; (f) \citet{Rupke13}.}
\label{tab:mol} 
\end{table}

 If we compare the total masses of warm molecular gas, shown in Table \ref{tab:mol}, J1430 has the lowest, and J0945 has a similar warm molecular gas mass as the two QSO2s with warm molecular outflows detected. Therefore, we cannot conclude that QSO2s with larger warm molecular gas reservoirs are more likely to show molecular outflows. 
 New NIR observations of a larger number of QSO2s would be needed to assess if the detection of warm molecular outflows could be related to the content of warm molecular gas or AGN luminosity. From the results summarized in Table \ref{tab:mol} we do not find any tentative trend.

 Other factors that could contribute to launch warm molecular outflows might be the geometry of the ionized outflows and/or jets, and jet power (\citealt{Ramos21}). Molecular gas tends to be distributed in discs, with the warm molecular gas having larger velocity dispersion (\citealt{2019ApJ...884..171H}). This molecular gas is more likely to be entrained by the jets/ionized winds when the latter are coplanar with them, than when they subtend a larger angle (see Fig. 4 in \citealt{Ramos21}). Detailed studies of the multi-phase outflow geometries are needed to tackle this complex parameter space.
 
 

\subsection{Mass outflow rate and kinetic power} \label{fiore}

\begin{figure}
\centering
\includegraphics[width=0.49\textwidth]{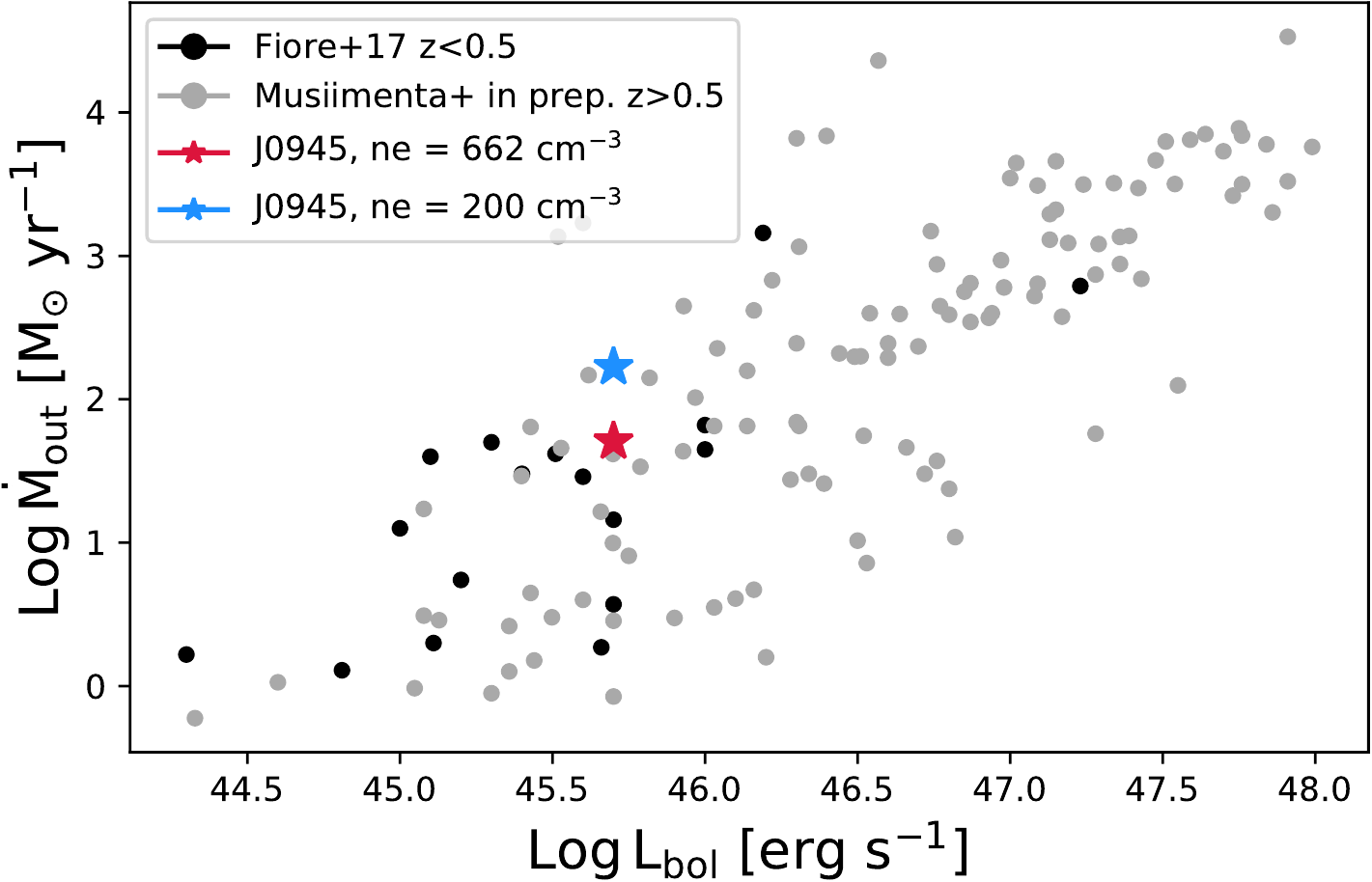}
\caption { Ionized mass outflow rates versus bolometric luminosities compiled from the literature by~\citet{Fiore17} for a sample of AGN with z<0.5 (black points), and by Musiimenta et al. (in prep.) for AGN at z>0.5 (grey points). The red and blue stars correspond to J0945 using the electron density calculated in Section~\ref{density} (n$_e$ = 662 cm$^{-3}$) and the value of 200 cm$^{-3}$ assumed by \citet{Fiore17}.}
\label{fig:fiore_mass}
\end{figure}

In our work we detected the outflow just in the ionized phase of the gas. To understand its impact on the host galaxy we measured its mass outflow rate, kinetic power, and coupling efficiency (see Table \ref{tab:dens}).
For the intermediate component, we measure an outflow rate ranging from 
6.6 to 7.9 $\text{M}_{\odot}~\text{yr}^{-1}$, depending on the density (i.e., calculated from the [SII] lines or using the trans-auroral lines), and a maximum outflow rate of 42--51 $\text{M}_{\odot}~\text{yr}^{-1}$. In spite of these high values, the kinetic power of the outflow represents a maximum of 0.1\% of the QSO2 bolometric luminosity. This coupling efficiency might appear small compared with the values considered in fiducial AGN feedback models (e.g., 5\% in \citealt{2005Natur.433..604D}). However, more recent simulations take into account that part of the wind energy has to be invested in escaping the galaxy's gravitational potential, leaving $\sim$0.5\% of L$_{\rm bol}$ in kinetic form (\citealt{2010MNRAS.401....7H,2018MNRAS.478.3100R}). 
Because of this and other considerations, there is no reason for the observed coupling efficiencies to match the values adopted in cosmological simulations (see \citealt{2018NatAs...2..198H} for further discussion).  Observed coupling efficiencies of $\sim$0.1\% have been reported in literature for different quasar samples, as shown in Fig.~7 in \citet{Bischetti17}. In any case, the kinetic power that we measure for the intermediate component of the outflow represents a considerable proportion of the quasar luminosity.

For the broad component, the outflow rate is 0.31$\pm$0.04 \text{M}$_{\odot}~\text{yr}^{-1}$, with $\dot{\text{{M}}}_{\text{{out, max}}}^{}$ =2.1$\pm$0.9 \text{M}$_{\odot}~\text{yr}^{-1}$. In this case, the maximum kinetic power of the outflow is 0.02\% of L$_{\rm bol}$.
Since the measurements obtained for the broad component have larger uncertainties, and the outflow geometry is not as clear as for the intermediate component, in the following we focus on the measurements obtained for the intermediate component.

The maximum outflow rate, of 42--51 $\text{M}_{\odot}~\text{yr}^{-1}$, is larger than the value measured by \citet{Oliveira21} from the GMOS IFU optical spectra. They found $\dot{\text{{M}}}_{\text{{out, max}}} = 28$ M$_\odot$ yr$^{-1}$, considering the same definition of the maximum outflow velocity we use in this work, and the same density, measured from the [S~II] doublet. 
Nevertheless, \citet{Oliveira21} showed that $\dot{\text{{M}}}_{\text{{out, max}}}$ can vary from 7.3 to 360 M$_\odot$ yr$^{-1}$ depending on different assumptions. 

Comparing $\dot{\text{{M}}}_{\text{{out, max}}}$ with the values  of other ionized outflows compiled by \citet{Fiore17} for AGN with redshifts z<0.5, we find that our target matches the highest values measured for AGN of similar bolometric luminosities. 
In Fig.~\ref{fig:fiore_mass} we show $\dot{\text{{M}}}_{\text{{out}}}$ vs L$_{\text{bol}}$, where the red star represents our target computing $\dot{\text{{M}}}_{\text{{out, max}}}$ with a gas density of 662 cm$^{-3}$  (as measured for the intermediate component, see Fig.~\ref{fig:denpy} and Table~\ref{tab:dens}), and the blue star the measurement obtained using n$_{\rm e}$ = 200 cm$^{-3}$ as in \citet{Fiore17}. We also plot as a comparison the data points obtained for a sample of $\sim$120 AGN at z>0.5 with ionised outflows detected in the [O~III] emission line (Musiimenta et al., in prep.). In this case n$_e$=200 cm$^{-3}$ has also been adopted. The value of $\dot{\text{{M}}}_{\text{{out, max}}}$ of J0945 is also among the highest values measured for AGN of similar bolometric luminosities in this higher-z sample. 
It is worth mentioning that the mass outflow rates that we are measuring for J0945 have been obtained from resolved measurements of the outflow properties, whereas in the case of \citet{Fiore17} and Musiimenta et al., in prep., most of the outflow mass rates come from unresolved measurements. This could lead to higher outflow rates in the case of J0945. For comparison, we can use the nuclear flux reported in Table \ref{tab:nuc} for the intermediate component of Pa$\alpha$ to calculate the outflow rate. If we use this flux, we measure $\dot{\text{{M}}}_{\text{{out, max}}}$=8 and 26 $\text{M}_{\odot}~\text{yr}^{-1}$ for the two densities considered in Fig. \ref{fig:fiore_mass}. These values are more similar to the average outflow rates of AGN of the same bolometric luminosity as J0945 compiled by \citet{Fiore17}.

 We found that half of the ionized gas that we detect in J0945 is outflowing, as shown in Table~\ref{tab:dens}.  
To investigate what fraction of this outflowing gas can escape the gravitational potential of the host galaxy (i.e., ejective feedback), we can estimate the escape velocity (v$_{\rm esc}$).
To do so, we assume a singular isothermal sphere potential, following \citet{Rupke02}, so the escape velocity at a given radius, r, is
\begin{equation}
v_{\rm esc}(r) = \sqrt{2}\text{v}_c[1 + \text{ln}(\text{r}_{\rm max}/\text{r})]^{0.5} 
\end{equation}

\noindent
v$_c$ is the circular velocity, and following \citet{Desay04}, it can be calculated from the line-of-sight stellar velocity dispersion as v$_c \approx 1.54 \sigma$. For J0945 we measure a maximum velocity dispersion of $\sim$145 \kms~for the rotating gas (i.e., we assume that the gas and stellar velocity dispersions are similar), that corresponds to a circular velocity of $\sim$223 \kms. Considering r=3.4 kpc (i.e., the outflow radius), and a dark matter halo of $\sim$100 kpc, r$_{\rm max}/r\sim$30, and v$_{\rm esc}\sim$660 \kms. 
This escape velocity is comparable with the values reported by \citet{Rupke02} for a sample of ULIRGs at z$<$0.3. Since we measure a maximum velocity of -840 \kms~for the intermediate component of Pa$\alpha$ in the nuclear spectrum, part of the ionized gas can potentially escape the galaxy's gravitational potential.

To quantify the amount of escaping gas we measured the escape fraction (f$_{\rm esc}$) as the fraction of the outflow with v>v$_{\rm esc}$, following \citet{Fluetsch19} and using our Pa$\alpha$ maps. Since most of the voxels in Fig. \ref{fig:vorpa} (third row) have v$_{\rm max}>v_{\rm esc}$, we find a high escape fraction, of $\sim$78\%. This value is higher than the 40-50\% escape fractions reported by \citet{Rupke02} for the ULIRGs. However, they used a more conservative outflow velocity (v$_{\rm max}$=$\Delta v_s$+FWHM/2) than ours (v$_{\rm max}$=$\Delta v_s$+2$\sigma$). Using the more conservative v$_{\rm max}$ definition, we obtain an escape fraction of $\sim$7\%, more similar to the maximum value of f$_{\rm esc}$ = 10\%  reported by \citet{Fluetsch19} for another sample of nearby AGN. Therefore, we conclude that part of the ionized gas in the outflow will escape the galaxy's gravitational potential, but depending on the velocity definition this fraction can vary from 7 to 78\%, with estimated uncertainties as large as 50\% according to \citet{Fluetsch19}.

 We measured the momentum outflow rate as $\dot{\text{P}}_{\text{out}} = \dot{\text{M}}_{\text{max}} \times \text{v}_{\text{out}}$. This value, compared with the radial pressure force ($\dot{\text{P}}_{\text{rad}} = \text{L}_{\text{bol}}/\text{c}$), can be used to estimate whether the outflow is energy- or momentum-driven. In the case of momentum-driven outflows the shocked inner wind radiates away its thermal energy, whilst in energy-driven outflows this thermal energy is preserved (\citealt{Costa14}). Values of $\dot{\text{P}}_{\text{out}}/\dot{\text{P}}_{\text{rad}} \approx 1$ are associated with  momentum driven outflows, 
whereas for the more efficient (i.e., faster, more extended) energy-driven outflows,  
$\dot{\text{P}}_{\text{out}}/\dot{\text{P}}_{\text{rad}} > 10$ are expected (\citealt{Faucher12}). For J0945, we obtained $\dot{\text{P}}_{\text{out}}/\dot{\text{P}}_{\text{rad}}=1.61$ considering  $\dot{\text{M}}_{\text{out}} = 51 \,\text{M}_{\odot} \text{yr}^{-1}$ and $\text{v}_{\text{max}} = -840$ \kms, measured for the intermediate component. According to this result, the outflow in J0945 is momentum conserving.

Focusing now on the  cold molecular gas component, if we use the empirical relation found by \citet{Fiore17},  i.e., a linear regression with a slope of $\sim$0.76, at the bolometric luminosity of J0945 (L$_{\rm bol}$=45.7 erg s$^{-1}$), we expect to have an outflow mass rate larger than 100 M$_\odot$ yr$^{-1}$. However, in a recent work by \citet{Ramos21} they found molecular outflow mass rates in the range 8-16 M$_\odot$yr$^{-1}$ for a small sample of QSO2s observed in CO(2-1) with ALMA. The difference between these values could be related to more or less conservative assumptions about the outflow properties, e.g., the outflow velocity, extension and geometry. Cold molecular gas observations of J0945 at high angular resolution would be necessary to prove if a cold molecular outflow is present in this QSO2 and to obtain reliable estimations of its mass outflow rate.




\subsection{An AGN-driven outflow} \label{origin}
We detect outflowing gas in the low-ionization (Pa$\alpha$ and Br$\delta$) and high-ionization ([Si~VI]) lines. These emission lines show broad and blue-shifted components in the nuclear and extended regions of the target. Measuring the mass-loading factor ($\eta = \dot{\text{{M}}}_{\text{{out}}}$/SFR) we found a maximum value of 0.70 for the intermediate component. Normally, mass-loading factors <1 are associated with SF-driven outflows instead of AGN-driven (e.g., \citealt{Schreiber19}), but considering the large uncertainties associated with the outflow and star formation rates, the mass loading factor is consistent with $\sim$1. Regardless, there are several reasons for which we believe that the outflow is AGN-driven. 

First, here we are just considering the ionized component of the outflow. 
If we use the \citet{Fiore17} relation to estimate the molecular outflow rate from the bolometric luminosity, we would have $\eta = 100/73 > 1$.
 In addition, the SFR is computed across the whole extension of the galaxy, while the region involved in the measurement of $\dot{\text{{M}}}_{\text{{out}}}$ does not extend more than R$_{\text{out}}$ (i.e. $\sim$ 3.37 kpc). This means we are comparing a mass outflow rate with a SFR which is not co-spatial.
 
Another argument in favour of an AGN-driven outflow is that we are observing very high blue-shifted velocities, of $-$840 and $-$1700 \kms~in the permitted (Pa$\alpha$ and Br$\delta$) and forbidden lines ([Si~VI]). Such high velocities can be hardly associated with SF-driven outflows since, for this driving mechanism, we generally expect velocities below 500 \kms~(e.g., \citealt{Chisholm15, Heckman15, Schreiber19}). Furthermore, the outflowing gas is extended (up to 3.37 kpc), with a clear direction towards the south-east  and it is detected in [Si~VI], which is a coronal line. This line is univocally associated with AGN activity, being either photoionized or excited by shocks produced by the interaction of jets with the surrounding gas. The latter mechanism is expected for gas at kpc-scale and with densities $>$100 cm$^{-3}$ (see e.g., \citealt{Ardila17}). 

\begin{figure}
\centering
\includegraphics[width=0.49\textwidth]{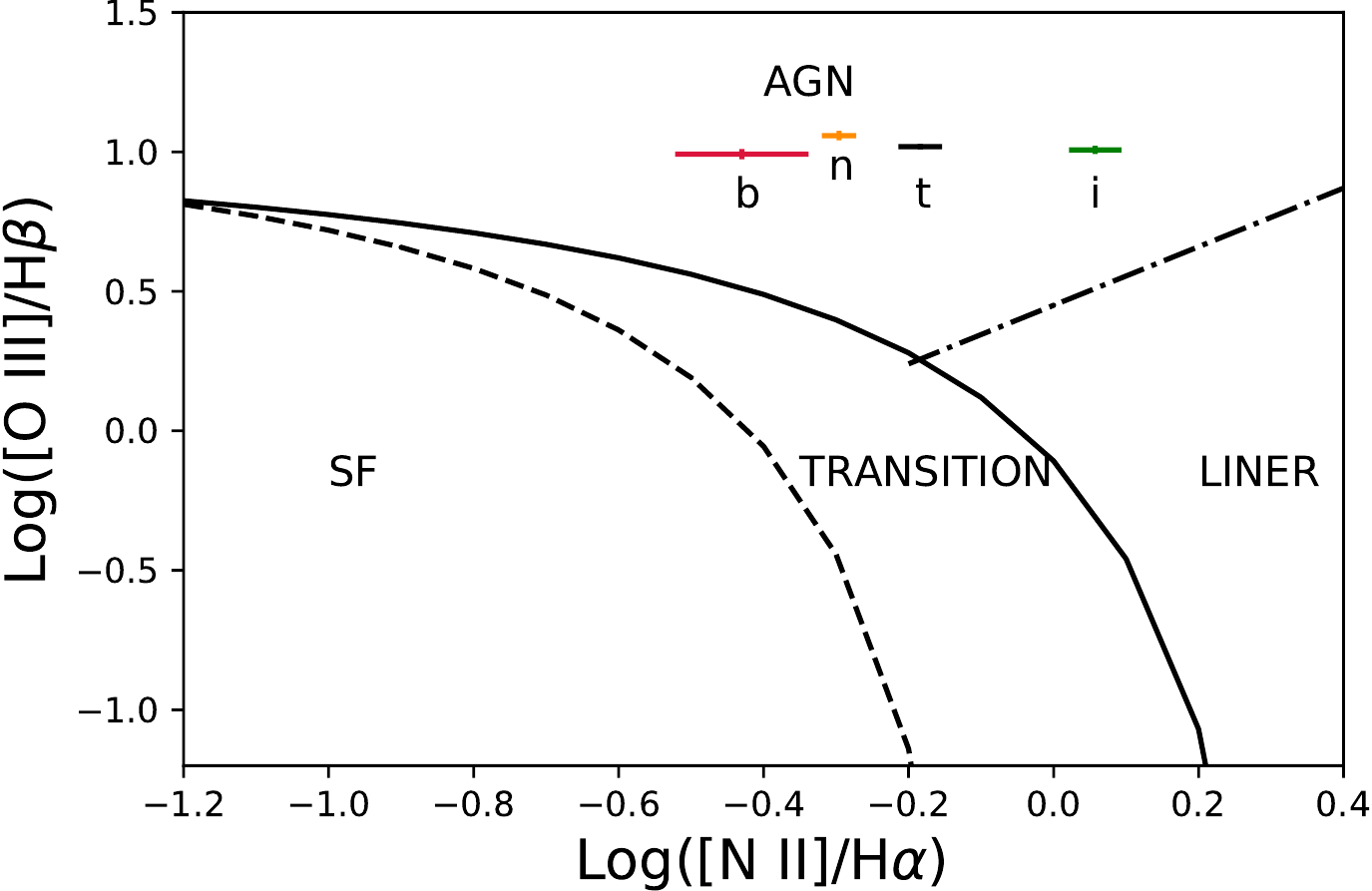}
\caption {Diagnostic diagrams for each component fitted from the SDSS spectrum of J0945, where the color code for the components is the same as in Fig.~\ref{fig:nuc}.
The lines separating AGN, transition objects, LINERs, and star-forming
galaxies  are from \citet{Kewley01} (solid curve), and \citet{Kauffmann03} (dashed curve), and \citet{Kewley06} (dash-dotted curve).
Error bars have been calculated from a series of 100 Monte Carlo simulations.
} 
\label{fig:bpt_diag}
\end{figure}

Finally, we computed flux ratios of the [O~III]/H$\beta$ and [N~II]/H$\alpha$ emission lines for each components (narrow, intermediate, and broad) and the total emission. The ratios were obtained by fitting the SDSS spectrum  (see Fig.~\ref{fig:nucop} for the fits of the [O~III] and H$\beta$ lines and Table~\ref{tab:nuc} for the measured fluxes) and can be used to see where the target is located in the Baldwin–Phillips–Terlevich diagnostic diagram (BPT diagram, \citealt{Baldwin81}). 
We obtain log$_{10}$([O~III]/H$\beta$) around 1
and log$_{10}$(N~II/H$\alpha$) from $\sim -$  0.4 to $\sim$ 0.06, thus lying in the AGN-dominated region of the diagram (see Fig.~\ref{fig:bpt_diag}). 

\begin{figure}
\centering
\includegraphics[width=0.49\textwidth]{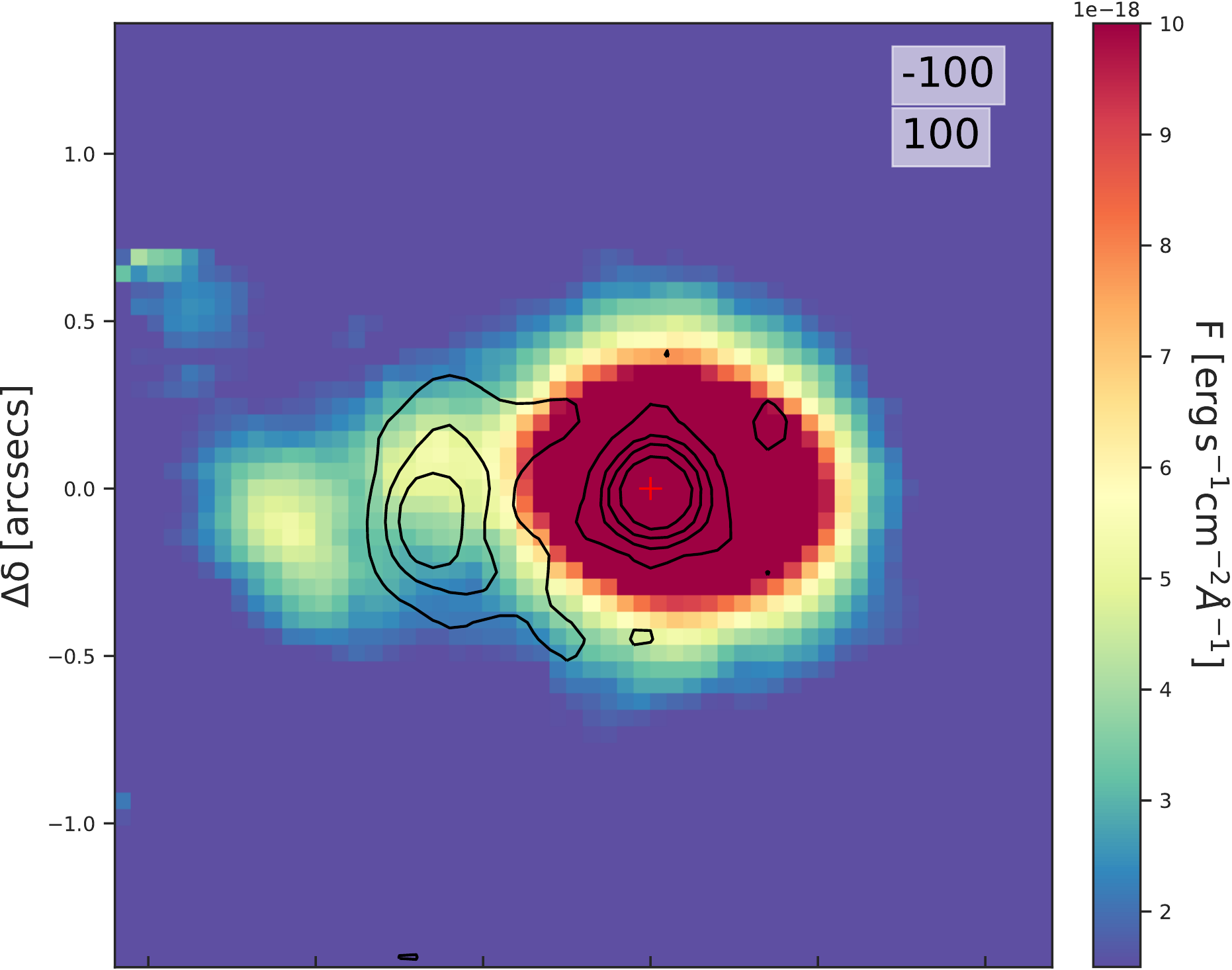}
\includegraphics[width=0.486\textwidth]{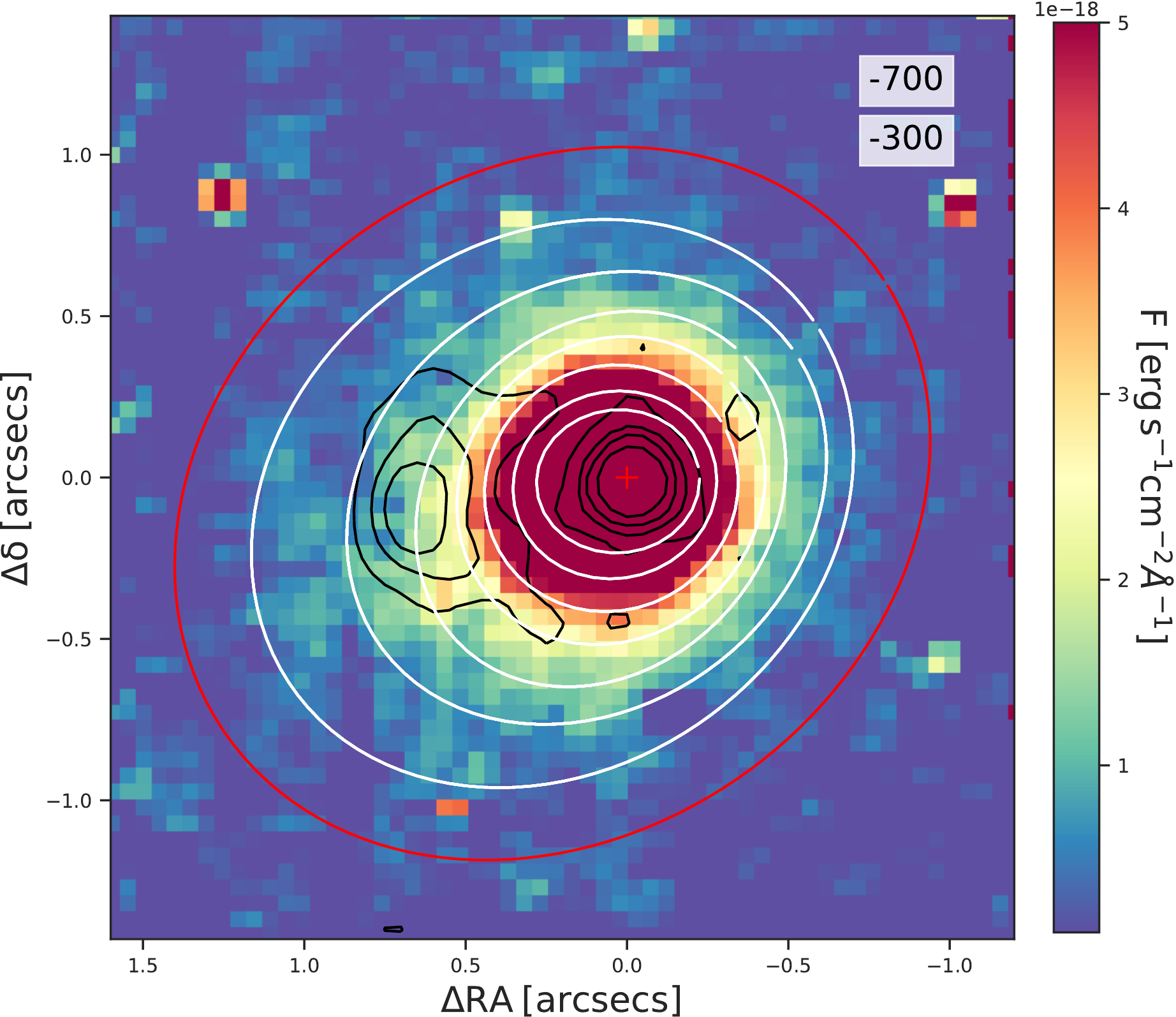}
\caption {Flux maps of the Pa$\alpha$ emission with contours at 3, 15, 30, 60 $\sigma$ of the radio VLA 6 GHz high resolution ($\sim$ 0.25 arcsec beam) image superimposed (black contours). Top panel: Systemic Pa$\alpha$ emission between $-$ 100 and 100 \kms. Bottom panel: flux map of the Pa$\alpha$ emission between $-$ 700 and $-$ 300 \kms with the fitted ellipses, at different distances from the central peak, in white and the ellipse used to measure the outflow extent in red. }
\label{fig:Pa_outflow_radio}
\end{figure}

\subsection{The role of the radio emission in J0945} \label{feedback}

The ionized outflow detected in J0945 is extending up to a distance of $\sim3.37$ kpc from the nucleus. The dominant feedback mechanism in a radio-quiet source as J0945 should in principle be the so-called ``quasar-mode'', e.g., shocks generated by the radiation pressure of the AGN accelerating the gas (\citealt{Fabian12}). Nonetheless, \citet{Jarvis19} reported the existence of a compact jet-like structure based on VLA data at 6 GHz ($\sim0.25$\arcsec~beam), and they suggested that it could be driving the outflowing gas.
This claim is based on the fact that the radio axis (PA $\sim 100^{\circ}$) is well aligned with the [O~III] emission (PA $\sim 75^{\circ}$). Indeed, these two PAs are even more similar if we consider the orientation of 90$^{\circ}$ reported by \citet{Storchi18} for the extended NLR of J0945, based on the HST [O~III] data. This alignment can also be seen when we compare the Pa$\alpha$ emission that is rotating (see top panel of Fig.~\ref{fig:Pa_outflow_radio}) with the VLA radio contours. The line-emitting gas
 exhibits some clumps in the eastern region, at > 0.5$\arcsec$ from the nucleus, and the outer edge of the radio jet coincides with a region between clumps or knots. This Pa$\alpha$ morphology could be produced by compression induced by the jet. At larger scales, of $\sim$11 kpc to the west, \citet{Storchi18} reported the presence of a line-emitting filamentary structure that, according to the comparison with new VLA observations presented in \citet{2021A&A...650A..84V}, could be also reminiscent of a bubble inflated by a radio source or a wide-angle AGN wind.

However, we find that this is not sufficient for assessing whether the outflowing gas is driven by the jet. The PA of the [O III] emission mentioned above ($\sim$90-100\degr) mostly corresponds to NLR emission. This gas is under the galaxy's gravitational potential, and it is ordinary rotating around the nucleus (see scan maps in Fig. \ref{fig:cutmap} of Appendix \ref{Appendix C}). Indeed, there is a well-known alignment between the line-emitting gas in the NLR and the radio emission in AGN (see e.g. \citealt{1996ApJ...469..554C}).

In order to gather additional evidence that the ionized outflow is driven by the radio jet we superimposed the VLA radio contours to the Pa$\alpha$ flux map extracted from the velocity range between $-$700 and $-$300 \kms. As explained in Section~\ref{properties}, these velocities are associated with the outflowing gas, so the flux map should be dominated by the outflow. This is shown in the bottom panel of Fig.~\ref{fig:Pa_outflow_radio}, where we also include the ellipses fitted using the \textit{Photutils} package to indicate how the PA of the high-velocity gas varies from the center outwards. In the innermost part, up to $\sim$0.4\arcsec~(0.91 kpc) from the peak of Pa$\alpha$, the PA is coincident with the axis of the radio emission ($\sim$95-120\degr), then it increases up to $\sim$135\degr~at 0.5-0.7\arcsec~(1.4 kpc), and it finally stabilizes at PA$\sim$125\degr~up to a distance of $\sim$1.3\arcsec~(3 kpc). Thus, the radio emission is elongated in the same direction as the high-velocity gas in the innermost region, and at the south-eastern edge of the radio jet, the PA of the outflow shows higher values. As mentioned above, this line-emitting gas morphology can be due to compression and acceleration induced by shocks created by the passage of the radio jet. 
 Based on this, we argue that the radio jet could have an impact in the kinematics of the outflowing gas, as it has been claimed for other nearby radio-quiet QSO2s (e.g., \citealt{2014MNRAS.440.3202V,2021A&A...650A..84V,Jarvis19,2022MNRAS.tmp..170G}).

 \section{Summary and conclusions} \label{conclusion}
 
We have analyzed the NIR spectrum of the radio-quiet QSO2 J0945, observed with the NIFS integral field spectrograph. The main goal of this work is to analyze  the kinematics of the ionized and warm molecular gas. Taking advantage of the capabilities of NIFS IFU data we explored, in addition to the nuclear emission within the central $\sim0.3\arcsec$ ($\sim 0.69$ kpc), the extended emission across the whole FOV of $3\arcsec \times 3\arcsec$~($\sim6.8 \times 6.8$ kpc$^2$). The main conclusions are the following. 

\begin{itemize}

\item We measure two blue-shifted components of FWHM $\sim$800 and $\sim$1700 \kms~in the permitted (Pa$\alpha$ and Br$\delta$) and forbidden lines ([Si~VI]) in both the nuclear spectrum and across the NIFS FOV. This confirms the detection of the ionized outflow first observed in [O~III] from optical spectra by \citet{Harrison14} and reveals, for the first, time its coronal counterpart.

\item We interpret the two blue-shifted components as two distinct outflow phases due to their different characteristics. They are blue-shifted by $-$130 and $-$260 \kms~from the narrow component and they have densities of n$_{\rm e}$ $\sim$ 700 and 12600 cm$^{-3}$.  
The broader and more blue-shifted counterpart shows a complex geometry with a maximum velocity of $-$1700 \kms. The other outflow component is preferentially orientated in the SE direction (PA$\sim125^{\circ}$),  having a maximum velocity of $-$840 \kms. 

\item For the two outflow components we measure a maximum extension of 3.37 kpc. We find that the radial outflow size, which in the case of J0945 is unresolved, can lead to severely underestimated outflow size measurements when the outflow emission is faint compared with the nuclear emission of the quasar.

\item The line-emitting gas moving in the gravitational potential of the galaxy and not participating in the outflow shows signatures of morphological disturbance.
We see a tail of emission at $\sim$1.3\arcsec~(3 kpc) east of the nucleus, and another tail extending to the north-east. Additionally, the velocity field studied here show signatures of kinematically disrupted gas that most likely correspond to the innermost part of the merging event that this galaxy is undergoing.

\item The extended radio emission shown by VLA data is elongated in the same direction as the clumpy gas in the NLR up to $\sim 1.6$ kpc, and it is also aligned with the high-velocity gas in the central 0.9 kpc. Beyond the central kpc, coinciding with the south-eastern edge of the radio jet, the PA of the outflow increases. We then argue that there is likely  a connection between the radio jet and the ionized outflow in J0945.


\item Considering the maximum value of the outflow mass rate, we find a mass-loading factor of $0.70$ ($\dot{\text{{M}}}_{\text{{out}}}/$SFR = 51/73) for the ionized outflow. We claim that the outflow is AGN-driven because of 1) the high velocities of the gas, of up to $-$840 and $-$1700 \kms; 2) the collimated structure of the outflow,  which shows an extension of $\sim 3.4$ kpc to the south-east, as shown in Fig.~\ref{fig:outflow}; 3) the detection of broad components in the coronal line of [Si VI]; and 4) the position of the outflow components in  the AGN-dominated region of BPT diagram.


\item We measure a maximum outflow rate of 42--51 $\text{M}_{\odot}~\text{yr}^{-1}$ for J0945, which is among the highest values compiled by \citet{Fiore17} for AGN at z<0.5 with similar bolometric luminosities. The outflow kinetic power corresponds to $\sim$0.1\% of the QSO2 bolometric luminosity.

\item The warm molecular gas traced by the H$_2$ lines can be characterized with a single narrow component. From this we measure a mass of ($5.9 \pm 3.1)\times 10^4$ M$_{\odot}$, which gives a low warm-to-cold mass ratio of M$_{\text{H}_2}/\text{M}_\text{cold} = (5.9 \pm 3.3) \times 10^{-6}$ when we compare with the amount of cold gas measured from APEX CO observations.

\end{itemize}

J0945 constitutes another example of radio-quiet QSO2 in which radio jets might be contributing to drive ionized outflows, reaching a maximum outflow rate of 51 $\text{M}_{\odot}~\text{yr}^{-1}$. We have not detected a warm molecular gas counterpart of this outflow, but data in the mm range are important to confirm/discard the presence of a  cold molecular outflow (see e.g., \citealt{Ramos21} for the case of the Teacup QSO2), more relevant in terms of mass than the ionized gas phase (see e.g., \citealt{2021MNRAS.505.5753F}).  Detailed multi-wavelength studies are important to have a comprehensive view of AGN feedback and its impact on the host galaxy. 


\begin{acknowledgements}
Based on observations obtained at the international Gemini Observatory, a program of NSF’s NOIRLab, which is managed by the Association of Universities for Research in Astronomy (AURA) under a cooperative agreement with the National Science Foundation. on behalf of the Gemini Observatory partnership: the National Science Foundation (United States), National Research Council (Canada), Agencia Nacional de Investigaci\'{o}n y Desarrollo (Chile), Ministerio de Ciencia, Tecnolog\'{i}a e Innovaci\'{o}n (Argentina), Minist\'{e}rio da Ci\^{e}ncia, Tecnologia, Inova\c{c}\~{o}es e Comunica\c{c}\~{o}es (Brazil), and Korea Astronomy and Space Science Institute (Republic of Korea).
This action has received funding from the European Union’s Horizon 2020 research and innovation programme under Marie Sk\l odowska-Curie grant agreement No 860744 (BID4BEST). GS and CRA acknowledge the project ``Feeding and feedback in active galaxies'', with reference
PID2019-106027GB-C42, funded by MICINN-AEI/10.13039/501100011033. CRA also acknowledges support from the projects ``Quantifying the impact of quasar feedback on galaxy evolution'', with reference EUR2020-112266, funded by MICINN-AEI/10.13039/501100011033 and the European Union NextGenerationEU/PRTR, and from the Consejer\' ia de Econom\' ia, Conocimiento y Empleo del Gobierno de 
Canarias and the European Regional Development Fund (ERDF) under grant ``Quasar feedback and molecular gas reservoirs'', with reference ProID2020010105, ACCISI/FEDER, UE.
RAR acknowledges the support from Conselho Nacional de Desenvolvimento Cient\'ifico e Tecnol\'ogico  and Funda\c c\~ao de Amparo \`a pesquisa do Estado do Rio Grande do Sul. GS and CRA thank Santiago García Burillo for useful advice on determining the near/far side of the galaxy and Anelise Audibert for alternative calculations of the SFR. JP acknowledges support from the Science and Technology Facilities Council (STFC) via grant ST/V000624/1.


\end{acknowledgements}

\medskip
\bibliography{./sample}
 
\appendix
\section{Fits of optical emission lines}
\label{Appendix A}

\begin{figure}[!h]
\centering
\includegraphics[width=0.50\textwidth]{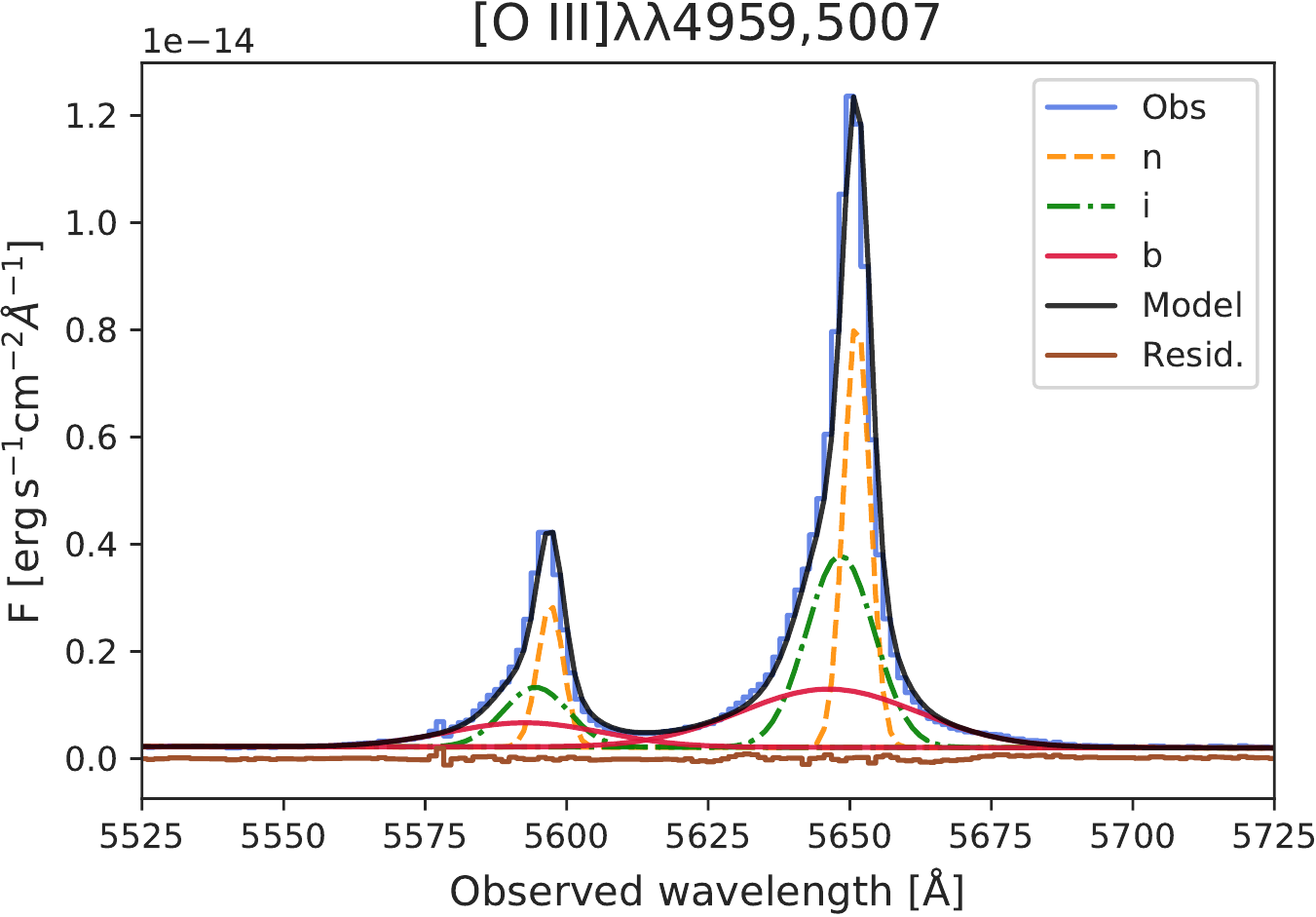}
\includegraphics[width=0.49\textwidth]{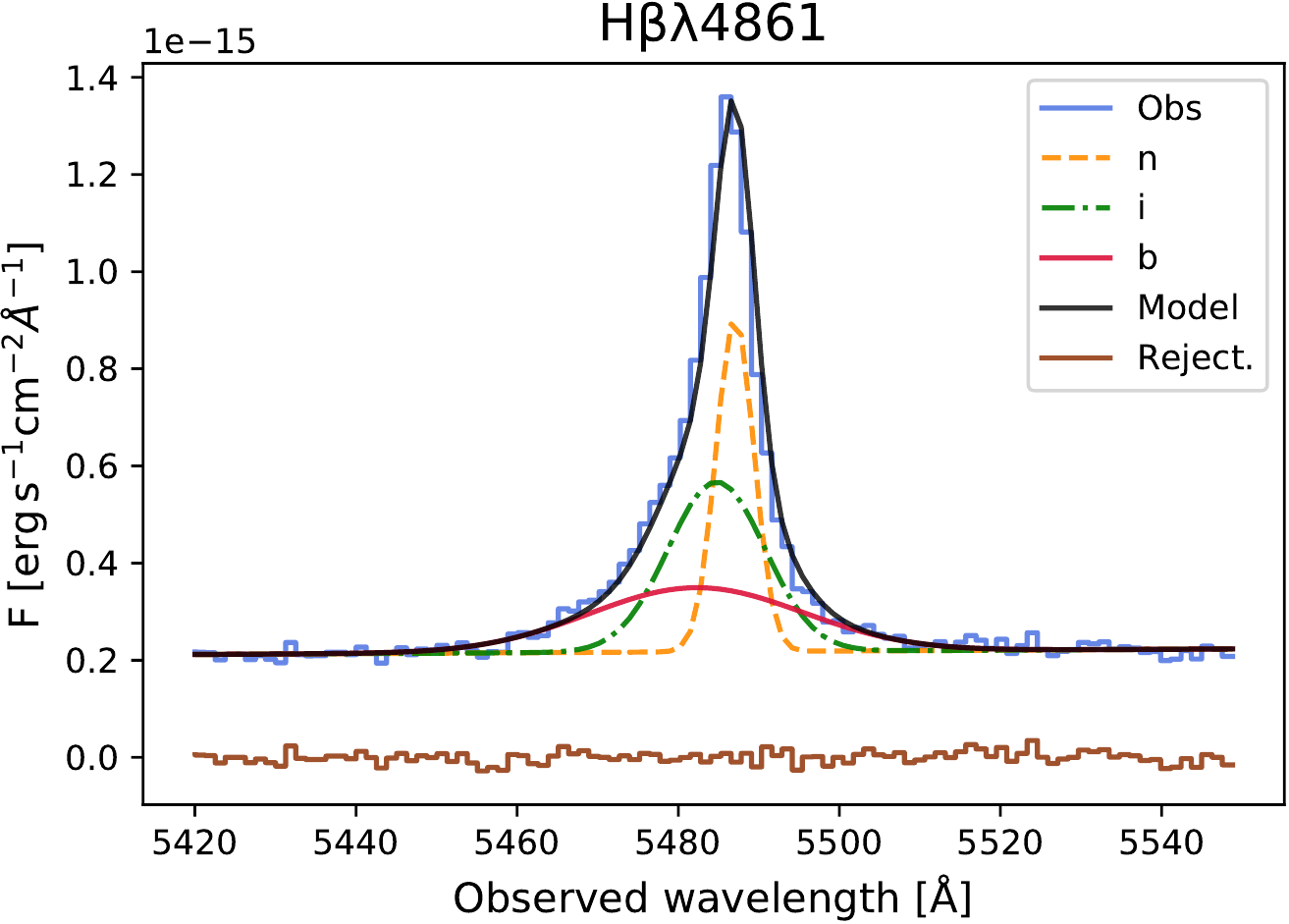}
\caption {Fits of emission lines in the optical SDSS spectrum. Top panel: the [O~III]$\lambda\lambda$4959,5007 \AA~doublet modelled using three Gaussians per line, which correspond to a narrow, intermediate and broad component. Bottom panel: fit of H$\beta$, which is also characterized by the three kinematic components.}
\label{fig:nucop}
\end{figure}

\begin{figure*}
\centering
\includegraphics[width=0.49\textwidth]{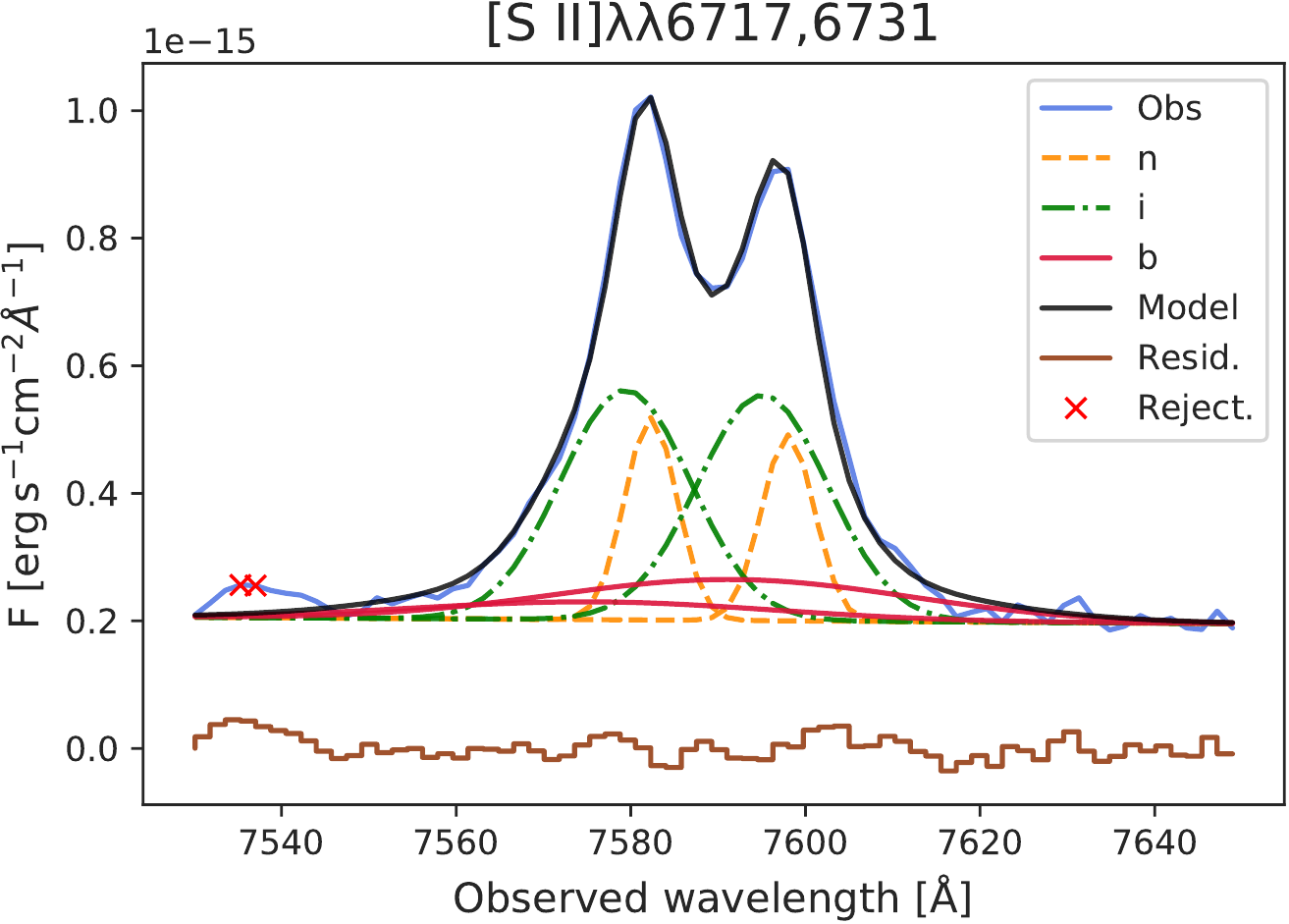}
\includegraphics[width=0.49\textwidth]{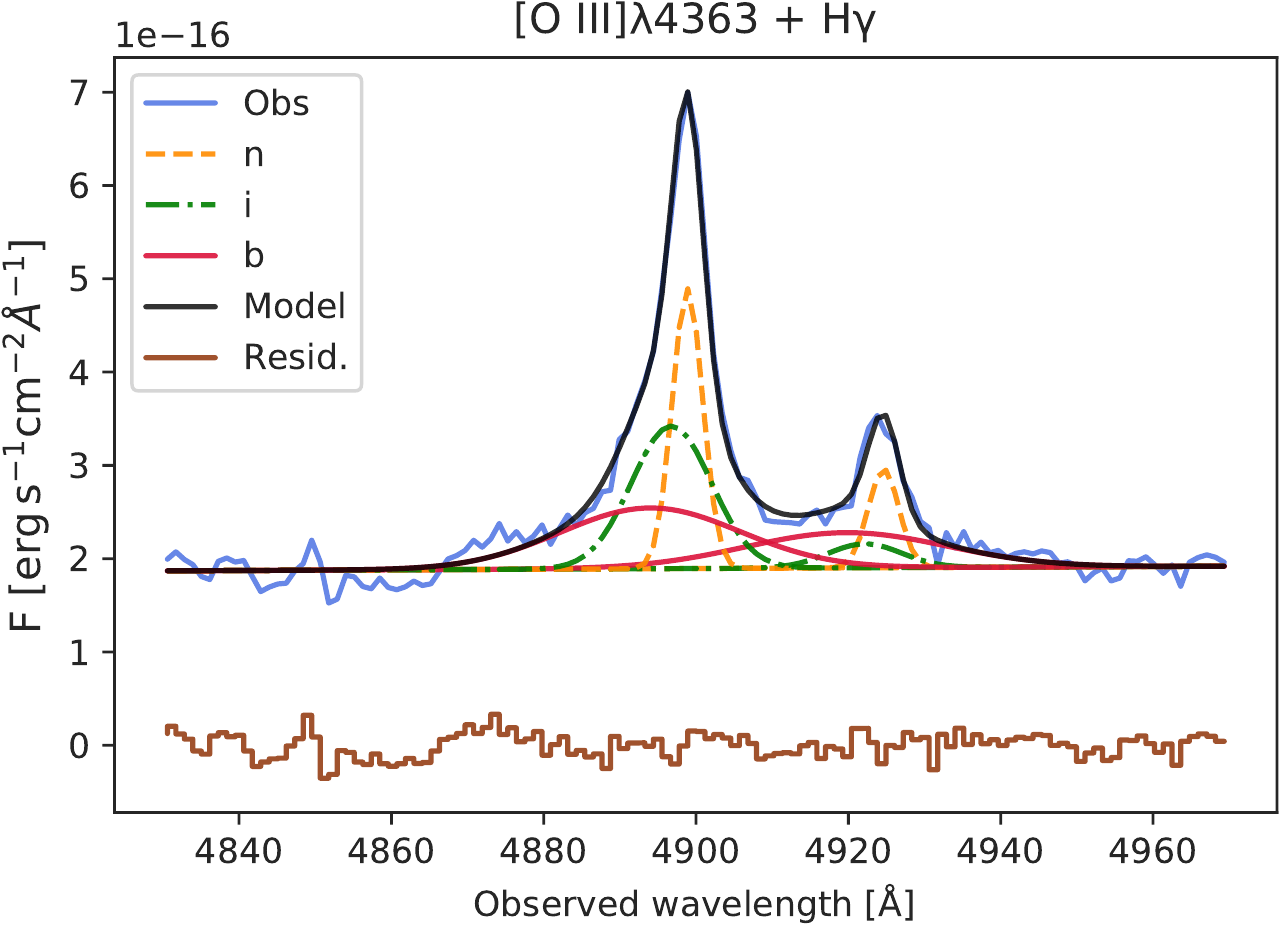}
\caption {Same as in Fig.~\ref{fig:nucop}, but for the [S~II]$\lambda\lambda$ 6717,6731 \AA~doublet (left panel) and the [O~III] and H$\gamma$ lines (right panel). Red crosses correspond to points that have been removed from the fit. }
\label{fig:fitpy}
\end{figure*}

\begin{figure*}
\centering
\includegraphics[width=0.49\textwidth]{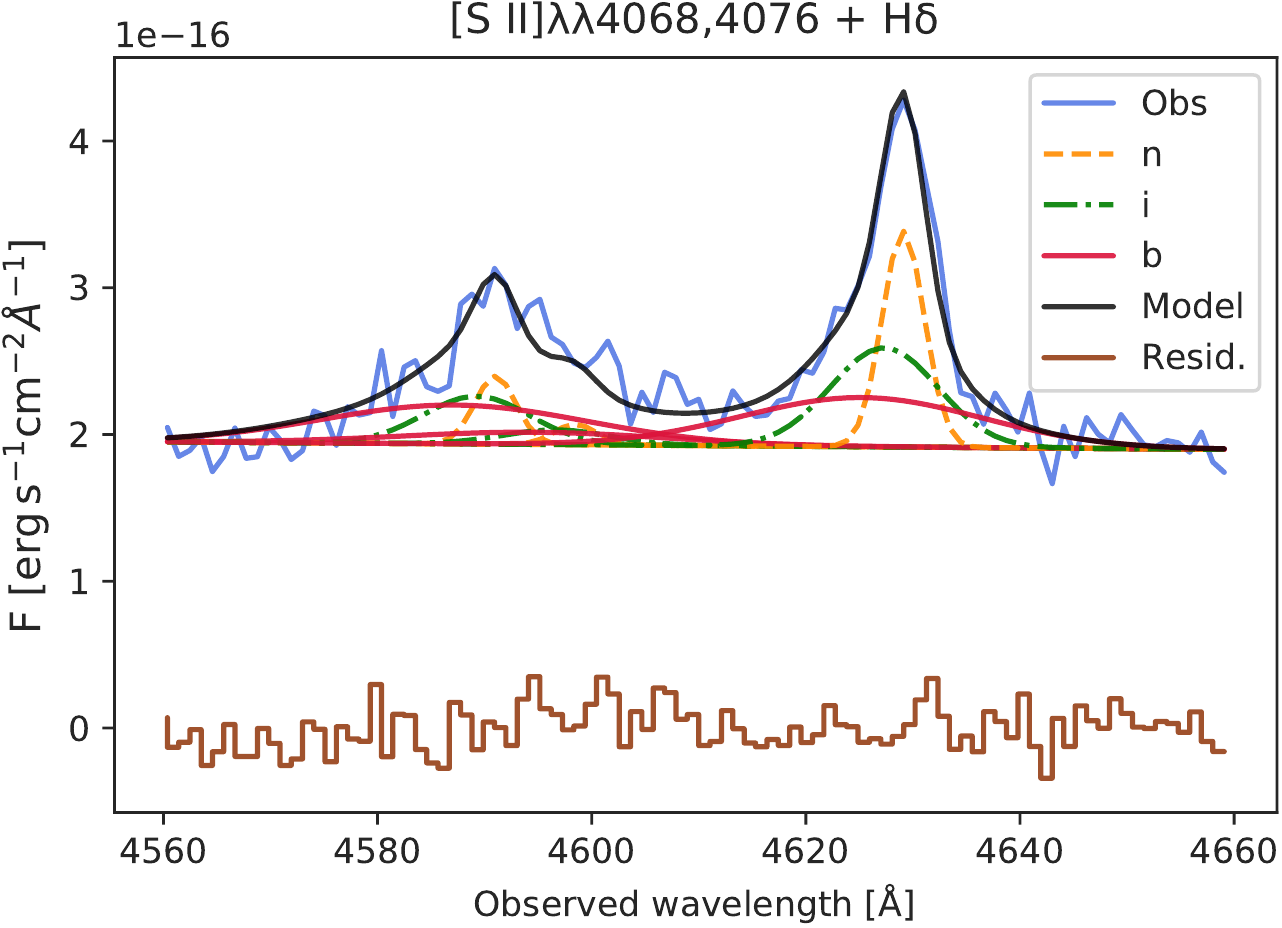}
\includegraphics[width=0.50\textwidth]{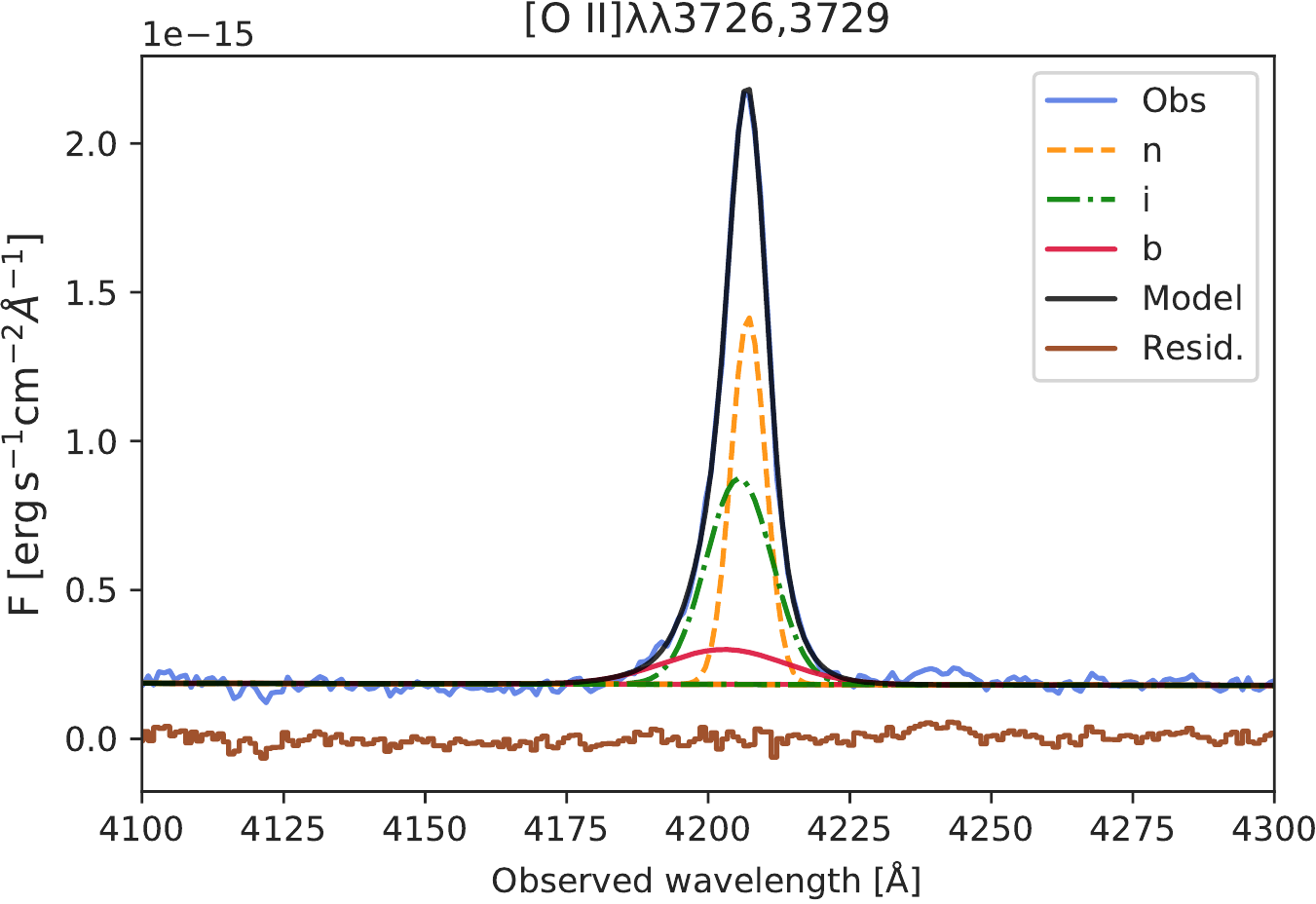}
\includegraphics[width=0.49\textwidth]{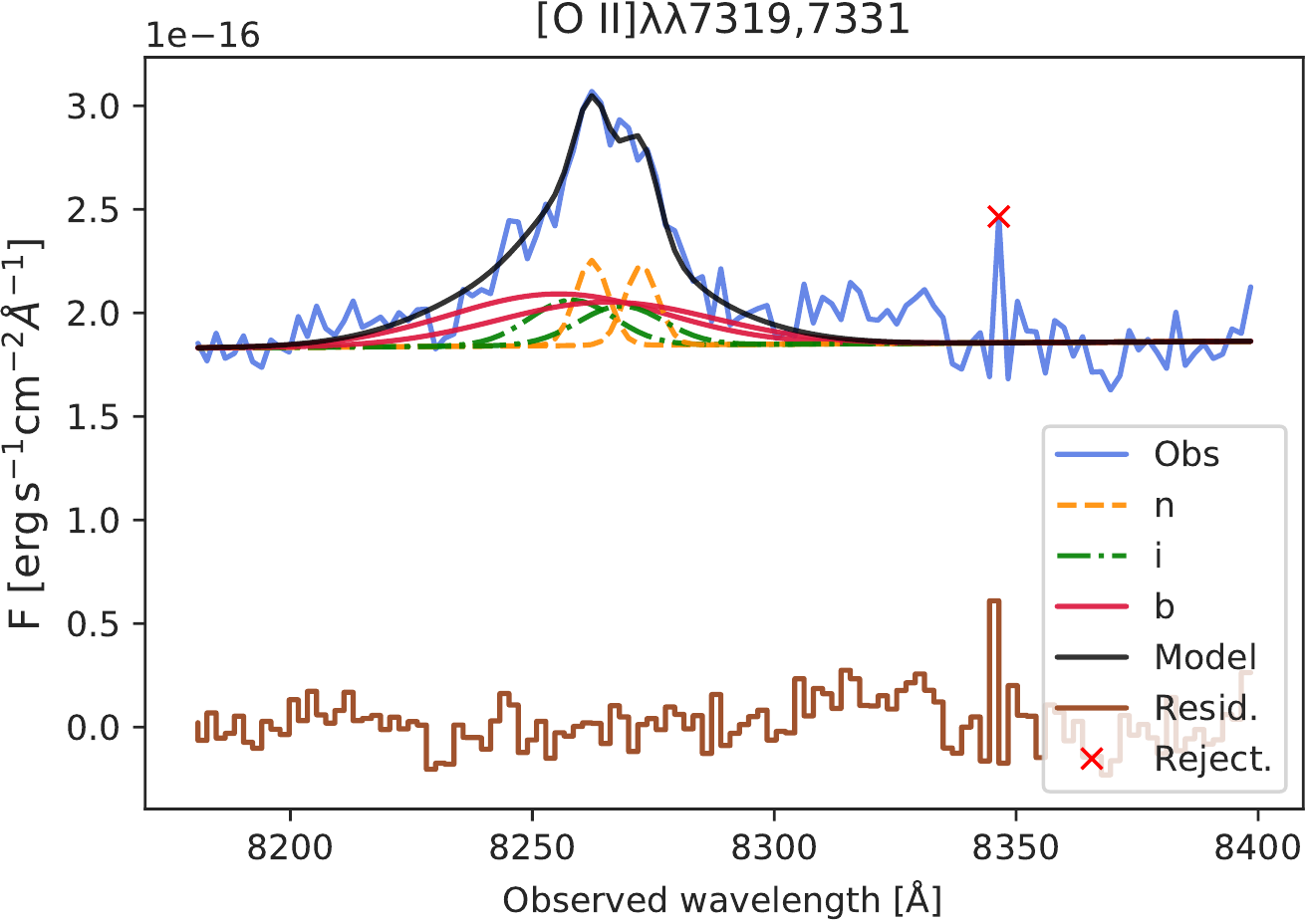}
\caption {Same as in Fig.~\ref{fig:nucop}, but for the [S~II]$\lambda\lambda$ 4068,4076 \AA~doublet and H$\delta$ profile (top left panel), the [O~II]$\lambda\lambda$3736,3729 \AA~doublet (top right panel), and the [O~II]$\lambda\lambda$7319,7331 \AA~doublet (bottom panel). Red crosses correspond to points that have been removed from the fit. }
\label{fig:fittrans}
\end{figure*}

Here we show the fits of the optical emission lines detected in the SDSS spectrum of J0945 (3$\arcsec$ of aperture) that we used as reference for the fits of the K-band emission lines (see Fig.~\ref{fig:nucop}). Three components are needed to reproduce the [O~III] and the H$\beta$ profiles. Corresponding fluxes, v$_{\rm s}$, and FWHMs are included in Table~\ref{tab:nuc} of Section \ref{nucleus}. Next, we show the fits of the different optical emission lines detected in the SDSS spectrum, necessary to obtain the electron densities reported in  Table~\ref{tab:dens}. The left panel of Fig.~\ref{fig:fitpy} displays the fit of the [S~II]$\lambda\lambda$6717,6731 \AA~doublet, and the right panel of the same figure, the fit of the [O~III]$\lambda$4363 \AA~line. 
To model [O~III]$\lambda$4363 \AA, we simultaneously fitted the H$\gamma$ profile. All these emission lines are fitted with three Gaussian components. 
In Fig.~\ref{fig:fittrans} we show the fits of the trans-auroral lines, i.e., the [S~II]$\lambda\lambda$4068,4076 \AA~and [O~II]$\lambda\lambda$7319,7331 \AA~doublets, and also of the [O~II]$\lambda\lambda$3726,3729 \AA~doublet. Their profiles were also fitted with three Gaussian components, i.e., six components in total for each doublet. The exception is the [O~II]$\lambda\lambda$3726,3729 \AA~doublet, which we cannot resolve with the spectral resolution of the SDSS spectrum. In this case, we used three components in total. We also modelled the H$\delta$ profile on the red side of the [S~II] doublet.
Since the trans-auroral lines are faint, we used as reference the fits of the [O~III]$\lambda$5007 \AA~and H$\beta$ profiles. Table~\ref{tab:sdss} shows the fluxes, v$_{\rm s}$, and FWHMs of all the kinematic components here described. The only exception are [O~III] and H$\beta$, for which they are shown in Table~\ref{tab:nuc}.  
 
\begin{table}[!h]
\caption{Emission line properties measured from the SDSS spectrum.}
\centering
\small
\begin{tabular}{lccc}
\hline
\hline
Line   &  FWHM  & $\Delta$$\Delta$v$_s$ & Flux $\times10^{-15}$  \\
      &  [km s$^{-1}$]   & [km s$^{-1}$]   &   [erg cm$^{-2}$s$^{-1}$]  \\
\hline
$[\text{S~II}]\lambda\lambda 6717,6731 \AA$ (n) & $278$     & $0$   & $~~4.68\pm0.19$ \\
$[\text{S~II}]\lambda\lambda 6717,6731 \AA$ (i) & $692$    & $-114$   & $13.30\pm0.43$  \\
$[\text{S~II}]\lambda\lambda 6717,6731 \AA$ (b) & $1887$ & $-250$ & $~~4.69\pm0.76$ \\
$[\text{O~III}]\lambda 4363 \AA$         (n) & $288$ & $1$ & $~~0.53\pm0.09$ \\
$[\text{O~III}]\lambda 4363 \AA$         (i) & $693$ & $-150$ & $~~0.31\pm0.20$ \\
$[\text{O~III}]\lambda 4363 \AA$         (b) & $1900$ & $-300$ & $~~1.25\pm0.32$ \\
$[\text{S~II}]\lambda\lambda 4068,4076 \AA$ (n) & $294$     & $0$   & $~~0.29\pm0.12$ \\
$[\text{S~II}]\lambda\lambda 4068,4076 \AA$ (i) & $760$    & $-114$   & $~~0.52\pm0.33$  \\
$[\text{S~II}]\lambda\lambda 4068,4076 \AA$ (b) & $1905$ & $-250$ & $~~1.08\pm0.36$ \\
$[\text{O~II}]\lambda\lambda 3726,3729 \AA$ (n) & $511\pm22$     & $~~~~~~69\pm10$   & $~~5.78\pm0.98$ \\
$[\text{O~II}]\lambda\lambda 3726,3729 \AA$ (i) & $979\pm56$    & $~~~~-90\pm22$   & $12.53\pm1.68$  \\
$[\text{O~II}]\lambda\lambda 3726,3729 \AA$ (b) & $1810\pm9$ & $-332\pm2$ & $~~4.36\pm0.59$ \\
$[\text{O~II}]\lambda\lambda 7319,7331 \AA$ (n)  & $290$     & $0$   & $~~0.68\pm0.22$ \\
$[\text{O~II}]\lambda\lambda 7319,7331 \AA$ (i) & $748$    & $-150$   & $~~0.90\pm0.17$  \\
$[\text{O~II}]\lambda\lambda 7319,7331 \AA$ (b) & $1900$ & $-260$ & $~~2.55\pm0.39$ \\

\hline
\end{tabular}
\tablefoot{Column description: (1) emission line name followed by the kinematic component: narrow (n), intermediate (i) or broad (b); (2) corresponding FWHM; (3) velocity shift ($\Delta$v$_s$) relative to the central $\lambda$ of the narrow component of Pa$\alpha$; (4) integrated flux. Measurements without errors correspond to parameters that have been fixed. In the case of emission-line doublets, column (4) corresponds to the sum of the flux of the single emission lines.}
\label{tab:sdss} 
\end{table}

\section{[Si~VI] and H$_2$1-0S(3) maps}
\label{Appendix B}

\begin{figure*}
\centering
\includegraphics[width=0.65\textwidth]{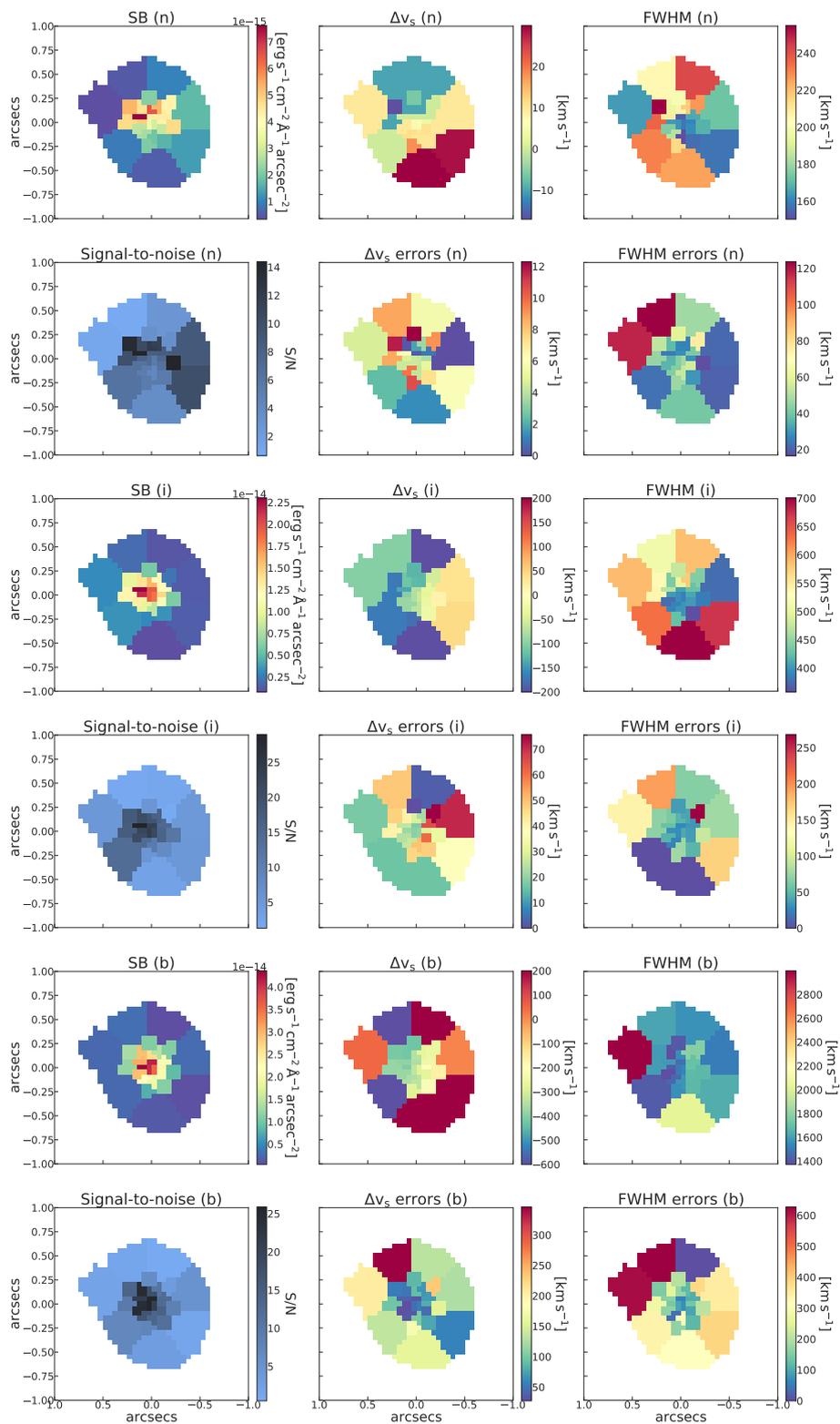}
\captionsetup{width=.65\linewidth}
\caption {Same as in Figure~\ref{fig:vorpa}, but for the [Si~VI] emission line. We imposed the narrow component of [Si~VI] to have a velocity shift consistent with the one measured for Pa$\alpha$. }
\label{fig:vorsi}
\end{figure*}

\begin{figure*}
\centering
\includegraphics[width=0.8\textwidth]{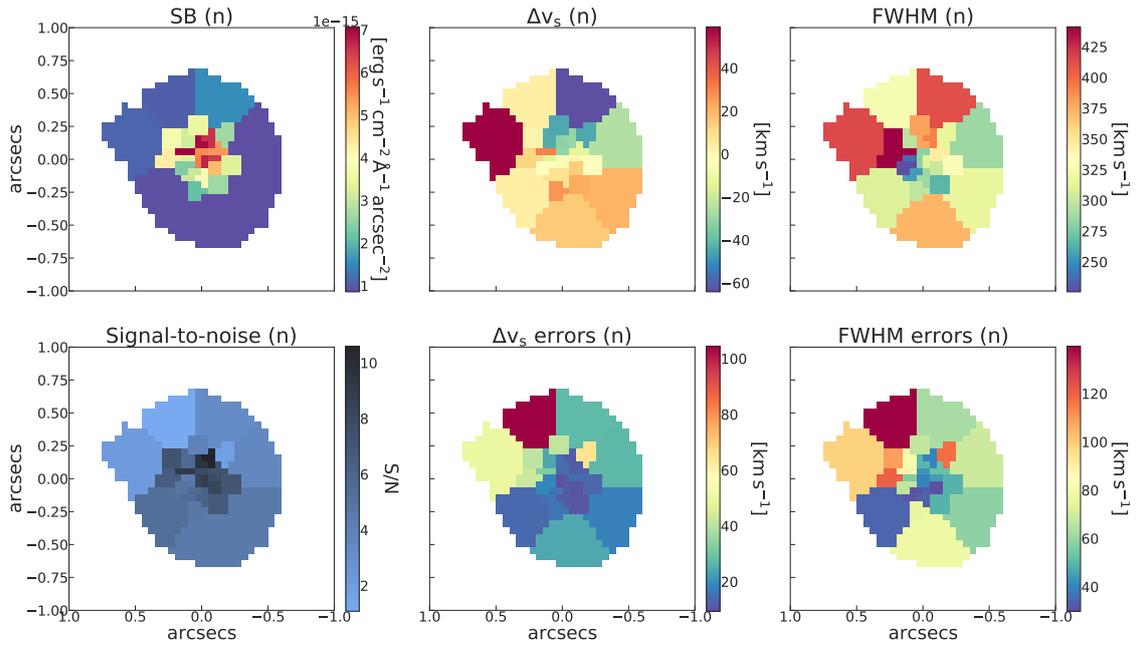}
\captionsetup{width=.8\linewidth}
\caption {Same as in Figure~\ref{fig:vorpa}, but for the H$_2$1-0S(3) emission line. Only one kinematic component was used to reproduce the emission line profiles across the whole FOV.}
\label{fig:vorh3}
\end{figure*}

In Section~\ref{voronoi} we show the surface brightness (SB), velocity and FWHM maps for each of the kinematic components fitted to the Pa$\alpha$ and and H$_2$1-0S(1) profiles. Here we show the maps of the [Si~VI] and H$_2$1-0S(3) lines (see Figs.~\ref{fig:vorsi} and ~\ref{fig:vorh3}). In the case of [Si~VI] we forced the velocity shifts of the narrow component to be consistent with those of Pa$\alpha$. These constraints were necessary to obtain reliable measurements, since the [Si~VI] emission line is blended with H$_2$1-0S(3). For the same reason, we also constrained the width of the intermediate component to be $\ga$350 \kms~and the width of the broad component to be between $\sim$1400 and 3000 \kms. The velocity field of the intermediate component, shown in the third row of Fig.~\ref{fig:vorsi}, is very similar to that of Pa$\alpha$, showing negative velocities towards the east, and the highest fluxes and velocities to the south-east, of up to $\sim -$ 200 \kms. The FWHM is higher to the south. The maps of the broad component of [Si~VI] are different from those of the broad component of Pa$\alpha$, likely due to the lower S/N and the challenge that these fits represent. In Fig.~\ref{fig:vorh3} we show the maps of H$_2$1-0S(3), which are consistent with those of H$_2$1-0S(1) (see Fig.~\ref{fig:vorh1} in Section \ref{voronoi}). No restrictions on the parameters have been applied in this case.

\section{Pa$\alpha$ scan maps}
\label{Appendix C}

\begin{figure*}
\centering
\includegraphics[width=0.95\textwidth]{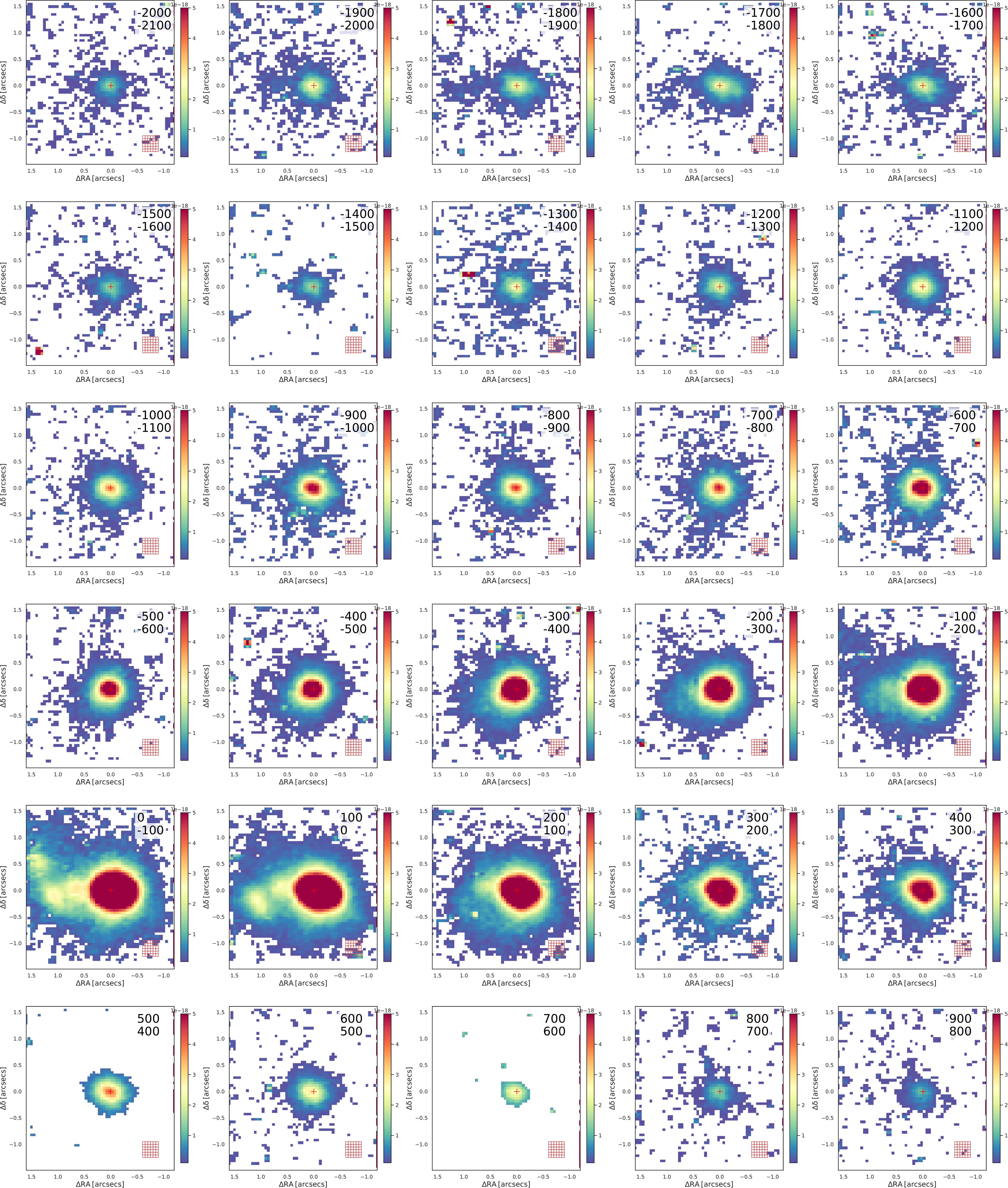}
\caption {Continuum-subtracted Pa$\alpha$ flux maps extracted in consecutive velocity steps of 100 \kms. We measure the standard deviation from the red square box at the bottom right corner of each panel, and we use that to show only emission above 3$\sigma$. The velocity interval used to construct the maps is indicated in the top right corner of each panel, in \kms. The peak of the nuclear Pa$\alpha$ emission is indicated with a red cross.}
\label{fig:cutmap}
\end{figure*}

In Fig.~\ref{fig:cutmap} we show the gas morphology of J0945 at different velocities obtained by integrating the Pa$\alpha$ emission in slices  of 100 \kms~from $-$2100 to 900 \kms, as shown in Fig.~\ref{fig:slice_100}. The maps between $-$2000 and $-$1500 \kms~and between $-$1500 and $-$700 \kms~show the gas around the peaks of He~II and He~I respectively. Instead, the maps between $-$700 and $-$300 \kms~correspond to the Pa$\alpha$ blue wing that is free of helium emission. The maps within this velocity range show that the gas is more elongated in the south-east direction, as in Fig. \ref{fig:outflow}. This is the same direction along which the Voronoi maps reach the highest velocities and widths (see third row of Fig.~\ref{fig:vorpa}). For velocities higher than $-$200 \kms~the gas morphology changes, and around the emission peak of Pa$\alpha$ (between $-$100 and 100 \kms) it becomes irregular, showing clumps eastwards of the nucleus, which is represented by a red cross in all the panels of Fig. \ref{fig:cutmap}. This disrupted gas is likely a product of the interaction/merger that J0945 is undergoing (see Section~\ref{morphology}). At larger veloties, on the red wing of the Pa$\alpha$ profile (from 200 \kms~onwards), the emission becomes compact.


\section{GTC/EMIR K-band spectrum}
\label{Appendix D}

J0945 was observed with the NIR multi-slit spectrograph EMIR (\citealt{Garzon06,Garzon14}) installed at the Naysmith-A focal station on the Gran Telescopio Canarias (GTC) at the Roque de los Muchachos Observatory, in La Palma. EMIR is a 2048x2048 Teledyne HAWAII-2 HgCdTe NIR-optimized chip with a pixel size of 0.2$\arcsec$. To perform the observations, the K-grism was used, covering a spectral range of 2.03-2.37 $\mu$m. The slit width was 0.8\arcsec.
The instrumental width measured from the spectra of the HgAr, Ne, and Xe lamps is $\sim$6.1 \AA~($\sim$ 87 \kms). The source was observed on 2018 December 19th with a total on-source time of 1920 s, following a nodding pattern ABBA (program ID: GTC62-18B, PI: Ramos Almeida). During the observation the airmass was 1.02 and the observing conditions spectroscopic.
The slit was oriented with a PA = 25$^{\circ}$ and it was positioned one arcmin to the left from the center of the detector, in order to expand the wavelength coverage to longer wavelengths.
The seeing measured from the average FWHM of 8 stars in the combined J-band acquisition image was $\sim$1.4\arcsec. 

The data were reduced using the \textit{lirisdr} software within the IRAF environment. Consecutive
pairs of AB two-dimensional spectra were subtracted to remove the sky background. Resulting frames were then wavelength-calibrated and flat-fielded before registering and co-adding all frames to provide the final spectra. The wavelength calibration was done using the spectra of the HgAr, Ne, and Xe lamps. We extracted a nuclear spectrum in an aperture of 1.4\arcsec, centred at the peak of continuum emission. This spectrum was then flux-calibrated and corrected from atmospheric transmission by using the spectrum of the A2 star BD+162159, observed immediately after J0945, and the IRAF task {\it telluric}.

\begin{figure*}
\centering
\includegraphics[width=0.485\textwidth]{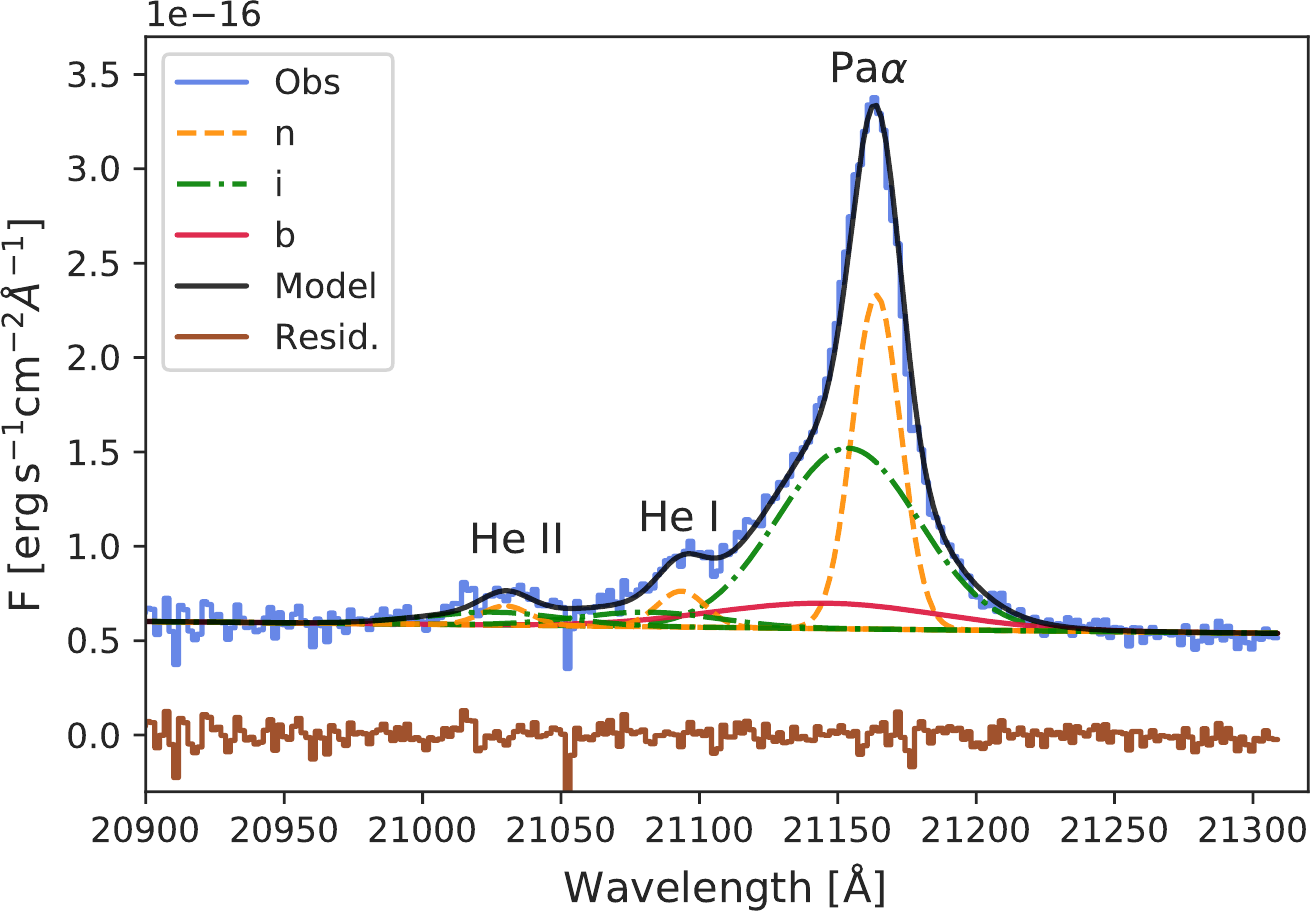}
\includegraphics[width=0.495\textwidth]{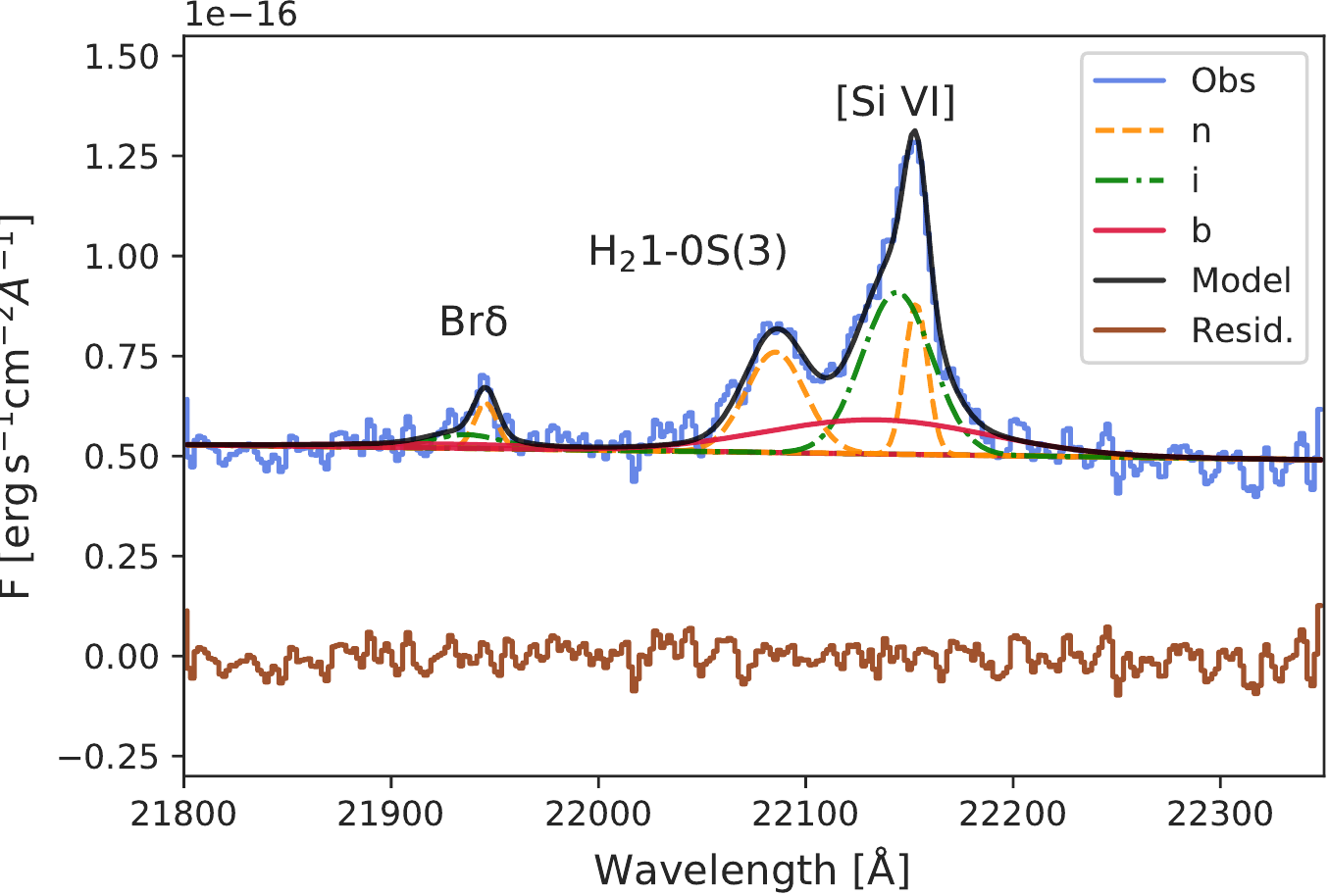}
\caption {Same as in Fig. \ref{fig:nuc}, but for the GTC/EMIR K-band nuclear spectrum, extracted in aperture of 1.4\arcsec. }
\label{fig:emir}
\end{figure*}

In Fig.~\ref{fig:emir} we show the fits of the Pa$\alpha$, [Si~VI], Br$\delta$ and H$_2$1-0S(3) emission lines detected in the EMIR spectrum. As in the case of the NIFS nuclear spectrum, three components are needed to characterize the profiles. The values corresponding to the parameters of the Gaussians used to model the profiles are shown in Table~\ref{tab:emir}. These values are comparable within the errors with the parameters obtained from the fit of the NIFS nuclear spectrum (see Table~\ref{tab:nuc}). We used the same constraints described in Section~\ref{nucleus} to obtain reliable fits, except for the Br$\delta$ line, whose parameters were forced to be match those of [Si~VI]. We note that the total on-source time of the EMIR spectrum is only 1920 s, whereas in the case of the NIFS data is 4000 s.


\begin{table}
\caption{Emission line properties measured from the GTC/EMIR K-band spectrum.}
\centering
\small
\begin{tabular}{lccc}
\hline
\hline
Line   &  FWHM  & $\Delta$v$_s$ & Flux $\times10^{-15}$  \\
      &  [km s$^{-1}$]   & [km s$^{-1}$]   &   [erg cm$^{-2}$s$^{-1}$]  \\
\hline
Pa$\alpha$ (n) & $290\pm16$     & $~~~~0\pm3$   & $~~3.85\pm0.32$ \\
Pa$\alpha$ (i) & $852\pm64$    & $-145\pm33$   & $~~6.12\pm0.83$  \\
Pa$\alpha$ (b) & $1360\pm333$ & $-260$ & $~~1.38\pm0.34$ \\
He~I        (n) & $290$ & $~~~~29\pm62$ & $~~0.41\pm0.12$ \\
He~I        (i) & $852$ & $~~-39\pm33$ & $~~0.49\pm0.43$ \\
He~II       (n) & $290$ & $~~~~67\pm52$ & $~~0.22\pm0.13$ \\
He~II       (i) & $852$ & $-104\pm33$ & $~~0.44\pm0.24$ \\
$[\text{Si~VI}] $ (n) & $176\pm75$    & $~~~~67\pm20$       & $~~0.52\pm0.27$  \\
$[\text{Si~VI}]$  (i) & $~~540\pm192$   & $~~-125\pm100$  & $~~1.72\pm0.81$ \\
$[\text{Si~VI}]$  (b) & $1717\pm740$  & $-260\pm29$  & $~~1.15\pm1.00$ \\
H$_2$1-0S(3)      (n) & $~~458\pm200$    & $~~~~65\pm54$    & $~~0.90\pm0.33$ \\
\hline
\end{tabular}
\tablefoot{Same columns description as in Table \ref{tab:sdss}.
}
\label{tab:emir} 
\end{table}

\newpage

\end{document}